\documentclass[aps,prb,twocolumn,groupedaddress,showpacs]{revtex4-1}

\usepackage{graphicx,color,url,hyperref}

\begin{document}

\title{
   (Sm,Zr)Fe$_{12-x}$M$_x$ (M=Zr,Ti,Co) for permanent-magnet applications:\\
   {\it Ab initio} material design integrated with experimental characterization
}

\author{
  Munehisa~Matsumoto, Takafumi~Hawai, Kanta~Ono
}
\affiliation{
  Institute of Materials Structure Science,
  High Energy Accelerator Research Organization (KEK), Oho 1-1, Tsukuba, Ibaraki 305-0801, Japan
}

\date{\today}

\begin{abstract}
  In rare-earth permanent magnets (REPM's),
  trade-off's between intrinsic magnetic properties are often encountered.
  A recent example is SmFe$_{12}$ where excellent magnetic properties
  can be achieved at the sacrifice of bulk structure stability.
  Bulk structure stability is sustained by the presence
  of the third substitute element as is the case with SmFe$_{11}$Ti, where Ti degrades magnetic properties.
  It is now in high demand to find out with which chemical composition
  a good compromise in the trade-off between structure stability and strong ferromagnetism is reached. We inspect the effects
  of representative substitute elements, Zr, Ti, and Co in SmFe$_{12}$
  by combining {\it ab initio} data with experimental data from neutron diffraction.
  The trend in the intrinsic properties
  with respect to the concentration
  of substitute elements are monitored and a systematic way to search the best
  compromise is constructed.
  A certain minimum amount of Ti is identified
  with respect to the added amount of Co and Zr.
  It is found that Zr brings about a positive effect on magnetization, in line with recent experimental developments,
  and we argue that this can be understood as an effective doping of extra electrons.
\end{abstract}


%
%
%
%
%

\maketitle

\section{Introduction}
\label{sec::intro}

\begin{figure}
  \scalebox{0.25}{\includegraphics{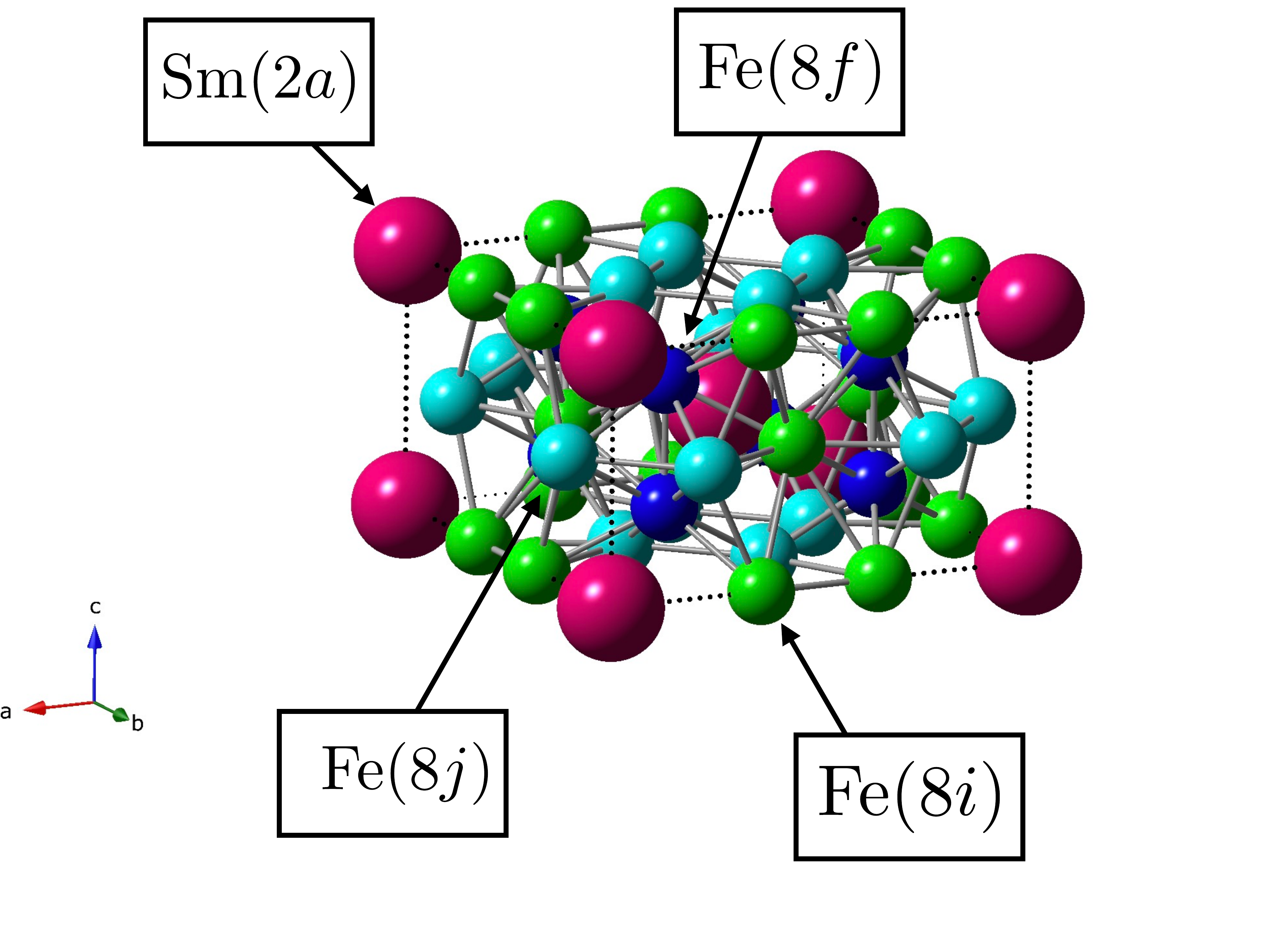}}
  \caption{\label{fig::lattice} (Color online)
    Crystal structure of SmFe$_{12}$. Large and red balls represent Sm($2a$) sites
    and green/cyan/blue and small
    balls represent Fe($8i$)/Fe($8j$)/Fe($8f$) sits, respectively. In the tetragonal box shown here,
    two formula units are included.
  }
\end{figure}

Rare-earth permanent magnet (REPM)
based on Nd$_2$Fe$_{14}$B~\cite{sagawa_1984,croat_1984,oesterreicher_1984,rmp_1991}
have been in commercial use in the past several decades. Nd-Fe-B ternary alloys
based on R$_2$Fe$_{14}$B (R=rare earth including Nd)
make excellent permanent magnets
except for a caveat on relatively low Curie temperature: the Curie temperature
of Nd$_2$Fe$_{14}$B is $585~\mbox{K}$, which is only
marginally beyond the typical high-temperature edge at $450~\mbox{K}$
of practical use in traction motors and power generators. Thus a way
to supplement the high-temperature properties has been pursued in various respects,
most notably addition of heavy-rare-earth elements to help high-temperature coercivity
via enhancing the high-temperature anisotropy field~\cite{hono_2012}.
Along the line of
searches for
alternative materials
with higher Curie temperature
or/and improved temperature coefficient of magnetization and anisotropy field,
NdFe$_{12}$N~\cite{miyake_2014,hirayama_2014,hirayama_2015}
and Sm(Fe,Co)$_{12}$~\cite{hirayama_2017}
recently
have triggered renewed interest
in ferromagnets with
the ThMn$_{12}$-type crystal structure as shown in Fig.~\ref{fig::lattice}.
It has been shown both theoretically~\cite{miyake_2014} and experimentally~\cite{hirayama_2014,hirayama_2017}
that the intrinsic magnetism of 100\% Fe-based ferromagnets with the ThMn$_{12}$ structure
may potentially be superior to Nd$_{2}$Fe$_{14}$B if the bulk structure stability
is guaranteed, which is so far achieved only on a special substrate in laboratory.
Originally it was only a few years after the discovery of Nd$_2$Fe$_{14}$B
that the material based on the ThMn$_{12}$ crystal structure had been found~\cite{ohashi_1987},
but the drawback that the particular ThMn$_{12}$ crystal structure is only metastable for RFe$_{12}$
(R=rare earth)
had hindered further developments. The problem is not yet entirely eliminated even today~\cite{scr_2018,hadjipanayis_2019},
but persistent efforts to bring the structure stability closer~\cite{sakurada_1992, sakurada_1996}
and recent renewed efforts~\cite{suzuki_2014,sakuma_2016,kuno_2016,hagiwara_2018,tozman_2018,tozman_2019}
together with the advent of various ways of data exploitation may change the perspective.
This work presents an attempt to combine theoretical and experimental data
to work with
the severe trade-off that
is almost always encountered
among the prerequisites for a ferromagnet to make a good
main phase of REPM, namely, strong magnetization, accordingly strong uni-axial magnetic anisotropy, high Curie temperature, and good structure stability, for the particular case of SmFe$_{12}$.

This paper is organized as follows.
In the next section we outline our methods that incorporate
both of experimental data and theoretical data. For the theory part,
the details of {\it ab initio} calculations
are given in Sec.~\ref{sec::calc_details}. Main results are shown
in Sec.~\ref{sec::results}: structure stability is inspected on the basis
of calculated
formation energy in Sec.~\ref{sec::results::deltaE}, and
one of the main messages therefrom
for the site preference of the substitute element Zr is confirmed
via data integration between experiment and theory in Sec.~\ref{sec::results::integration}.
Trends in the magnetic properties are inspected in Sec.~\ref{sec::results::mag}
and an optimal concentration in the middle of the trade-off between the structure
stability and the magnetic properties is identified.
In Sec.~\ref{sec::disc} we discuss the results in the light
of old and recent experimental findings.
Conclusions and outlook are described in Sec.~\ref{sec::conc}.
Details of calculations and data analyses are encapsulated in Appendix
so that the presentation can be followed straightforwardly
and also in a self-contained way.

\section{Methods and target materials}
\label{sec::methods}

\subsection{
  {\it Ab initio} inspection on the effects of substitute elements in SmFe$_{12}$}
Effects of substitute elements, Zr, Ti, and Co on pristine SmFe$_{12}$ are investigated
by calculating the formation energy and intrinsic magnetic properties from first principles.
Formation energy is obtained through {\it ab initio} structure optimization based
on generalized gradient approximation (GGA)~\cite{pbe_1996},
which is known to predict reasonable lattice parameters for most of the Fe-based ferromagnets.
Lattice parameters out of the structure optimization for SmFe$_{12}$ is summarized in Table~\ref{table::1-12}
together with counterpart numbers from previous
theoretical~\cite{harashima_2015} and experimental~\cite{hirayama_2017} works.
Calculated energy of SmFe$_{12}$ is subtracted by with the summation of calculated total energy
of the ingredient elements and we inspect how much formation energy is gained by the substitute elements.
Details of these {\it ab initio} calculations are given in Appendix~\ref{sec::calc_details}.
\begin{table}
  \begin{tabular}{ccccc} \hline
    & \multicolumn{3}{c}{Pristine} & Alloy \\ \hline
    & \multicolumn{2}{c}{Calc.} & Expt. & Calc.+Expt. \\ \hline
    &   this work & Ref.~\onlinecite{harashima_2015} & Ref.~\onlinecite{hirayama_2017} & this work\\ 
    &             &                            &                           & Zr($2a$) / Zr($8i$) \\ \hline
    $a$~(\AA) &   $8.569$ & $8.497$  & $8.35$  & $8.507$ \\ \hline
    $c$~(\AA) &   $4.735$ & $4.687$  & $4.81$  & $4.770$ \\ \hline
    $c/a$     &  $0.5525$ & $0.5516$ & $0.576$ & $0.5607$ \\ \hline
    $x_i$ &    $0.3590$ & $0.3588$ & & $0.3562$ / $0.3553$ \\ \hline
    $x_j$ &    $0.2712$ & $0.2696$ & & $0.2764$ / $0.2768$ \\ \hline
  \end{tabular}
  \caption{\label{table::1-12} Our {\it ab initio} lattice parameters
    from structure optimization for the pristine material SmFe$_{12}$,
    compared to a few previous works~\cite{harashima_2015,hirayama_2017},
    and our lattice parameters for SmFe$_{12}$ (Zr-substituted Sm(Fe,Co,Ti)$_{12}$)
    derived from the self-consistent analysis between {\it ab initio} calculations
    and Rietveld analysis of neutron diffraction data. The internal coordinates
    $x_i$ and $x_j$ are defined in Appendix.~\ref{sec::calc_details::str_opt}.
  }
\end{table}

\begin{figure}
  \begin{center}
  \scalebox{0.25}{\includegraphics{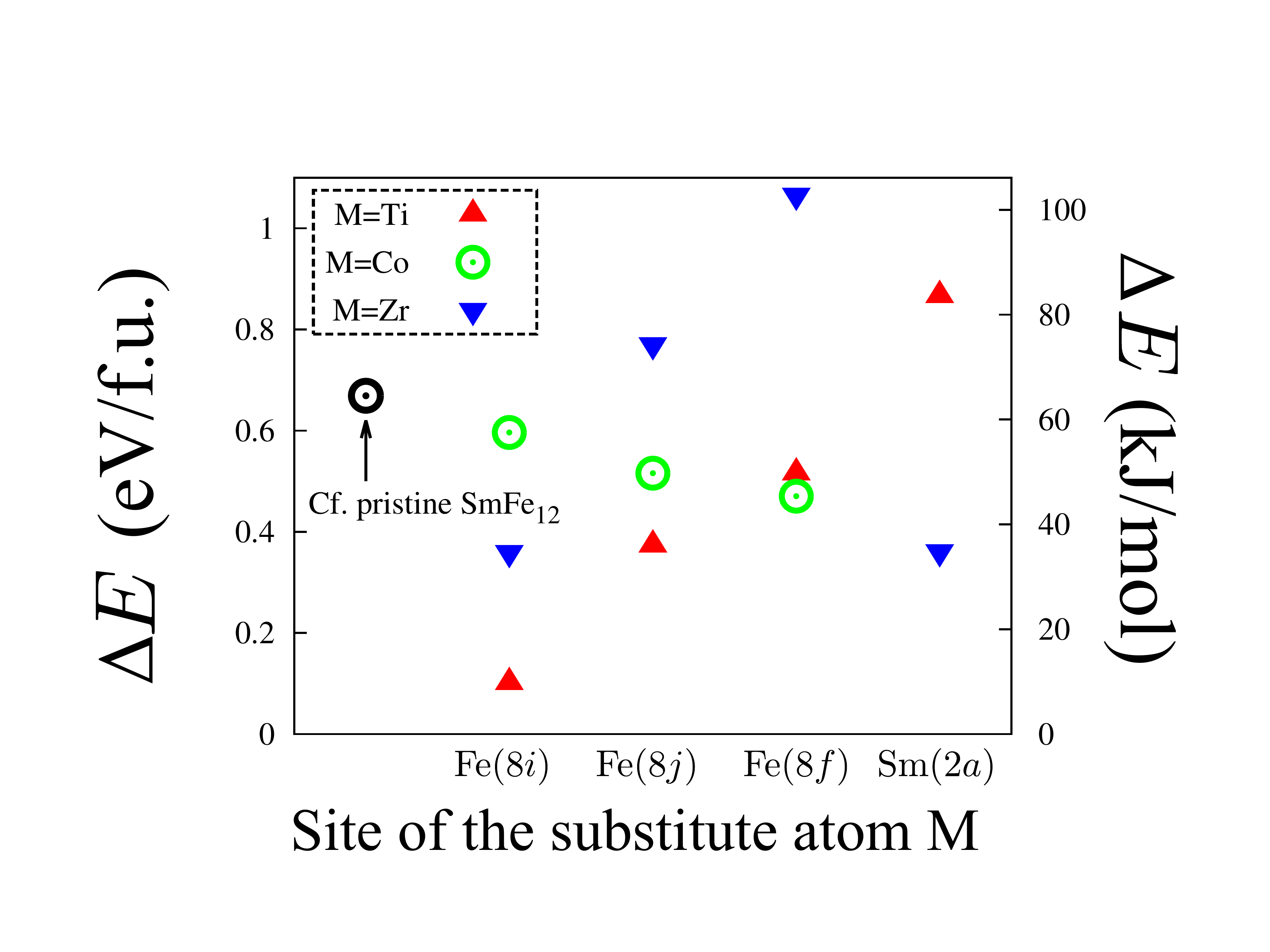}}
  \end{center}
  \caption{\label{fig::doped_SmFe12} (Color online)
    Calculated formation energy, $\Delta E$, of SmFe$_{12}$ with substitute elements Ti, Co, or Zr.
    In the tetragonal unit shown in Fig.~\ref{fig::lattice}
    containing two formula units of SmFe$_{12}$, one substitute atom, either Ti, Co, or Zr,
    replaces one host atom and {\it ab initio} structure optimization is done
    to extract the formation energy per formula unit (f.u.).}
\end{figure}
As has been shown in Fig.~\ref{fig::lattice}, the crystal structure of SmFe$_{12}$ is characterized
by multiple sublattices, namely, Sm($2a$), Fe($8i$), Fe($8j$), and Fe($8f$). {\it Ab initio} studies
showed the relative trends in magnetic moment $m[r]$ on site $r$
as $m[\mbox{Fe($8i$)}] > m[\mbox{Fe($8j$)}] > m[\mbox{Fe($8f$)}$] in RFe$_{12}$~\cite{miyake_2014}.
A guiding principle for the possible design of an optimal material would be
to keep the magnetization from Fe($8i$)
as much as possible while gaining structure stability, but
unfortunately the preference of the substitute Ti atom
goes for the Fe($8i$) site from $T=0$ all the way to higher temperatures~\cite{gino},
which we also confirm for $T=0$ as shown in Fig.~\ref{fig::doped_SmFe12}.
Thus the dominant magnetic moment from Fe($8i$)
is sacrificed while achieving the bulk structure stability.
Given this trade-off, control of the chemical composition toward a better compromise has been
pursued in the following way~\cite{sakurada_1992, sakurada_1996}:
the structure stability can be gained by Zr partly replacing Sm($2a$)
and thus the amount of Ti to stabilize the crystal structure might be able to be
reduced, leading to an improved magnetization. Now a question can arise
concerning the nature of substitute Zr, which should be chemically similar to Ti,
being on the same family on the periodic table of elements: how can the preference
of host sublattice be so drastically different between Ti and Zr?
Indeed our calculations of formation energy of Zr-substituted SmFe$_{12}$, exploring all possible sublattices
for the substitute Zr atom, show that Zr energetically favors Fe($8i$) site as well as Sm($2a$) site
as shown in Fig.~\ref{fig::doped_SmFe12}. In contrast,
recent investigations on (Sm,Zr)(Fe,Co,Ti)$_{12}$ are in progress
presuming
that Zr atom mostly replaces Sm($2a$)~\cite{tozman_2018}.
Precise understanding on the roles of Zr, Co, and Ti in SmFe$_{12}$
seems to be in acute need.

\subsection{Integration between {\it ab initio} data and Rietveld analysis
  of neutron diffraction experiment}
\begin{figure}
  \scalebox{0.25}{\includegraphics{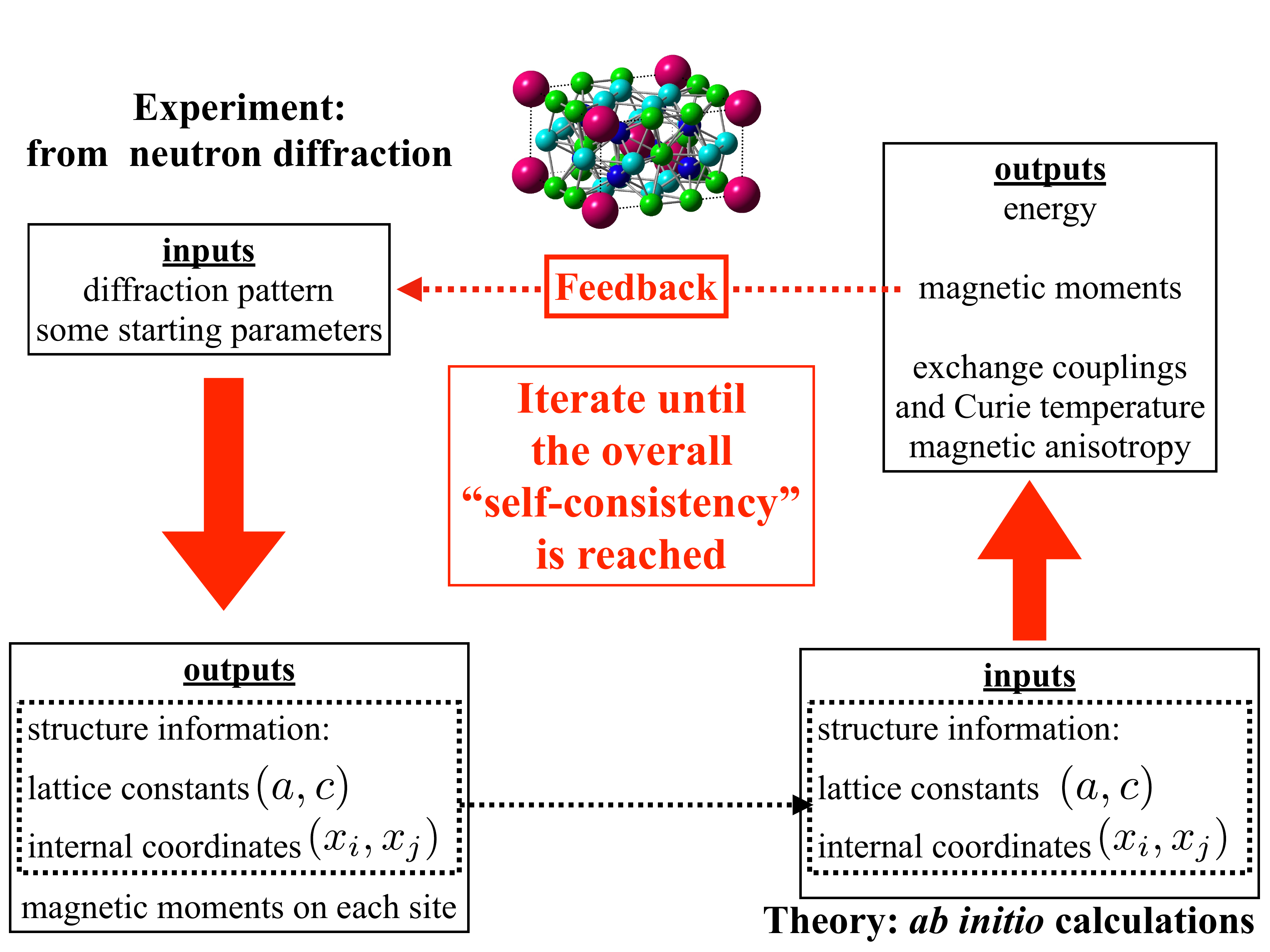}}
  \caption{\label{fig::scheme} (Color online)
    The scheme of the ``LDA+Rietveld'' self-consistent iterations.
    }
\end{figure}

\begin{figure}
  \scalebox{0.25}{\includegraphics{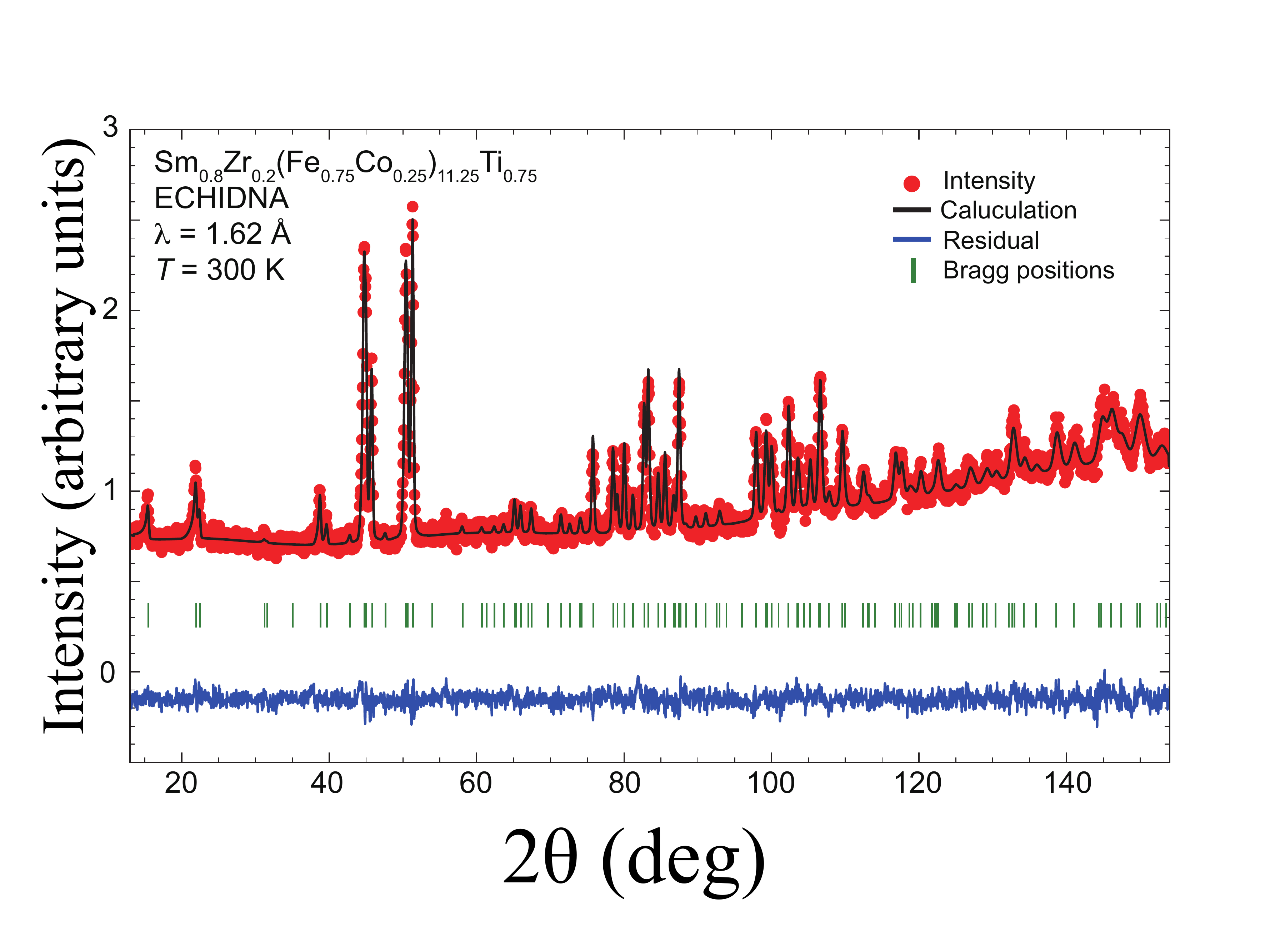}}
  \caption{\label{fig::rietveld} (Color online)
    The initial shot of the Rietveld analysis of which output makes an initial
    input to {\it ab initio} KKR-CPA
    in the overall ``LDA+Rietveld'' iteration.
    }
\end{figure}

In order to take a closer look into the experimental facts for Zr-substituted SmFe$_{12}$,
we combine {\it ab initio} inputs and outputs
with our experimental data from neutron diffraction.
Here the powder sample of
Sm$_{0.8}$Zr$_{0.2}$(Fe$_{0.75}$Co$_{0.25}$)$_{11.25}$Ti$_{0.75}$
was provided by Toyota Motor Corporation
and the powder neutron diffraction measurements
at room temperature were performed on ECHIDNA
at Australian Nuclear Science and Technology Organisation (ANSTO)~\cite{avdeev_2018}.
Rietveld analysis of diffraction data~\cite{rietveld} gives the lattice constants
and the internal coordinates of Fe($8i$) and Fe($8j$) in the unit cell. These are
plugged into {\it ab initio} calculations using coherent potential approximation (CPA)~\cite{shiba_akai}
based on local density approximation (LDA) following Vosko, Wilk, and Nusair~\cite{vosko_1980}
which is known to give reasonable magnetic moments on a given crystal structure.
We make a few steps further: calculated magnetic moment on each atom is fed back
into the Rietveld analysis of diffraction data
to obtain the refined input data consisting of lattice constants and internal coordinates~\cite{ito_2018}.
We observe that this overall self-consistent iteration loop between Rietveld-analysis of experimental data
{\it ab initio} calculations converges in quite a fast and robust way.
We would tentatively
refer to this particular combination of theory and experiment as ``LDA+Rietveld''. Detailed data
during the overall iteration procedure can be
found in the Appendix~\ref{sec::integrated_data_details}.
Combination of the Rietveld analysis
with {\it ab initio} calculation to verify the structure stability has been widely
done recently~\cite{ishikawa,todo}, while the way to reinforce the convergence of the data
via the feedback between theory and experiment is new to the best of the authors' knowledge. Here
the scope of the problem imposed on the Rietveld analysis is slightly different from the conventional one
for the structure analysis: out lattice structure space is limited, restricting ourselves on the given prototype
of ThMn$_{12}$ structure, while the details of the sublattice-resolution in the multiple-sublattice ferromagnetism is
the present problem. In such a restricted working space, the feedback between theory and experiment
can be implemented directly and easily.
The scheme for this type of self-consistent iteration is shown in Fig.~\ref{fig::scheme}
and the initial shot of the Rietveld analysis is shown in Fig.~\ref{fig::rietveld}.

For the assessment of magnetic anisotropy in the typical working temperature range,
one of the most influential factors is actually an indirect exchange coupling
between rare earth atom and Fe atom~\cite{mm_2016}. We use Sm($5d$)-Fe($3d$)
exchange coupling as a key descriptor for the leading-order
of the finite-temperature
anisotropy field which intrinsically controls
the finite-temperature
coercivity. Thus we actually do
without spin-orbit interaction in our {\it ab initio} calculations. Even though
we do not directly address the uni-axial magnetic anisotropy, inspection of linear
trends in the leading order contribution
around a reasonable limit at the pristine SmFe$_{12}$ would do.
This simplifies the theory part and enable a wide coverage of parameter space
spanned by the chemical composition. Our target observables consist of
the formation energy, magnetization, and inter-atomic exchange couplings
out of which Fe-Fe couplings control the Curie temperature and Fe-Sm couplings control
the room-temperature anisotropy field.

\section{Results}
\label{sec::results}

\begin{figure}
  \scalebox{0.25}{\includegraphics{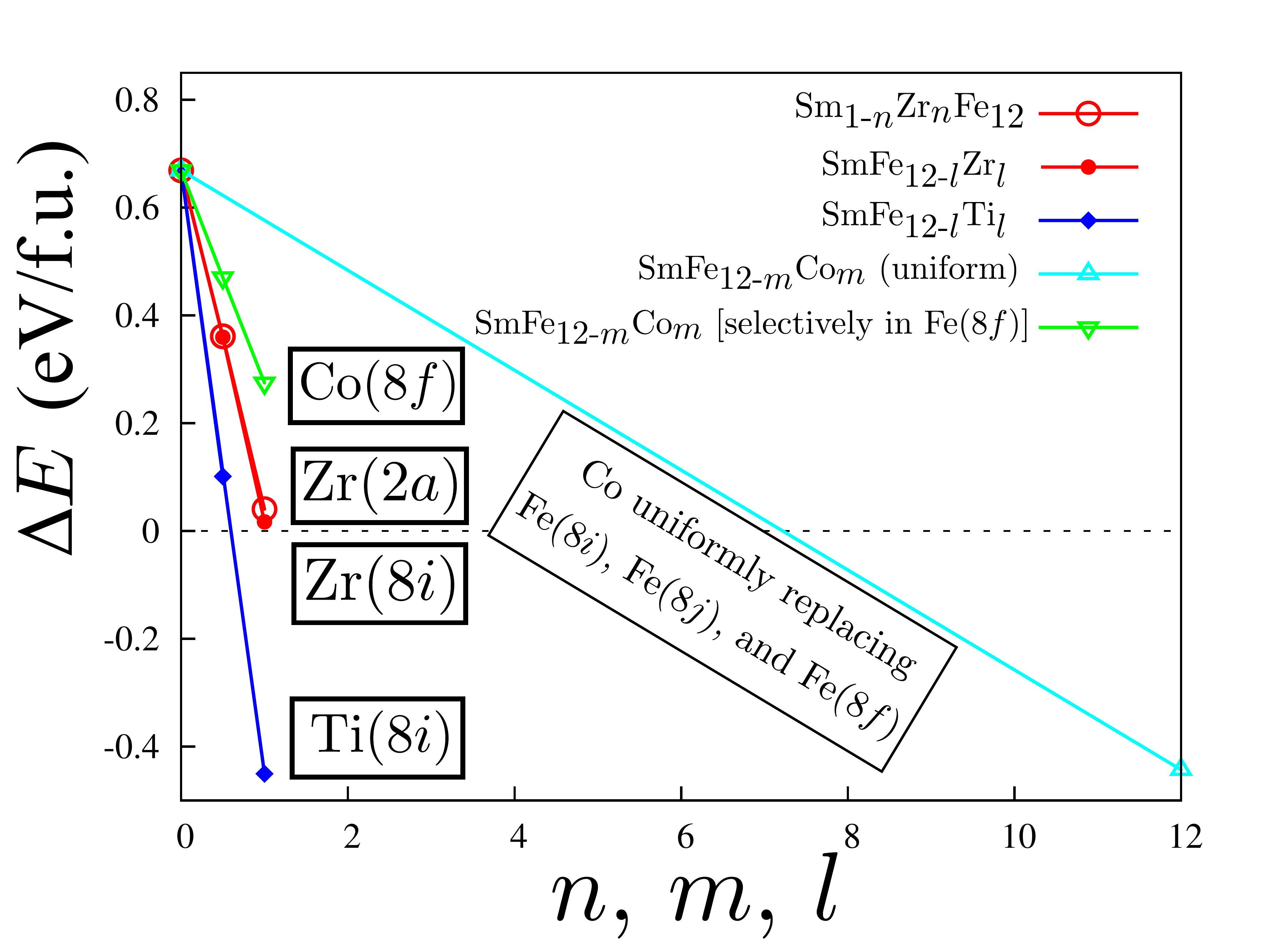}}
  \caption{\label{fig::doped_SmFe12_delta_E_trends} (Color online)
    Calculated formation energy of SmFe$_{12}$
    per formula unit (f.u.)
    as a function of the concentration of substitute elements. In these calculations,
    the substitute elements are selectively put
    into the most energetically favorable sublattice,
    unless otherwise stated.
    }
\end{figure}
\begin{table}
  \begin{tabular}{cc} \hline
    substitute &  $U_{\Delta E}\equiv(-\partial\Delta E)/\partial n_{\rm M}$ \\ 
    & (eV/f.u.)  \\ \hline
    %
    Zr in Sm($2a$) & $0.63$  \\ \hline
    Zr in Fe($8i$) & $0.62$  \\ \hline
    Ti in Fe($8i$) & $1.13$  \\ \hline
    Co (uniformly substituting) & $0.093$  \\ \hline
    Co in Fe($8f$) & $0.4$  \\ \hline
  \end{tabular}
  \caption{\label{table::coeff_delta_E} Derivative of the formation energy
    of SmFe$_{12}$
    per formula unit (f.u.)
    around the pristine limit with respect the
    number
    of substitute elements per formula unit.}
\end{table}
\subsection{Calculated formation energy of SmFe$_{12}$}
\label{sec::results::deltaE}
We inspect the influence of each substitute element, Ti, Co, and Zr, in SmFe$_{12}$
on the structure stability
based on the derivative of calculated formation energy $\Delta E$ with respect to
the concentration of substitute elements and see how the amount of Ti can be reduced
with the possible help from Zr and Co to be on a par with SmFe$_{11}$Ti
concerning the structure stability.
Negative and large absolute value of the formation energy
is beneficial. The exact formula and values of calculated energy
are summarized in Appendix~\ref{sec::calc_details::str_opt}.
The results for SmFe$_{12}$
have been shown in Fig.~\ref{fig::doped_SmFe12}.
Observing the minimum energy among the multiple choices for the substitute elements,
the preference of substituting Ti goes for Fe($8i$), Co does for Fe($8f$),
both in agreement with the claims of past works~\cite{gino,harashima_2016},
while the conclusion for Zr from this data set
may not be entirely consistent with recent experimental developments for SmFe$_{12}$:
Zr in Fe($8i$) and Zr in Sm($2a$) look almost degenerate energetically in our data
while Zr substituting mostly Sm($2a$) is seen in the literature~\cite{tozman_2018}.
Formation energy is summarized
as a function of concentration as shown in Fig.~\ref{fig::doped_SmFe12_delta_E_trends}.
Utility of a substitute element M may be characterized by a differential coefficient, $(-\partial\Delta E/\partial n_{\rm M})$,
where $n_{\rm M}$ is the number
of substitute element M per formula unit.
From the data shown in Fig.~\ref{fig::doped_SmFe12_delta_E_trends}
the coefficients are extracted as shown in Table.~\ref{table::coeff_delta_E}.
Ti indeed works for the structure stability most effectively.
For our target compound
(Sm$_{1-n}$Zr$_n$)(Fe$_{12-m-l-l'}$Co$_m$Ti$_l$Zr$_{l'}$), where $n$ is the number of Zr atoms
substituting Sm($2a$), $m$ is the number of substituting Co atoms per formula unit,
$l$ is the number of substituting Ti atoms in the Fe($8i$) sublattice per formula unit,
and $l'$ is the number of substituting Zr atoms in the Fe($8i$) sublattice per formula unit,
to be better or on a par with SmFe$_{11}$Ti
concerning the formation energy, the following condition must be met:
\begin{eqnarray*}
  && n U_{\Delta E}(\mbox{Zr($2a$)})
  + m U_{\Delta E}(\mbox{Co})+ l U_{\Delta E}(\mbox{Ti}) \\
  && + l' U_{\Delta E}(\mbox{Zr($8i$)})\\
  & \stackrel{>}{\sim} & (1/12)U_{\Delta E}(\mbox{Ti}).
\end{eqnarray*}
Assuming uniform substitution by Co considering
the relatively minor preference of Co substituting in SmFe$_{12}$,
the following relation is imposed.
\begin{equation}
0.63 n + 0.093 m + 1.13 l + 0.62 l' \stackrel{>}{\sim} 1.13
\label{eq::stability_condition}
\end{equation}
With our sample
having 0.2 Zr atoms, thus the relation
$n+l'=0.2$ is imposed, and
  $\stackrel{<}{\sim}25\% $ of Co atoms
per formula unit,
we end up with the following condition,
\begin{equation}
  0.279
  + 0.63n+0.62(0.2-n)\stackrel{>}{\sim} 1.13(1-l),
\end{equation}
which gives us the lower bound on $l$
as $0.643-0.009n$ ($0\le n\le 0.2$):
almost independently of
how Zr atoms are distributed over Sm($2a$) or Fe($8i$) sublattices,
the minimum amount of Ti can be reduced to the smaller number than SmFe$_{11}$Ti by 35\%.
Thus we quantitatively confirm that Zr in Sm($2a$), partly also in Fe($8i$),
and uniformly substituting Co indeed enables the reduction of Ti
concerning the formation energy.

\subsection{LDA+Rietveld results for Zr-substituted Sm(Fe,Co,Ti)$_{12}$}
\label{sec::results::integration}
Having confirmed the assumed utility of Zr for the structure stability from first principles,
we inspect the real sample of Sm$_{0.8}$Zr$_{0.2}$(Fe$_{0.75}$Co$_{0.25}$)$_{11.25}$Ti$_{0.75}$
to see how the calculated
site preference may be reflected in experimental reality.
Rietveld analysis is performed to extract the localized magnetic moments,
concentration distribution of component elements occupying the same sublattice, 
internal coordinates, and lattice constants.
Out of these outputs from the experimental data analysis,
the lattice structure information typically make the inputs to {\it ab initio} calculations
in order to do the simulations of the complicated multiple-sublattice material as realistically as possible.
{\it Ab initio} structure optimization does illustrate the relative trends but it does
not always pin-point the quantitative
results from experimental measurements, as is seen in Table~\ref{table::1-12}.
Detailed and reliable inputs for the internal coordinates as well as the lattice constants, such as reported
in Ref.~\onlinecite{moze_1988} for YFe$_{11}$Ti, have been in great demand to address the subtle interplay among
various contributions from different sublattices and the trade-off between prerequisite properties
for multiple-sublattice magnets.

Given the structural inputs, {\it ab initio} calculations
yield electronic structure including magnetic moment on each atom:
these outputs can now be recycled as a renewed input
to the Rietveld analysis so that the solution of the inverse problem to decode the neutron
diffraction pattern would be more robust~\cite{ito_2018}. We have iterated such feedback process
from {\it ab initio} outputs to Rietveld analysis in the next step
until self-consistency is reached, now following
Korringa-Kohn-Rostoker (KKR)
Green's function method combined with coherent potential approximation (CPA)
for alloys~\cite{shiba_akai}.
The details of the computational
setup and the specific way the overall iteration proceeds are described
in Appendix~\ref{sec::integrated_data_details}.
Remarkably, the convergence down to 5-6 digits is achieved within a few iteration steps,
counting a set of Rietveld analysis and {\it ab initio} calculation as one step.
Here the self-consistency is identified by reaching a fixed point of the iterative loop
where the input and output information become identical within the numerical precision.

As long as the calculated energy within KKR-CPA is concerned, the energy of the electronic structure
is not minimized as the consistency between theory and experiment is approached
as seen in Table.~\ref{table::results} in Sec.~\ref{sec::integrated_data_details} below.
Systematic deviation between experiment and theory concerning the lattice constant
is reasonable considering the fact the experiments are done at room temperature
while {\it ab initio} calculations are for zero temperature~\cite{andreev_1995}. The difference
of working temperatures between theory and experiment
should not be a problem as long as the intrinsic Curie temperature
is high enough to render the room-temperature properties
close to the ground state. It is to be noted that
the LDA+Rietveld analysis at this stage is not a variational framework
with respect to the intrinsic energy of the electronic state
but a generalized data fitting technique assisted by the guideline data provided from first principles.

We observe that either case of Sm($2a$) taking Zr or Fe($8i$) taking Zr is equally plausible,
in line with the site preference inspected with
the {\it ab initio} structure optimization. The resultant lattice constants
have been summarized in Table~\ref{table::1-12}. The detailed quantitative
distribution of Zr over Sm($2a$) and Fe($8i$) has not been entirely determined here
due to a problem with Zr in the Rietveld analysis of the neutron diffraction data
in that magnetic moment on Zr may be too small.
At least
we have seen that
there should be
some finite contribution from Zr atoms substituting the Fe($8i$) sublattice.

\subsection{Optimal chemical composition with SmFe$_{12}$}
\label{sec::results::mag}
Now the prerequisite magnetic properties for REPM
are inspected.
The observables are magnetization, Curie temperature as obtained on the basis of
mean-field approximation for calculated exchange couplings between $d$-electrons,
and the $5d$-$3d$ exchange couplings that indirectly
binds $4f$ and $3d$ electrons together as a key measure
for finite-temperature magnetic anisotropy~\cite{mm_2016}. The last observable
is denoted as ``$J_{\rm RT}$'', emphasizing that this is the coupling between
rare-earth atom (R) and transition-metal atom (T). For SmFe$_{12}$ there are three sublattices
for Fe and accordingly $J_{\rm RT}$ has three variants $J_{\rm RT}(8i)$, $J_{\rm RT}(8j)$,
and $J_{\rm RT}(8f)$.
Calculated results of them for doped SmFe$_{12}$,
here we denote as Sm$_{1-n}$Zr$_{n}$(Fe$_{12-m-l-l'}$Co$_{m}$Ti$_{l}$Zr$_{l'}$)
for each of the focus elements: Ti, Co, and Zr, are shown in
Appendix~\ref{sec::cpa_details}.

\begin{widetext}
The partial derivative coefficients of the target observables
$M$, $T_{\rm Curie}$, $J_{\rm RT}$ for Fe($8i$), Fe($8j$), and Fe($8f$),
with respect to the number of the substitute elements around the pristine limit,
$n$, $m$, $l$, and $l'$,
per formula unit,
SmFe$_{12}$~\cite{harashima_2015},
can be summarized as follows. The detailed derivation of this working matrix
is given in Table~\ref{table::derivative} in Appendix.
\begin{eqnarray}
&& \left(
\begin{array}{cccc}
 (\partial M/\partial n)/M &  (\partial M/\partial m)/M &  (\partial M/\partial l)/M &  (\partial M/\partial l')/M\\
  (\partial T_{\rm Curie}/\partial n)/T_{\rm Curie}  & (\partial T_{\rm Curie}/\partial m)/T_{\rm Curie} &
  (\partial T_{\rm Curie}/\partial l)/T_{\rm Curie}  & (\partial T_{\rm Curie}/\partial l')/T_{\rm Curie}\\
  \{\partial J_{\rm RT}(8f)/\partial n\}/J_{\rm RT}(8f) & \{\partial J_{\rm RT}(8f)/\partial m\}/J_{\rm RT}(8f)
  & \{\partial J_{\rm RT}(8f)/\partial l\}/J_{\rm RT}(8f) & \{\partial J_{\rm RT}(8f)/\partial l'\}/J_{\rm RT}(8f) \\
  \{\partial J_{\rm RT}(8i)/\partial n\}/J_{\rm RT}(8i) & \{\partial J_{\rm RT}(8i)/\partial m\}/J_{\rm RT}(8i)
  & \{\partial J_{\rm RT}(8i)/\partial l\}/J_{\rm RT}(8i) & \{\partial J_{\rm RT}(8i)/\partial l'\}/J_{\rm RT}(8i) \\
  \{\partial J_{\rm RT}(8j)/\partial n\}/J_{\rm RT}(8j) & \{\partial J_{\rm RT}(8j)/\partial m\}/J_{\rm RT}(8j) &
  \{\partial J_{\rm RT}(8j)/\partial l\}/J_{\rm RT}(8j) & \{\partial J_{\rm RT}(8j)/\partial l'\}/J_{\rm RT}(8j)
\end{array}
\right) \nonumber \\ 
 & = & \left(
\begin{array}{cccc}
0.0244238 & 0.0302537 & -0.189388  & -0.177486 \\
0.196759  & 0.234125  & -0.0047857 & -0.0697258 \\
0.0955673 & 0.0831466 & -0.0480785 & -0.0926378 \\
0.116806  & 0.0804769 & 0.0804248  & 0.0335155 \\
0.135327  & 0.0614006 & -0.036302  & -0.0497742
\end{array}
\right) \label{eq::dev_mat}
\end{eqnarray}
Each element in the above derivative matrix has been normalized
by the absolute values at the pristine limit.
\end{widetext}

The derivative matrix in Eq.~(\ref{eq::dev_mat})
shows that Zr atom substituting
the Sm($2a$) sublattice and Co atom substituting uniformly
the overall Fe sublattice works almost on a par, positively
for the intrinsic magnetic properties. The positive effect of Zr
seems to be consistent with the recent experimental observation~\cite{tozman_2019}
and the microscopic mechanism is discussed
below in Sec.~\ref{sec::disc::zr}.

On the other hand, Ti or Zr substituting
the Fe($8i$) sublattice is detrimental to almost all properties.
The only positive effect of Ti($8i$) and  Zr($8i$) is seen
on the Fe($8i$)-Sm($2a$) exchange coupling, which is the nearest-neighbor exchange path
between Sm and Fe.

\begin{figure}
  \begin{tabular}{l}
    (a)\\
    \scalebox{0.25}{\includegraphics{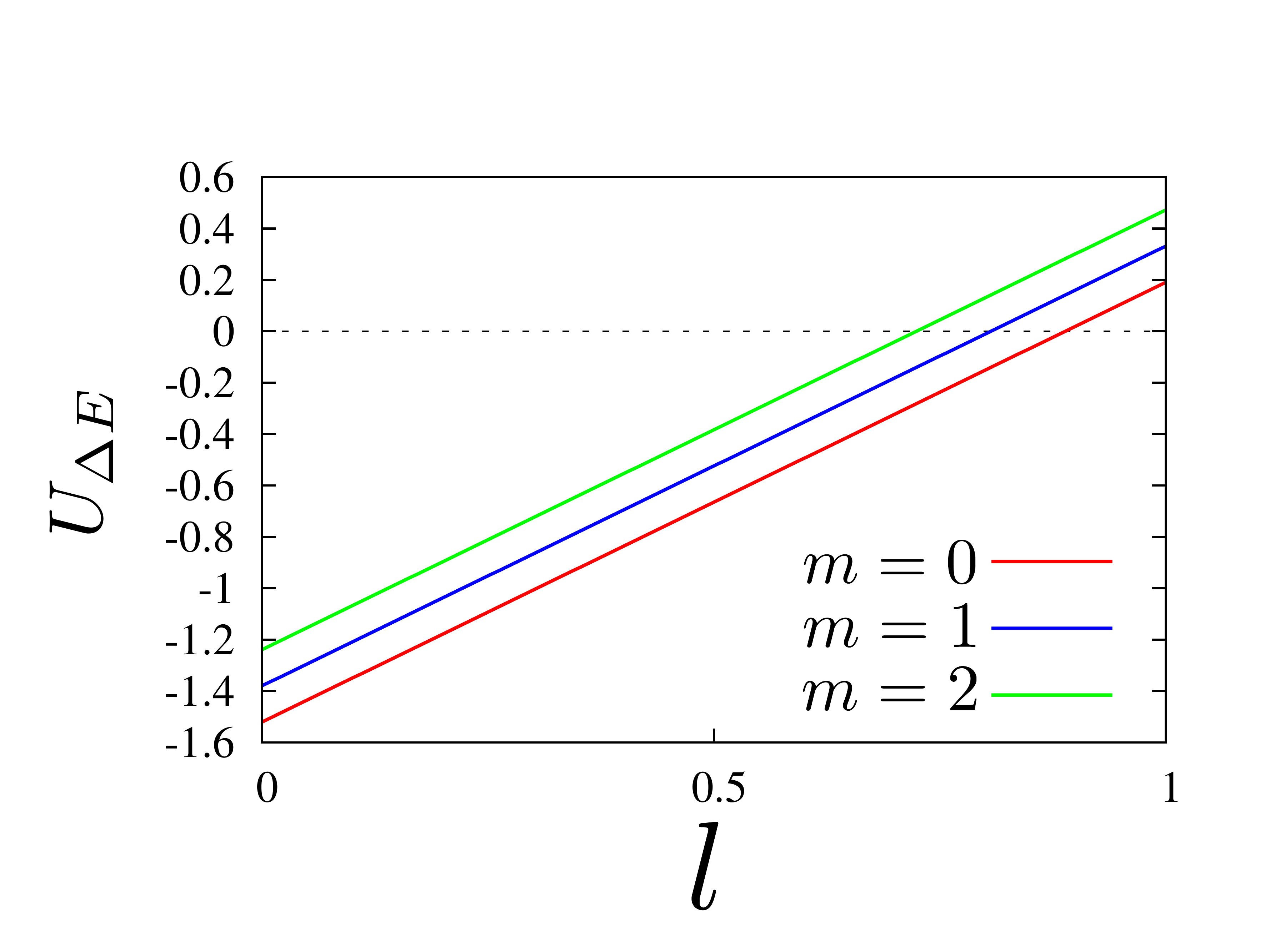}} \\
    (b) \\
    \scalebox{0.25}{\includegraphics{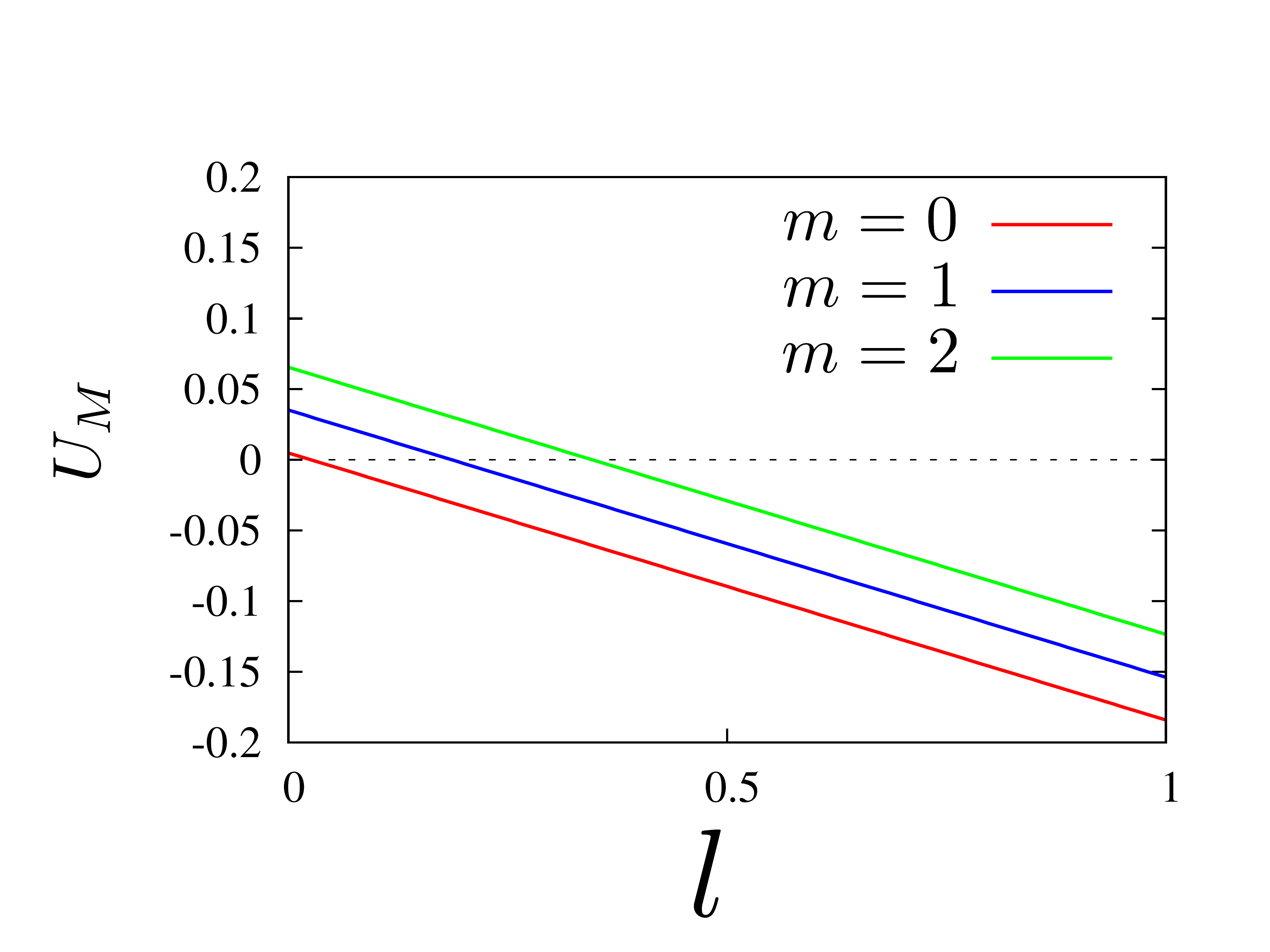}}
  \end{tabular}
  \caption{\label{fig::utilities} (Color online)
    Utility of the target material
    Sm$_{1-n}$Zr$_n$Fe$_{12-m-l-l'}$Co$_{m}$Ti$_{l}$Zr$_{l'}$ as a function of $l$, with $n=0.2$ and $l'=0$ fixed.}
\end{figure}

\begin{figure}
  \scalebox{0.25}{\includegraphics{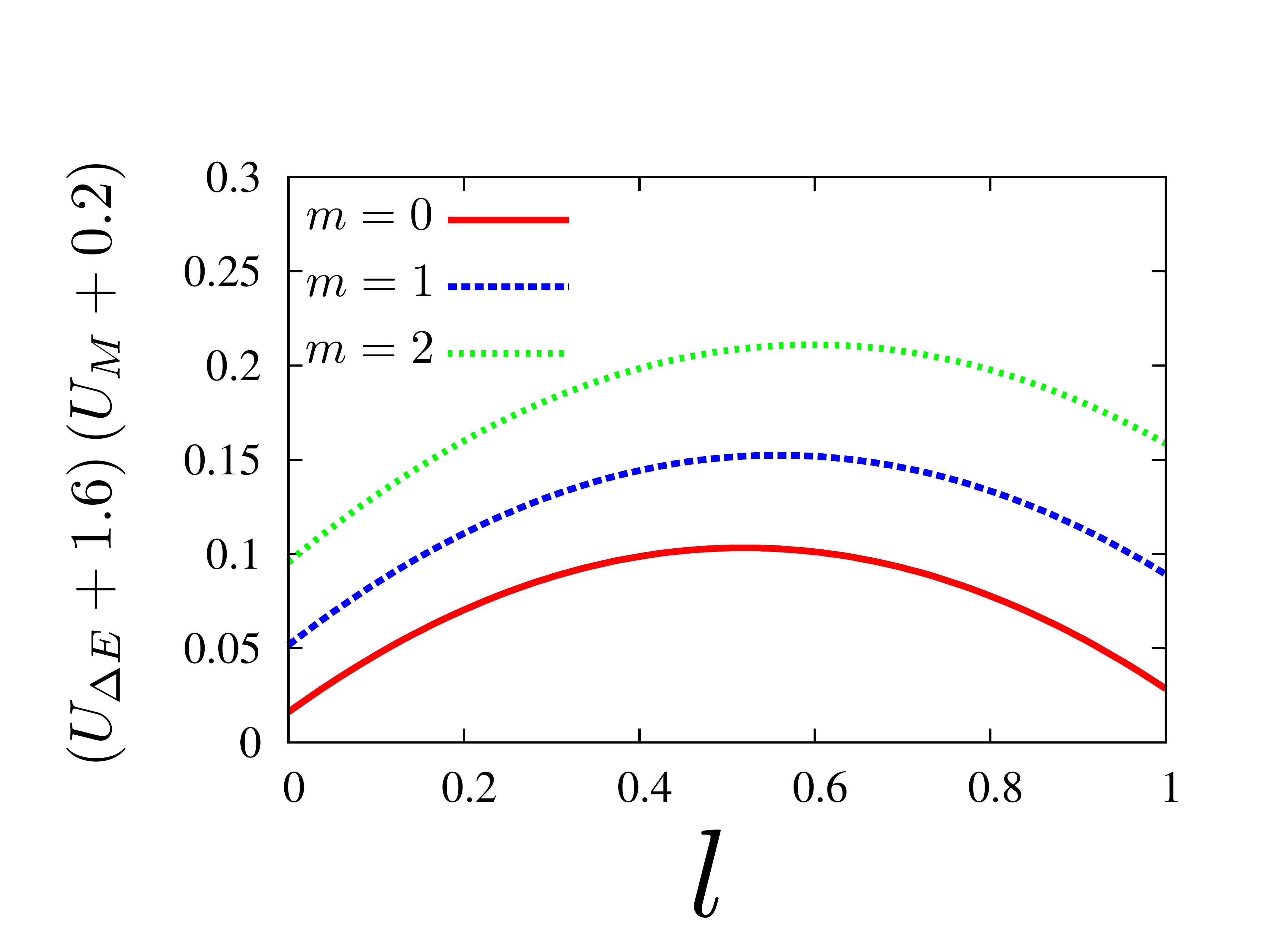}}
  \caption{\label{fig::compromise0}
    (Color online)
    Compromise between structure stability and magnetization
    inspected with the partial utility functions
    of the target material Sm$_{1-n}$Zr$_n$Fe$_{12-m-l-l'}$Co$_{m}$Ti$_{l}$Zr$_{l'}$
    as a function of $l$, with $n=0.2$ and $l'=0$ fixed.}
\end{figure}

With the quantified
effects of substitute elements Ti, Co, and Zr in SmFe$_{12}$,
we put forward a formulation of the trade-off problem described
in Sec.~\ref{sec::intro}.
The target material Sm$_{1-n}$Zr$_n$Fe$_{12-m-l-l'}$Co$_{m}$Ti$_{l}$Zr$_{l'}$
can be characterized by an overall target
function $U_{\rm all}(\{
U_{\Delta E},U_{M},U_{T_{\rm c}},U_{J_{\rm RT}(8f)},U_{J_{\rm RT}(8i)},U_{J_{\rm RT}(8j)}\})$
that is to be defined in terms of partial target function for each
of the observables
that can be defined referring to Eqs.~(\ref{eq::stability_condition})~and~(\ref{eq::dev_mat}):
\begin{eqnarray*}
  U_{\Delta E} & \equiv & (0.63 n + 0.093 m + 1.13l + 0.62l'-1.13)/(\Delta E_{0}) \\
  U_{M} & \equiv & 0.0244n+ 0.0303m -0.189l -0.177l' \\
  U_{T_{\rm c}} & \equiv & 0.197n +  0.234m -0.00479l -0.0697l' \\
  U_{J_{\rm RT}(8f)} & \equiv & 0.0956n + 0.0831m -0.0481l  -0.0926l' \\
  U_{J_{\rm RT}(8i)} & \equiv & 0.117n  + 0.0805m + 0.0804l + 0.0335l' \\
  U_{J_{\rm RT}(8j)} & \equiv & 0.135n  + 0.0614m -0.0363l  -0.0498l'.
  \end{eqnarray*}
Here $\Delta E_{0}$ is a reference formation energy which
we take as the number at the pristine limit
shown in Fig.~\ref{fig::doped_SmFe12}.
We note that the expansion around the pristine limit should be
used only within the linear regime in the overall
dependence on the concentration / number of substitute elements
of which overview is shown in Fig.~\ref{fig::deriv} in Appendix.
Motivated by our sample used in the neutron diffraction experiment,
let us fix $n+l'=0.2$ as a working cross
section in the multi-dimensional composition space.
Then
the trade-off problem within this scope can be illustrated
by comparing
the plots, $Z_{\Delta E}(l)$ and $Z_{M}(l)$, as shown in Fig.~\ref{fig::utilities}.
The conflicting trends between formation energy and magnetization
reaches a compromise in the middle around $0.5 \le l \le 0.6$
as seen with the product $\tilde{U}_{\Delta E}(l)\tilde{U}_{M}(l)$
shown in Fig~\ref{fig::compromise0}.
In $\tilde{U}_{\Delta E}\equiv U_{\Delta E}(l)+1.6$ and $\tilde{U}_{M}\equiv U_{M}(l)+0.2$,
an offset number has been incorporated
into each target function so that all factors would be positive definite
and a place
to control external preference for prioritizing some of the observables can be reserved.
The same procedure is applied to the target function whenever negative sign appears
in the working space.
Presence of such constants in order to have positive
definite weights can be justified
when the main interest lies in the relative trends among the target functions.
There might be a better way
to define the derivative coefficient
in such a way that
it is always positive definite
by construction and
introduce the external factors separately.
At this stage the overall trend
toward the optimal point is reasonable,
in the sense that a combination of a increasing merit
and a decreasing merit ends up with a ¡Èbest compromise¡É in between.
The procedure here is partly
in an analogy to adding a constant to the diagonal elements of density matrices involved
in quantum Monte Carlo methods~\cite{sandvik} in order to define positive definite weights.
Thus redefined target function is denoted by $\tilde{U}$.
We note that the best compromise always comes with more than
0.5 Ti atoms in the currently working cross section,
in agreement with the past experimental findings~\cite{suzuki_2014, suzuki_2016, kuno_2016}.

\begin{figure}
  \scalebox{0.25}{\includegraphics{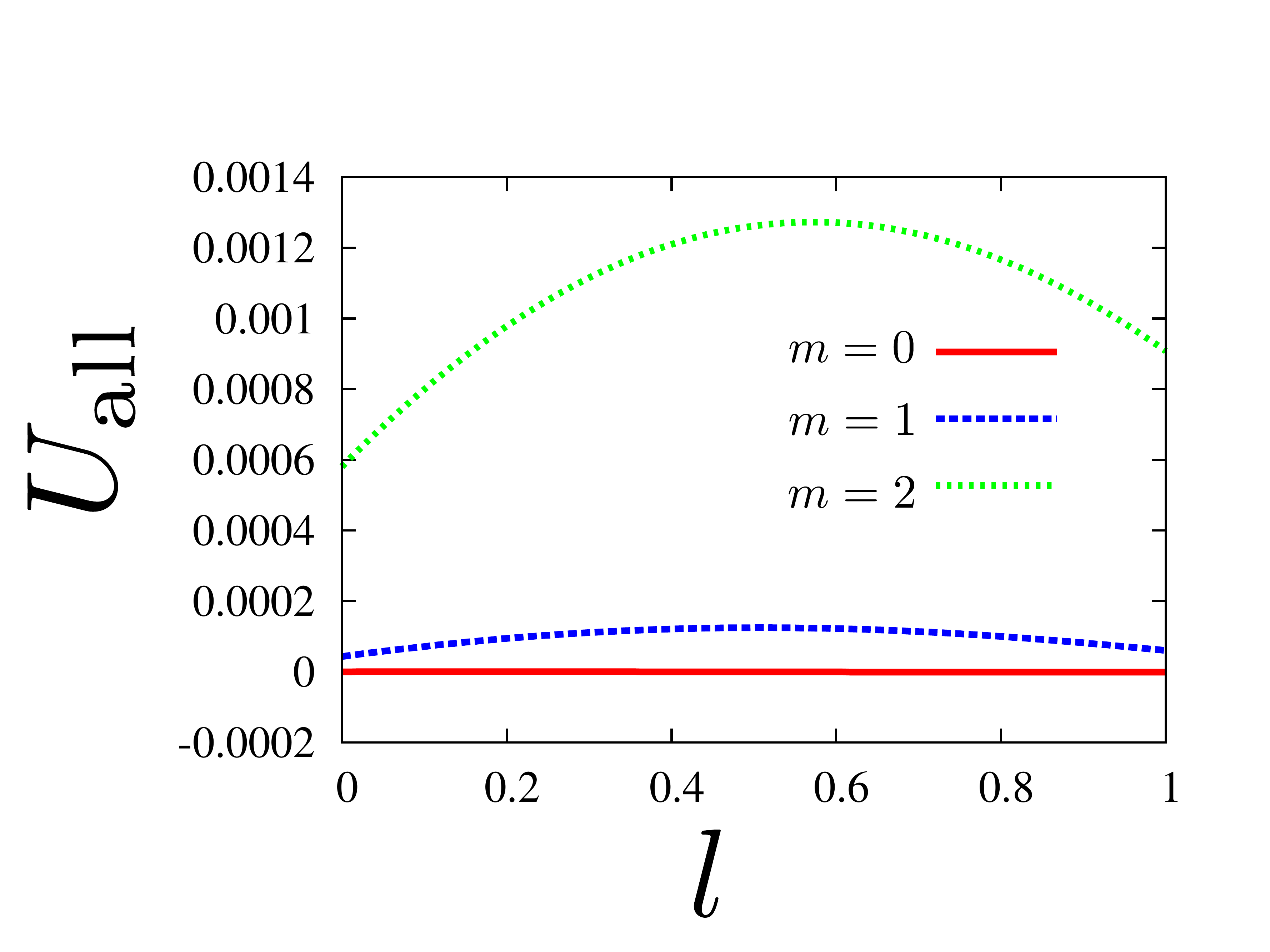}}
  \caption{\label{fig::compromise}
    (Color online)
    Compromise between structure stability, magnetization, Curie temperature,
    and room-temperature anisotropy field of which leading-order
    trend may be well captured by Sm-Fe exchange couplings.
    Overall product of the partial utility functions
    for the target material Sm$_{1-n}$Zr$_n$Fe$_{12-m-l-l'}$Co$_{m}$Ti$_{l}$Zr$_{l'}$
    as a function of $l$, with $n=0.2$ and $l'=0$ fixed.}
\end{figure}

It is the dimensionless parameter set as defined
in Eq.~(\ref{eq::dev_mat})
that enables the combination
of observables with different dimensions, such as the product
$\tilde{U}_{\Delta E}(l)\tilde{U}_{M}(l)$
as demonstrated above.
Multiple requirements are often simultaneously
imposed on practical materials, even to the level where non-physical
or extrinsic requirements, such as environmental friendliness
or prices of ingredient elements,
need to be incorporated with some parametrization
in order to implement a multiple-objective optimization.
What is implemented here constitutes
a starting point of such comprehensive optimization framework for materials design.
Equation~(\ref{eq::dev_mat}) may not be the unique way
to formulate the dimensionless parameters
but we would consider that this can be one of the most natural ways
in the spirit of constructing a manifold in the general materials space.
More mathematically rigorous construction might be possible.
As a crude starting point, the overall optimization can be inspected with an overall utility function
as we can define as follows
\[
U_{\rm all}\equiv \tilde{U}_{\Delta E}\tilde{U}_{M}\tilde{U}_{T_{\rm c}}
\tilde{U}_{J_{\rm RT}(8f)}\tilde{U}_{J_{\rm RT}(8i)}\tilde{U}_{J_{\rm RT}(8j)}
\]
for which the data within the working cross-section on $n=0.2$ and $l'=0$
are shown in Fig.~\ref{fig::compromise} for a few choices of the number of substituted Co atoms.
The amount of Ti around 0.5 in the overall optimization problem
seems to be a dead-end within the present working space.

\section{Discussions}
\label{sec::disc}

\subsection{Toward more optimal materials}
\label{sec::disc::Ti}

The lower bound identified
for the amount of Ti in the ferromagnet based on
the ThMn$_{12}$ structure
seems to be consistent with what has been established by experimental efforts
in the past three decades~\cite{sakurada_1992,sakurada_1996,sakuma_2016,tozman_2019}.
In principle our methodology can predict the limiting case to save
a lot of those experimental efforts. This is going to be a help in the upcoming
development of new materials with multiple relevant observables for a given utility.
Given that 0.5 Ti atoms per formula unit seems to be
the best compromise with SmFe$_{12}$ for permanent-magnet
utility, the working space needs to be extended in order to go beyond what has been achieved so far.

Furthermore, our formulation can straightforwardly
incorporate more information including lattice-structure variants
and external requirements on each element as long as they can be sufficiently well parametrized.
Not only concerning the intrinsic properties of the target materials,
but also various external factors could also be optimized altogether to design a good working bulk material.

\begin{figure}
  \scalebox{0.2}{\includegraphics{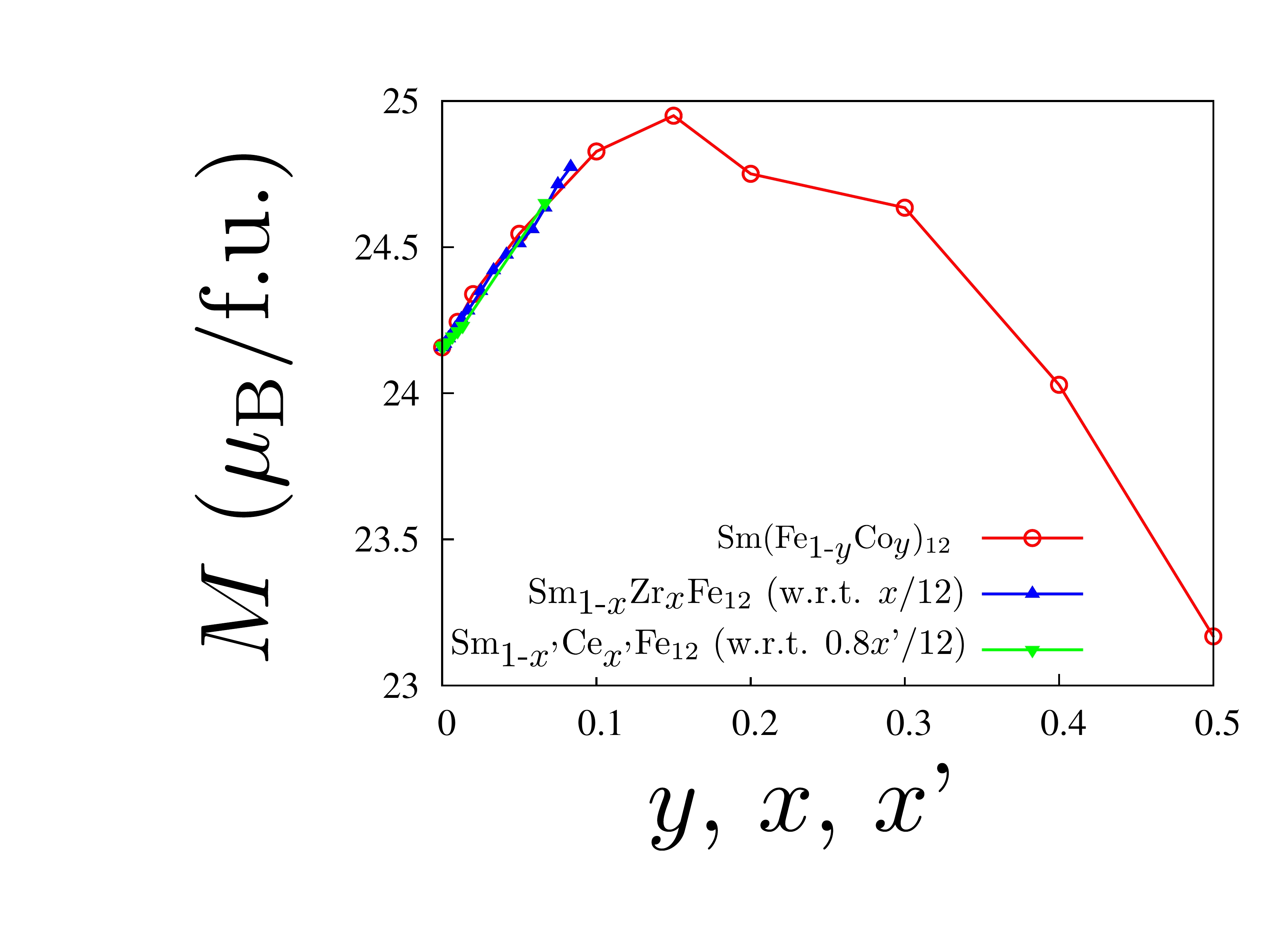}}
  \caption{\label{fig::mag_vs_Co_Ce_Zr}(Color online)
    Calculated magnetization of doped SmFe$_{12}$
    per formula unit (f.u.)
    on a fixed lattice of SmFe$_{12}$~\cite{harashima_2015}.
    It is seen that one extra electron from
    Zr($2a$) or
    Ce($2a$),
    with a proper rescaling of the concentration
    on the horizontal axis,
    works like one more electron in the minority spin band from Co and an
    analogue of the Slater-Pauling curve can be realized in the doped $2a$ site.}
\end{figure}
\subsection{Enhancing magnetization in SmFe$_{12}$ by substitution}
\label{sec::disc::zr}
Calculated magnetization for partly substituted SmFe$_{12}$ is summarized
in Fig.~\ref{fig::mag_vs_Co_Ce_Zr}
which show the celebrated Slater-Pauling curve~\cite{slater,pauling_1938} for Sm(Fe,Co)$_{12}$.
Remarkably, partly substituting Sm($2a$) by Zr shows
analogous enhancement of magnetization which is demonstrated in Fig.~\ref{fig::mag_vs_Co_Ce_Zr}
by a proper rescaling of the concentration of substitute elements. Experimentally,
it might be counter-intuitive to see an enhancement of magnetization triggered by 
non-magnetic elements like Zr, while we argue below that
this trend can be naturally interpreted in terms of electron number counting.
The mechanism is attributed to charge transfer from Zr($2a$) to Fe($8f$).
The same mechanism holds for Ce$^{4+}$($2a$)
where the delocalized $4f$-electron gets transferred to Fe($8f$).

\subsubsection{Slater-Pauling curve in Sm(Fe,Co)$_{12}$}
Calculated density of states for SmFe$_{12}$ in Fig.~\ref{fig::dos}
shown in Appendix points to a key role carried by Fe($8f$) sublattice
in realizing the Slater-Pauling curve. When a part
of the majority spin band of Fe has an overlap on the Fermi level,
partial substitution of Fe by Co leads to the electronic state
where the majority-spin state gets shifted
toward below the Fermi level, enhancing the intrinsic magnetization~\cite{bozorth,mott_1964,julie_1994,chikazumi,kubler}.
In Fig.~\ref{fig::dos}, majority-spin state overlapping the Fermi level is mostly carried by Fe($8f$)
and thus Fe($8f$) seems to be dominantly involved in the observed Slater-Pauling curve in Fig.~\ref{fig::mag_vs_Co_Ce_Zr},
even though the substituting Co has been put uniformly over all of the Fe sublattices without caring for the relatively minor
site preference.
We note in passing that the Nd-analogue, NdFe$_{12}$ as was investigated in Ref.~\onlinecite{miyake_2014},
does not have much of the majority-spin states overlapping the Fermi level and thus would not show the Slater-Pauling curve
for magnetization. The difference comes from the smaller lattice constant of the present Sm variant.

\subsubsection{Electron doping into Fe($8f$) by substituted Zr in Sm($2a$)}
We observe that Zr($2a$) actually helps
in enhancing magnetization and other observables
in partly substituted SmFe$_{12}$,
instead of diluting the magnetic properties of pristine SmFe$_{12}$,
within the approximation of the fixed lattice.
Care must be taken in assessing the nature of the electronic states of intermetallics
where a significant part of the electrons are delocalized, and naive expectation such as dilution of magnetic moments
which is rather oriented for a qualitative picture of insulators may not hold in some cases.
It is seen in the calculated density of states for ZrFe$_{12}$ in
Figs.~\ref{fig::dos_ce_and_zr_analogues}~(a), (b) and (c) shown in Appendix
that delocalized $4d$-electron states from Zr in ZrFe$_{12}$ contributes like Co,
in that both of Co and Zr($2a$) reduce the overlap of the majority-spin states on the Fermi level
for Fe($8f$).
Thus the Slater-Pauling curve with Co adding up one electron
on top of $3d$-electron band of Fe, most significantly Fe($8f$), to maximize the magnetization
can be simulated with Zr($2a$) also adding up electron via the delocalized $4d$-electrons.
The magnetic equivalent of Zr to Co is measured by a rescaling $12x\simeq y$
as can be inspected in Fig.~\ref{fig::mag_vs_Co_Ce_Zr}.
Here the intersite distance between Sm($2a$) and Fe($8f$)
of the host lattice seems to fall in a good range
in realizing the charge transfer from Zr($2a$) to Fe($8f$).

\subsubsection{Electron doping into Fe($8f$) by substituted Ce$^{4+}$}
Having seen the positive impact of Zr($2a$) in SmFe$_{12}$, the same effect
can be expected for Ce$^{4+}$ substituting the Sm($2a$) site
in providing the additional electron with the delocalized $4f$ electron.
A part of the delocalized $4f$-electrons, which seems to be 80\% as inspected
from a manual scaling to achieve the data collapse seen in Fig.~\ref{fig::mag_vs_Co_Ce_Zr},
hybridizes with $d$-electron band and adds up the filling,
leading to an analogue of the Slater-Pauling curve in the particular electronic structure
with the charge transfer from Ce$^{4+}$($2a$) to Fe($8f$).

Further details are given in Sec.~\ref{sec::e-doping} in Appendix.

\section{Conclusions and outlook}
\label{sec::conc}

We have inspected
an optimal chemical composition for a ferromagnet SmFe$_{12}$ presuming the possible utility
as a permanent magnet. Representative substitute elements, Ti, Co, Zr have been considered.
Combining experimental data and {\it ab initio} data in a self-consistent way,
we have seen that, in Zr-substituted SmFe$_{12}$, Zr occupies both Sm($2a$) and Fe($8i$) almost equally likely
in terms of energetics. While Ti as the stabilization element has been
found to be unavoidable, the lower bound found {\it ab initio} around $0.5$
is close to what has been achieved experimentally so far~\cite{tozman_2018}.

Concerning the intrinsic magnetization enhanced by Zr~\cite{tozman_2019},
we find that the electron doping effects brought by Zr can be exploited
to gain both of structure stability and magnetization. The similar effect is expected for Ce as well.
Both of these happens because of the particular electronic structure with the hybridization
between $3d$-electron band from the Fe sublattices and $5d$-electron band from rare earth sublattice.

Methodologically, the construction to iterate between macroscopically measured experimental data
and microscopically calculated {\it ab initio} data might make a step forward to
a multi-scale description of materials,
which should be an important part of the possible theoretical description of coercivity of REPM's.

\begin{acknowledgments}
This work is partly supported by JST-Mirai Program, Grant Number JPMJMI19G1, and is partly supported by the Elements Strategy Initiative Center for Magnetic Materials (ESICMM), Grant Number 12016013, through the Ministry of Education, Culture, Sports, Science and Technology (MEXT).
We gratefully acknowledge the financial support by Toyota Motor Corporation.
The sample preparation was performed under the future pioneering program ¡ÈDevelopment of magnetic material technology for high-efficiency motors¡É commissioned by the New Energy and Industrial Technology Development Organization (NEDO), Japan.
Part of this work was performed at  Australian Nuclear Science and Technology Organisation (ANSTO), and
J.~R.~Hester's help in the collection of experimental data
is gratefully acknowledged.
One of the authors (MM) gratefully acknowledges
helpful discussions with M.~Ito, N.~Sakuma, M.~Yano, T.~Shoji,
T.~Miyake, Y.~Harashima, T.~Ozaki, H.~Akai, M.~Hoffmann,
C.~E.~Patrick, J.~B.~Staunton in related projects.
He also benefited from
comments and suggestions
given by Y.~Kuramoto, K.~Ohashi, S.~Sakurada, X.~Tang, G.~Hrkac, T.~Ishikawa.
Numerical computations were executed on System B in ISSP Supercomputer
Center, University of Tokyo.
Images in Figs.~\ref{fig::lattice}~and~\ref{fig::scheme}
have been
generated using CrystalMaker$^{\textregistered}$: a crystal and molecular structures program for Mac and Windows.
CrystalMaker Software Ltd, Oxford, England (www.crystalmaker.com).
\end{acknowledgments}

\appendix

\section{Details of {\it ab initio} calculations}
\label{sec::calc_details}

\subsection{{\it Ab initio} structure optimization of stoichiometric compounds
and compounds on discrete points in chemical composition space}
\label{sec::calc_details::str_opt}

The structure optimization
has been done
utilizing the open-source package
for
{\it ab initio} electronic structure calculations, OpenMX~\cite{OpenMX,Ozaki2003,Ozaki2004,Ozaki2005,Duy2014,Lejaeghere2016}
which works on the basis of pseudopotentials~\cite{MBK1993,Theurich2001} and localized basis sets.
Present type of {\it ab initio} structure optimization
utilizing OpenMX to evaluate the formation energy referring to elemental systems
have been described elsewhere~\cite{mm_2018}.
We concisely describe what is extensively used in the present study.
The choice of the local basis set has been the followings:
\verb|Sm8.0_OC-s2p2d2f1|, 
\verb|Fe6.0S-s2p2d1|,
\verb|Co6.0S-s2p2d2f1|,
\verb|Ti7.0-s3p3d3f1|,
and
\verb|Zr7.0-s3p3d3f1|, within the generalized gradient approximation (GGA)
according to Perdew, Burke, and Enzerhof (PBE)~\cite{pbe_1996}.
Partial core correction in the open-core approximation for Sm
is set both for $\alpha$-Sm as the reference system and the target compound SmFe$_{12}$.
Convergence with respect to the number of $k$-points
and the cutoff energy is monitored. Given a material $\mbox{M}$, the optimized structure
comes with the calculated energy on the basis of the choice of the particular basis set as given above:
we will refer to this calculated energy as ``$U_{\rm tot}[\mbox{M}]$''
for the convenience of reference. We note that $U_{\rm tot}$ has been defined
up to the particular choice of the basis sets and the pseudopotentials
specified above and it is not entirely the true total energy.

In the lattice structure of SmFe$_{12}$ shown in Fig.~1 in the main text,
there are three sublattices for Fe, namely,
Fe($8i$), Fe($8j$), and Fe($8f$),
and one rare-earth sublattice, Sm($2a$).
The internal cartesian coordinates
of them can be defined as Sm($2a$)$(0,0,0)$, Fe($8i$)$(x_{i},0,0)$, Fe($8j$)$(x_{j},0.5,0)$,
and Fe($8j$)$(0.5-x_{8j},0,0.5)$.
It is instructive to note that
Fe($8i$) and Fe($8j$) atoms approximately form a regular hexagon,
corresponding to Co($2c$) atoms in the SmCo$_5$ prototype~\cite{li_and_coey_1991_}
on the side faces of the tetragonal box shown in Fig.~1 of the main text.
There is a relation $2(0.5-x_{i})\simeq x_{j}$ among the internal coordinates
as imposed by the local hexagonal symmetry of the CaCu$_5$ prototype.
We typically
set a set of starting internal coordinates, $(x_{8i},x_{8j})=(0.36,0.27)$ which
roughly satisfies
this relation. Referring to the recent experimental work on NdFe$_{12}$~\cite{hirayama_2014}
we set a set of starting lattice constants to be $(a,c)=(8.52,4.80)$ (\AA). With these starting structure parameters,
one of the host atoms in the tetragonal unit, with two formula units i.e. two Sm atoms and 24 Fe atoms,
are replaced by substitute atoms one by one, and the crystal lattice structure is optimized
to evaluate the total energy of the electronic state with the given chemical composition. For Zr-substituted SmFe$_{12}$,
we obtain the optimized lattice parameters
for each of (Sm,Zr)Fe$_{24}$ and Sm$_2$Fe$_{23}$Zr on the tetragonal unit with two formula units
and compare the relative trends
among them concerning magnetization and the energy of the electronic state.

\begin{table}
  \begin{tabular}{ccc}\\ \hline
    M & $N_{\rm atom}[\mbox{M}]$ 
    & $U_{\rm tot}$ (eV) \\ \hline
    hcp-Ti & $2$ & 
    $-3227.8$ \\ \hline
    hcp-Zr & $2$ & 
    $-2634.5$ \\ \hline
    bcc-Fe & $1$ & 
    $-2437.4$ \\ \hline
    hcp-Co & $2$ & 
    $-5830.3$ \\ \hline
    $\alpha$-Sm & $9$ & 
    $-11383$ \\ \hline
  \end{tabular}
  \caption{\label{table::reference_energy} Calculated energy for the reference elemental systems
    based on the local basis sets as described in the text. The data for bcc-Fe, hcp-Co, and $\alpha$-Sm
    are taken from Ref.~\onlinecite{mm_2018} and reproduced here for the convenience of reference.}
\end{table}

\begin{figure}
  \scalebox{0.2}{\includegraphics{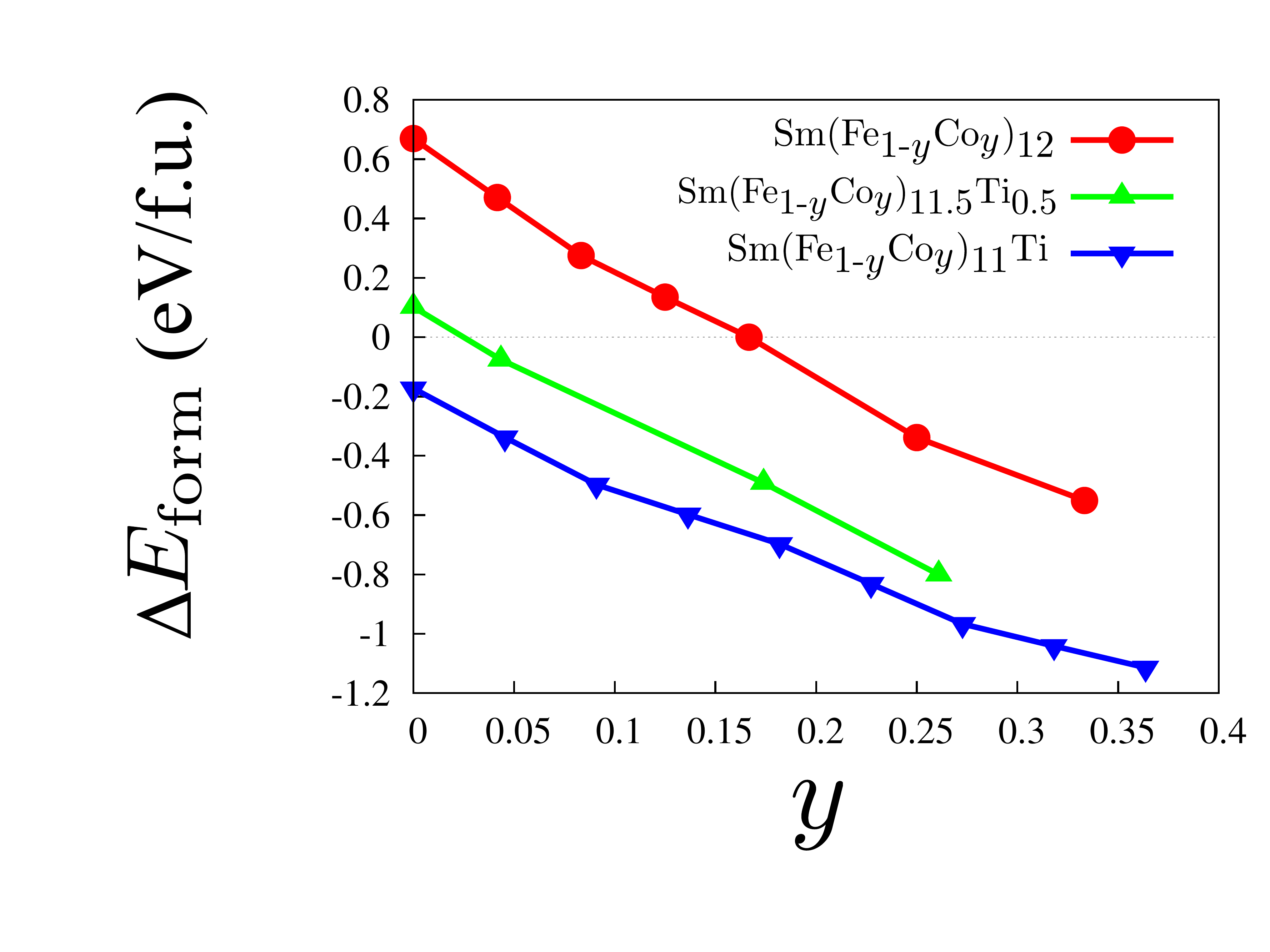}}
  \caption{\label{fig::SmFeCo12} (Color online)
    Calculated formation energy of Co-substituted SmFe$_{12}$, SmFe$_{11.5}$Ti$_{0.5}$, and SmFe$_{11}$Ti
    per formula unit (f.u.).
    The reference systems are the ingredient elements.}
\end{figure}  
The formation energy $\Delta E[\mbox{C}]$
of a compound C is
defined as a difference between the energy of the compound
and the sum of the energy of the constitute elemental systems as follows:
\begin{widetext}
\begin{eqnarray}
  &&  \Delta E_{\rm form}[\mbox{Sm$_{1-x}$Zr$_x$(Fe$_{1-y-z}$Co$_{y}$Ti$_z$)$_{12}$}] \nonumber \\
  & \equiv & U_{\rm tot}[\mbox{Sm$_{1-x}$Zr$_x$(Fe$_{1-y-z}$Co$_{y}$Ti$_z$)$_{12}$}] 
   -(1-x)\frac{U_{\rm tot}[\mbox{$\alpha$-Sm}]}{N_{\rm atom}[\mbox{$\alpha$-Sm}]} - x \frac{U_{\rm tot}[\mbox{hcp-Zr}]}{N_{\rm atom}[\mbox{hcp-Zr}]}\nonumber \\
  &&  -12(1-y-z)\frac{U_{\rm tot}[\mbox{bcc-Fe}]}{N_{\rm atom}[\mbox{bcc-Fe}]}-12y \frac{U[\mbox{hcp-Co}]}{N_{\rm atom}[\mbox{hcp-Co}]} 
   - 12 z \frac{U_{\rm tot}[\mbox{hcp-Ti}]}{N_{\rm atom}[\mbox{hcp-Ti}]}.
\end{eqnarray}
\end{widetext}
Here $N_{\rm atom}[\mbox{$\alpha$-Sm}]=9$, $N_{\rm atom}[\mbox{hcp-Zr}]=2$,
$N_{\rm atom}[\mbox{bcc-Fe}]=1$, $N_{\rm atom}[\mbox{hcp-Co}]=2$,
and $N_{\rm atom}[\mbox{hcp-Ti}]=2$, are the number atoms in the unit cell of
the elemental systems.
Calculated results for the formation energy of Co-substituted SmFe$_{12}$, SmFe$_{11.5}$Ti$_{0.5}$, and SmFe$_{11}$Ti
with respect to the concentration of Co are shown in Fig.~\ref{fig::SmFeCo12}.
Referring to Fig.~2 in the main text and
results of previous works, we have selectively put Ti in the Fe($8i$) sublattice
and Co in the Fe($8f$) sublattice in order to get the overall trend of the formation energy
with respect to the concentration of substituted Co.
It is again confirmed that
SmFe$_{12}$ substituted with Ti significantly
gains the formation energy. Also the presence of Co helps the formation energy to be negative and large:
Co substitution in SmFe$_{12}$ can be the rare case where both of the strong ferromagnetism and structure stability
can be gained at the same time, as long as possible other compounds with
different crystal structure do not compete severely in the formation.
\begin{figure}
  \scalebox{0.25}{\includegraphics{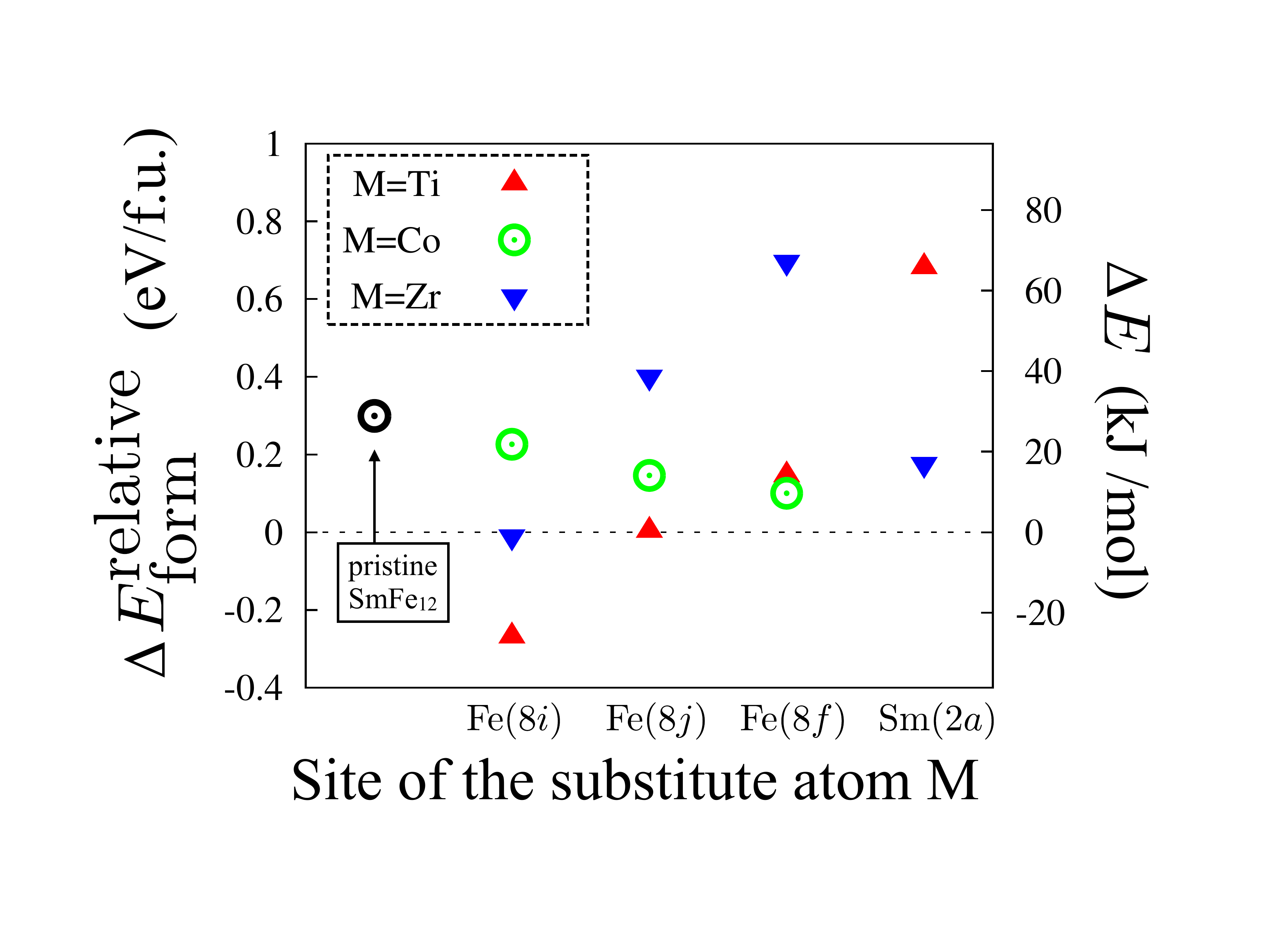}}
  \caption{\label{fig::doped_SmFe12_referring_to_Sm2Fe17} (Color online)
    Calculated formation energy of M-substituted SmFe$_{12}$ (M=Ti, Co, and Zr)
    per formula unit (f.u.).
    The reference system is taken to be
    Sm$_2$Fe$_{17}$ in the same spirit as is written in Eq.~(\ref{eq::SmFe12_relative_to_Sm2Fe17}).
    The same data as those shown in Fig.~2 in the main text
    are plotted with
    the reference systems set to be Sm$_2$Fe$_{17}$.}
\end{figure}

In the present case of the assessment
of formation of SmFe$_{12}$, the most probable competing phase would be Sm$_2$Fe$_{17}$.
The crystal structure of Sm$_2$Fe$_{17}$,
of the rhombohedral type belonging to the Space Group No.~166,
is common among $4f$-$3d$ intermetallic compounds in REPM's,
such as Sm$_2$Co$_{17}$
in the main cell phase of the Sm-Co magnet~\cite{strnat_1991}
that was the champion magnet in 1970's and Sm$_2$Fe$_{17}$N$_x$ that has been a candidate since early
1990's to potentially go beyond Nd$_2$Fe$_{14}$B-based REPM~\cite{handbook_1995}.
A relative formation energy of SmT$_{12}$ referring to Sm$_2$T$_{17}$ (T=Fe or Co),
$\Delta E_{\rm form}^{\rm relative}$,
can be defined as follows,
\begin{widetext}
\begin{eqnarray}
  \Delta E_{\rm form}^{\rm relative}[\mbox{SmFe$_{12}$}] 
  \equiv  \frac{U_{\rm tot}[\mbox{SmFe$_{12}$}] }{N_{\rm fu}[\mbox{SmFe$_{12}$}]}
  -\frac{1}{2}\frac{U_{\rm tot}[\mbox{Sm$_{2}$Fe$_{17}$}]}{N_{\rm fu}[\mbox{Sm$_2$Fe$_{17}$}]}
  -\frac{7}{2} \frac{U_{\rm tot}[\mbox{bcc-Fe}]}{N_{\rm atom}[\mbox{bcc-Fe}]} \label{eq::SmFe12_relative_to_Sm2Fe17}
\end{eqnarray}
and
\begin{eqnarray}
  \Delta E_{\rm form}^{\rm relative}[\mbox{SmCo$_{12}$}]
  \equiv \frac{U_{\rm tot}[\mbox{SmCo$_{12}$}]}{N_{\rm fu}[\mbox{SmCo$_{12}$}]}
  -\frac{1}{2}\frac{U_{\rm tot}[\mbox{Sm$_{2}$Co$_{17}$}]}{N_{\rm fu}[\mbox{Sm$_2$Co$_{17}$}]}
  -\frac{7}{2} \frac{U_{\rm tot}[\mbox{hcp-Co}]}{N_{\rm atom}[\mbox{hcp-Co}]} \label{eq::SmFe12_relative_to_Sm2Co17}.
\end{eqnarray}
\end{widetext}
Thus defined relative formation energy of doped
SmFe$_{12}$ referring to Sm$_2$Fe$_{17}$ is shown in Fig.~\ref{fig::doped_SmFe12_referring_to_Sm2Fe17}.
Compared to the data in Fig.~2 in the main text,
Zr in Fe($8i$) looks more energetically favorable than being in Sm($2a$) sublattice.
On the basis of {\it ab initio} electronic structure of doped SmFe$_{12}$,
we do not see any particular reason to
expect that Zr would be selectively substituting the Sm($2a$) site
except for the possible scenario that
Ti and Zr compete over the most preferable Fe($8i$) sites
and Ti atoms dominate with the larger energy gain than Zr atoms.
As a compromise, Zr atoms can go more into Sm($2a$) sites than into Fe($8i$) sites.

It is indeed true that the relative formation
energy of SmT$_{12}$ referring to Sm$_2$T$_{17}$ is slightly lower for the T=Co case,
but still both stoichiometric limits of SmT$_{12}$ (T=Fe or Co)
will be purged by the formation of Sm$_{2}$T$_{17}$ because the latter
is more favorable energetically. Thus the decreasing slope on the Co concentration
axis seen in Fig.~\ref{fig::SmFeCo12}
does not straightforwardly point to a true gain in the structure stability
when the intervening phases such as Sm$_2$Co$_{17}$
comes in. At least
in the middle of Fe-Co axis, most promising chemical composition concerning the stability
is seen on a slightly Fe-rich side
as we explore the chemical composition space
continuously as described below.
\begin{table}
  \begin{tabular}{ccc} \hline
    $M$ & $U_{\rm tot}[\mbox{M}]$ (eV) & $N_{\rm fu}[\mbox{M}]$ \\ \hline
    SmFe$_{12}$ & $-61026$ & $2$ \\ \hline 
    SmCo$_{12}$ & $-72494$ & $2$ \\ \hline
    Sm$_2$Fe$_{17}$ & $-43965$ & $1$ \\ \hline 
    Sm$_2$Co$_{17}$ & $-52089$ & $1$ \\ \hline
\end{tabular}
\caption{\label{fig::SmFe12_vs_Sm2Fe17} Calculated energy with the optimized structure
  for the target compounds
  SmFe$_{12}$ and SmCo$_{12}$ and the corresponding data
  for the reference systems Sm$_2$Fe$_{17}$ and Sm$_2$Co$_{17}$, respectively. Number of formula units
in the optimized structure is denoted as $N_{\rm fu}$.}
\end{table}

\subsection{{\it Ab initio} interpolation for alloys}
\label{sec::calc_details::alloys}

Fractional parameters of the chemical composition
can be continuously explored
with coherent potential approximation (CPA) for random alloys.
Korringa-Kohn-Rostoker (KKR)~\cite{KKR_the_original} Green's function method
combined with CPA (KKR-CPA)
provides a convenient way to obtain an interpolated electronic structure,
e.g. for Sm(Fe$_{1-y}$Co$_{y}$)$_{12}$ with $0<y<1$, between the
stoichiometric limits of SmFe$_{12}$ and SmCo$_{12}$. We have used
the implementation of KKR-CPA in AkaiKKR~\cite{AkaiKKR}.

\begin{figure}
  \scalebox{0.2}{
    \includegraphics{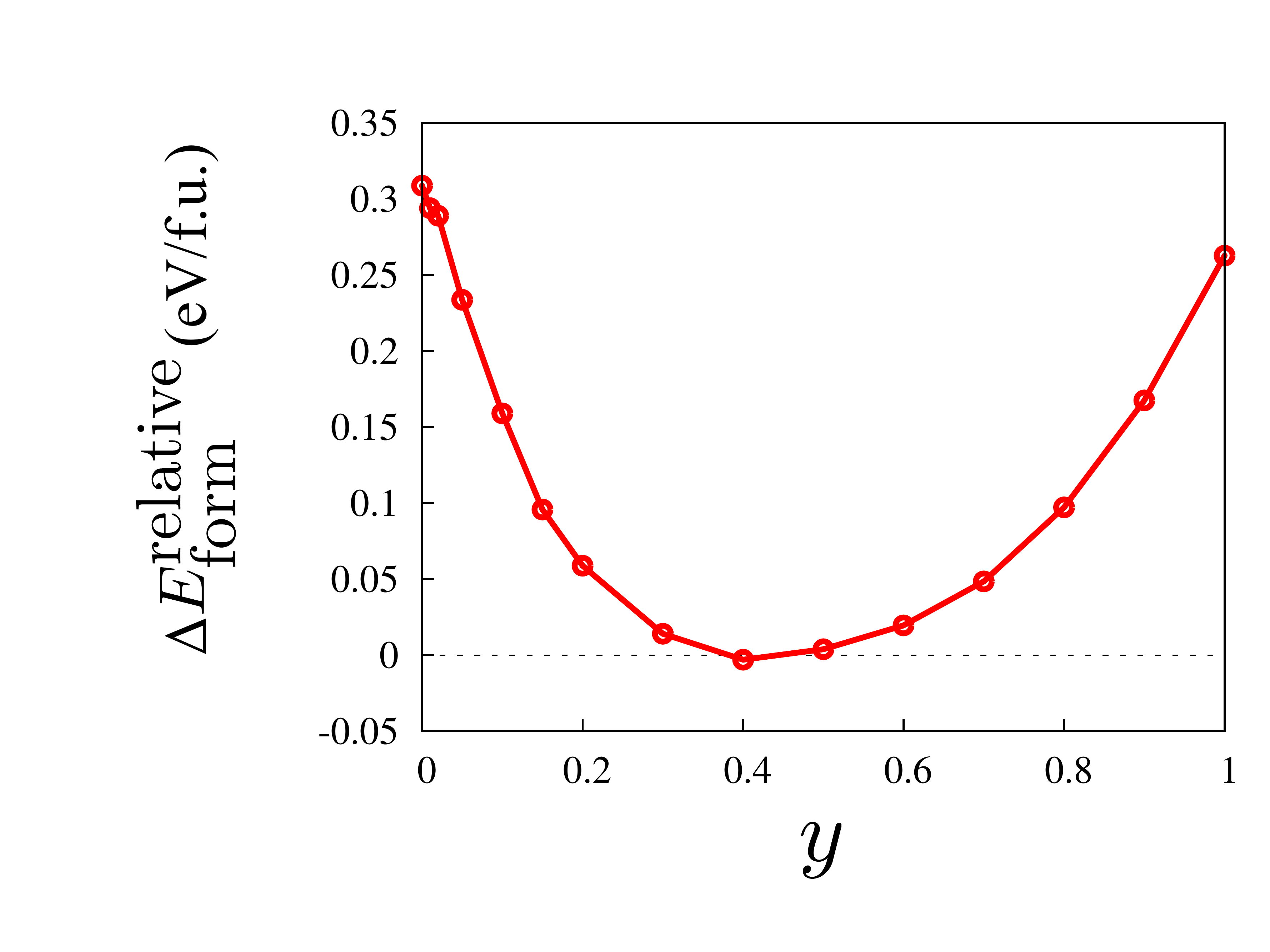}
    }
  \caption{\label{fig::SmFeCo12_KKR_CPA_with_stoich} (Color online)
    Interpolated formation energy of SmT$_{12}$
    per formula unit (f.u.)
    with the reference systems being Sm$_2$Fe$_{17}$
    and Sm$_2$Co$_{17}$ following
    Eqs.~(\ref{eq::SmFe12_relative_to_Sm2Fe17}),
    (\ref{eq::SmFe12_relative_to_Sm2Co17}),
    (\ref{eq::mixing_energy}), and~(\ref{eq::interpolated_delta_E}).}
\end{figure}  
KKR-CPA can yield a reliable estimate of magnetization and magnetic exchange couplings
for magnetic alloys as a continuous function of the composition parameters.
Unfortunately, the absolute value of calculated energy, which we denote here
by $E_{\rm tot}^{\rm KKRCPA}$, suffer from a systematic deviation
due to a cutoff parameter $l_{\rm max}$ in the expansion with respect to spherical harmonics
in solving the multiple scattering problem at the core of KKR Green's function method.
In the present calculations we set $l_{\rm max}=2$ for all elements
which should be well justified within the open-core approximation assuming well localized $4f$-electrons.
Comparison of calculated energy between different materials can be tricky within KKR-CPA
but it is feasible to observe the trend in the calculated energy as long as the range of the
target materials is restricted within the same type of crystal structure.
Thus mixing energy $\Delta E_{\rm mix}$ of an alloy
defined in Eq.~(\ref{eq::mixing_energy})
below can be combined 
with the formation energy calculated for the stoichiometric compounds
to assess an interpolated formation energy for alloys.
\begin{widetext}
\begin{eqnarray}
  && \Delta E_{\rm mix}[\mbox{Sm(Fe$_{1-x}$Co$_{x}$)$_{12}$}] \nonumber \\
    & \equiv  &
  E_{\rm tot}^{\rm KKRCPA}[\mbox{Sm(Fe$_{1-x}$Co$_{x}$)$_{12}$}] 
   -(1-x)E_{\rm tot}^{\rm KKRCPA}[\mbox{SmFe$_{12}$}] -x E_{\rm tot}^{\rm KKRCPA}[\mbox{SmCo$_{12}$}] \label{eq::mixing_energy}
\end{eqnarray}
Combining
the formation energy for the stoichiometric compounds
and the mixing energy for alloys, an interpolated formation energy for the alloy
can be estimated as follows:
\begin{eqnarray}
  &&  \Delta E_{\rm form}^{\rm relative}[\mbox{Sm(Fe$_{1-x}$Co$_{x}$)$_{12}$}] \nonumber \\
  & \simeq &
  (1-x) \Delta E_{\rm form}^{\rm relative}[\mbox{SmFe$_{12}$}]
  + \Delta E_{\rm mix}[\mbox{Sm(Fe$_{1-x}$Co$_{x}$)$_{12}$}]
  +x \Delta E_{\rm form}^{\rm relative}[\mbox{SmCo$_{12}$}]
 %
\label{eq::interpolated_delta_E}
\end{eqnarray}
\end{widetext}
The results for Sm(Fe$_{1-y}$Co$_{y}$)$_{12}$ are shown in Fig.~\ref{fig::SmFeCo12_KKR_CPA_with_stoich}.
It is seen that 40\% of Co can bring the formation energy
of the 1:12 phase almost on a par with the formation energy of 2:17 phase.

\section{Details of the
  ``LDA+Rietveld'' iterations}
\label{sec::integrated_data_details}

Rietveld refinement of the neutron diffraction data
for Sm$_{0.8}$Zr$_{0.2}$(Fe$_{0.7}$Co$_{0.3}$)$_{11.25}$Ti$_{0.75}$
has been done utilizing the FullProf program~\cite{fullprof_1993}.
AkaiKKR~\cite{AkaiKKR} was employed to calculate the magnetic moments of each atom
from first principles with the lattice structure information provided by Rietveld analysis as the input.
Then the output of AkaiKKR for calculated magnetic moments is fed back to the Rietveld analysis in the next stage,
and the overall process is iterated until the parameters in the inputs and the outputs converge.

It is to be noted that the experimental measurements are done at room temperature
and {\it ab initio} calculations are done at zero temperature. Since our Co-containing samples
come with sufficiently high Curie temperatures beyond 800K~\cite{prb_2001},
we regard that room temperature is close enough to zero temperature
for the present analysis. If the Curie temperature of the sample is not quite high,
we would turn to {\it ab initio} finite temperature calculations formulated
on the basis of KKR-CPA~\cite{pindor_1983,gyorffy_1985,julie_1985,julie_1986,julie_2004,julie_2006,chris_2019}
at the cost of some extra computational time. For Co-substituted SmFe$_{12}$, we can safely skip this.

Remarkably, the convergence of the overall LDA+Rietveld iteration
is achieved within only a few iteration steps
as are shown
in Table~\ref{table::results} with the input and output parameters of KKR-CPA.
Here we count a set of Rietveld analysis and {\it ab initio} calculation as one iteration step.
We observe that either Sm($2a$) or Fe($8i$) can accommodate the Zr atoms in an equally plausible way.
Thus the message from the data from the {\it ab initio} structure optimization for the site preference
of Zr is confirmed in the analysis of the neutron diffraction data for the real sample,
Sm$_{0.8}$Zr$_{0.2}$(Fe$_{0.75}$Co$_{0.25}$)$_{11.25}$Ti$_{0.75}$.
\begin{table*}
  \begin{tabular}{l}
    (a) Zr in Sm($2a$) \\
  \begin{tabular}{rrcrcrc}\hline
            & \multicolumn{2}{c}{step 1} & \multicolumn{2}{c}{step 2} & \multicolumn{2}{c}{step 3} \\
    KKR-CPA & inputs & outputs ($\mu_{\rm B}$)
    & inputs & outputs ($\mu_{\rm B}$)
    & inputs & outputs ($\mu_{\rm B}$) \\ \hline\hline
    $8f$ Fe &               $69.2\%$ & $1.89946$ & 70.4\%   & 1.89762  & 70.4\%   & 1.89768 \\ \hline
         Co &               $30.8\%$ & $1.32717$ & 29.6\%   & 1.32660  & 29.6\%   & 1.32665 \\ \hline\hline
    $8i$ Fe & $x_i=0.35617$ $73.6\%$ & $2.50941$ & $x_i=0.35619$ 72.4\%  & 2.50494  & $x_i=0.3562$ 72.4\%  & 2.50471\\ \hline
         Co &                $8\%$   & $1.66443$   & 8.8\%   &   1.66101 & 8.8\%  & 1.66083 \\ \hline
         Ti &              $18.8\%$  & $-0.75328$ & 18.8\%   &   -0.75300 & 18.8\%  & -0.75306 \\ \hline\hline
    $8j$ Fe & $x_j=0.27678$ $68\%$   & $2.28426$ & $x_j=0.27645$ 68\%     & 2.28343  & $x_j=0.27644$ 68\% & 2.28351 \\ \hline
         Co &               $32\%$   & $1.47931$ & 32\%     &   1.47928 & 32\%  & 1.47936 \\ \hline\hline
    $2a$ Sm &               $80\%$   & $-0.41385$ & 80\%    &  -0.41207 & 80\%  & -0.41197 \\ \hline
         Zr &               $20\%$  & $-0.41066$ & 20\%    &   -0.41000 & 20\%  & -0.40995 \\ \hline\hline
         ($a$~(\AA), $c/a$) & $(8.50758, 0.5607)$ & $21.3041$ & $(8.50705, 0.5607)$  & $21.2733$  & $(8.50705, 0.5607)$   & $21.2732$ \\
         for the unit cell  & & & & & & \\ \hline
  Calculated energy (Ry) & & $-47793.09020$ & & $-47787.10120$ & & $-47787.10106$ \\ \hline
  \end{tabular}\\
  \\
  (b) Zr in Fe($8i$) \\
  \begin{tabular}{rrcrcrc}\hline
            & \multicolumn{2}{c}{step 1} & \multicolumn{2}{c}{step 2} & \multicolumn{2}{c}{step 3} \\
    KKR-CPA & inputs & outputs ($\mu_{\rm B}$)
    & inputs & outputs ($\mu_{\rm B}$)
    & inputs & outputs ($\mu_{\rm B}$) \\ \hline\hline
    $8f$ Fe &               $69.17\%$ & $1.82528$     &              70.63\% & 1.82613     & 70.63\%   & 1.82611 \\ \hline
         Co &               $30.83\%$ & $1.26294$     &              29.37\% & 1.26432     & 29.37\%   & 1.26431 \\ \hline\hline
    $8i$ Fe & $x_i=0.35515$ $70\%$ &    $2.52271$     & $x_i=0.35532$ 68.77\% & 2.51612    & $x_i=0.35532$ 68.77\%  & 2.51594\\ \hline
         Co &                $6.324\%$   & $1.67340$  &               7.51\% &  1.66825    & 7.51\%  & 1.66809 \\ \hline
         Ti &              $18.577\%$  & $-0.71585$   &              18.57\% &   -0.71660  & 18.577\%  & -0.71663 \\ \hline
         Zr &               $5.138\%$   & $-0.43412$  &              5.138\% &   -0.43445  & 5.138\%  & -0.43446 \\ \hline\hline
    $8j$ Fe & $x_j=0.27717$ $68.379\%$   & $2.23359$  & $x_j=0.27685$ 68.25\%  & 2.23380   & $x_j=0.27683$ 68.25\% & 2.23387 \\ \hline
         Co &               $31.621\%$   & $1.42782$  &               31.75\%  & 1.42873  & 31.75\%  & 1.42880 \\ \hline\hline
    $2a$ Sm &               $100\%$  & $-0.40525$     &                $100\%$ & -0.40270  & $100\%$  & -0.40259 \\ \hline\hline
         ($a$~(\AA), $c/a$)  & $(8.50758, 0.5607)$ & $20.3790$ & $(8.50705, 0.5607)$        & $20.3529$  & $(8.50705, 0.5607)$   & $20.3527$ \\
         for the unit cell  & & & & & & \\ \hline
  Calculated energy (Ry) & &  $-51311.90695$ & & $-51310.68873$ & & $-51310.68847$ \\ \hline         
  \end{tabular}
  \end{tabular}
  \caption{\label{table::results} Proceedings of the ``LDA+Rietveld'' iteration.
    (a) assuming that Zr resides in Sm($2a$) and (b) Zr resides only in Fe($8i$).
    We find that either assumption works on a par as long as the Rietveld analysis combined with {\it ab initio} KKR-CPA
    is concerned.
    We note that the absolute values of the calculated energy via KKR-CPA
    should not be compared between (a) and (b), but only within (a) or (b).
  }
\end{table*}
Details of the Rietveld analysis
together with a wider range of target compounds will be reported elsewhere~\cite{hawai_et_al}.

\section{Extracting the derivative of the target observables with respect to the chemical composition parameters}
\label{sec::cpa_details}

\begin{figure*}
  \begin{tabular}{llll}
    (a) $M(n)$ & (b) $M(m)$ & (c) $M(l)$ & (d) $M(l')$ \\
    \scalebox{0.12}{\includegraphics{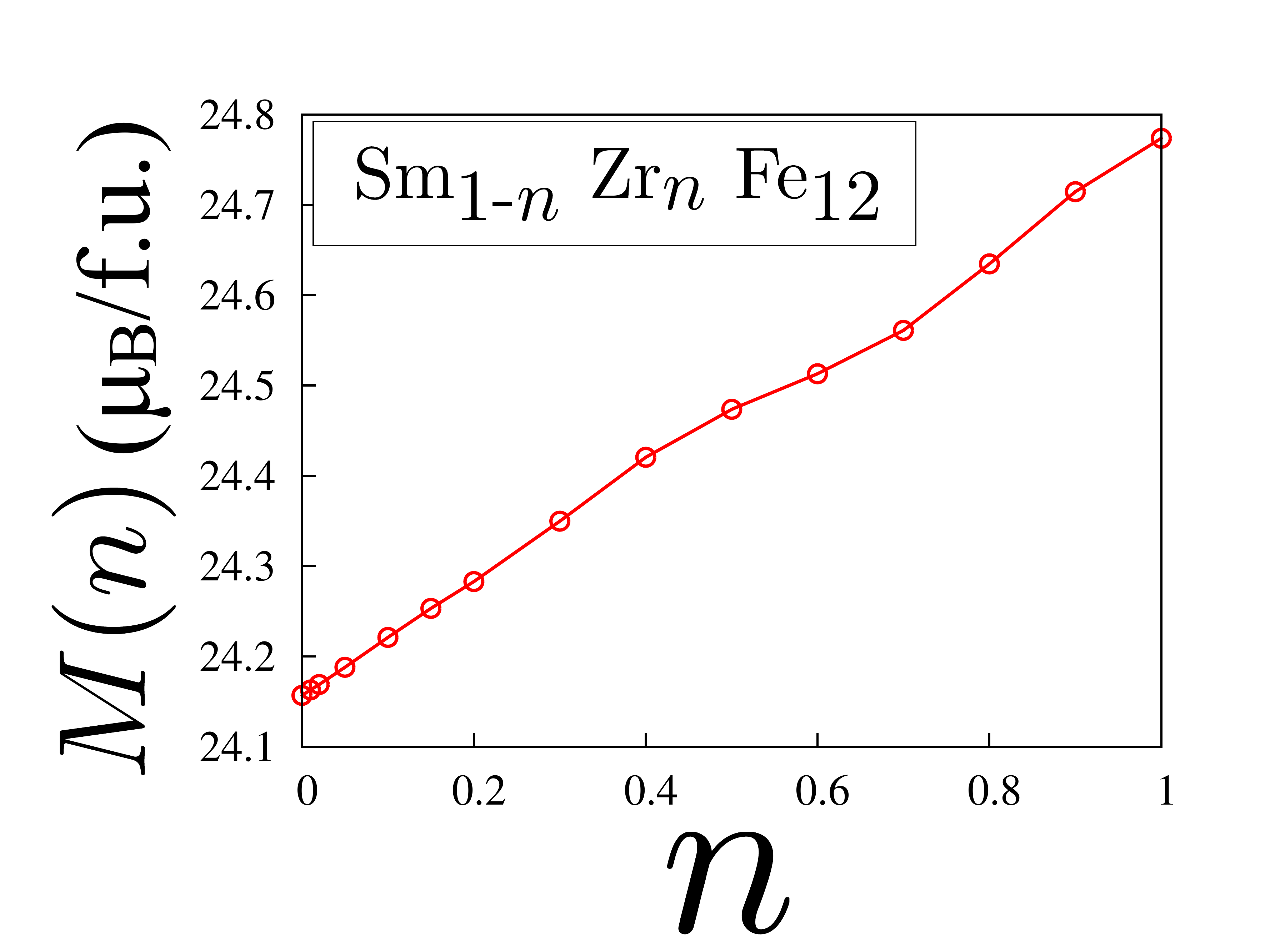}}
    &
    \scalebox{0.12}{\includegraphics{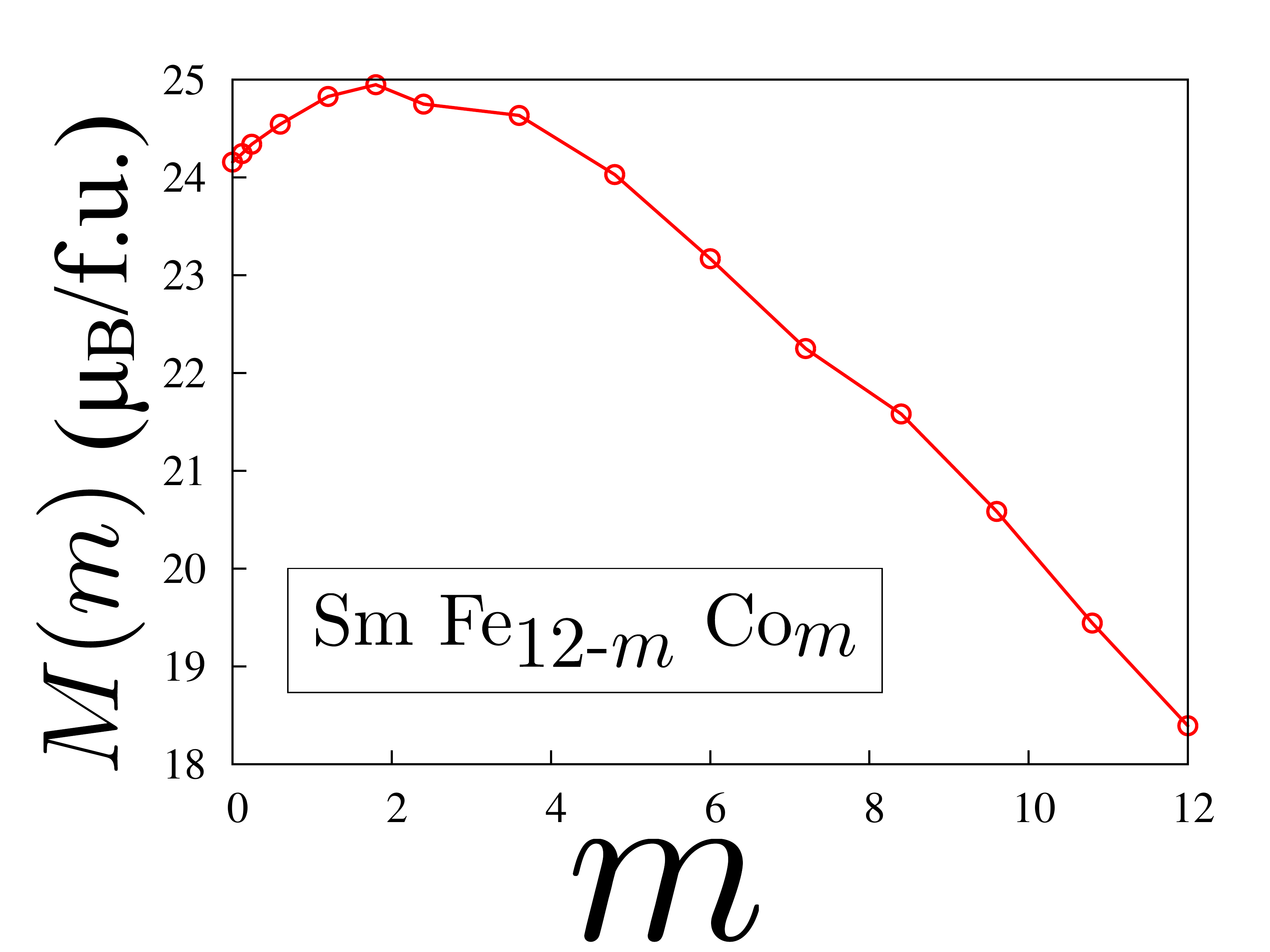}}
    &
    \scalebox{0.12}{\includegraphics{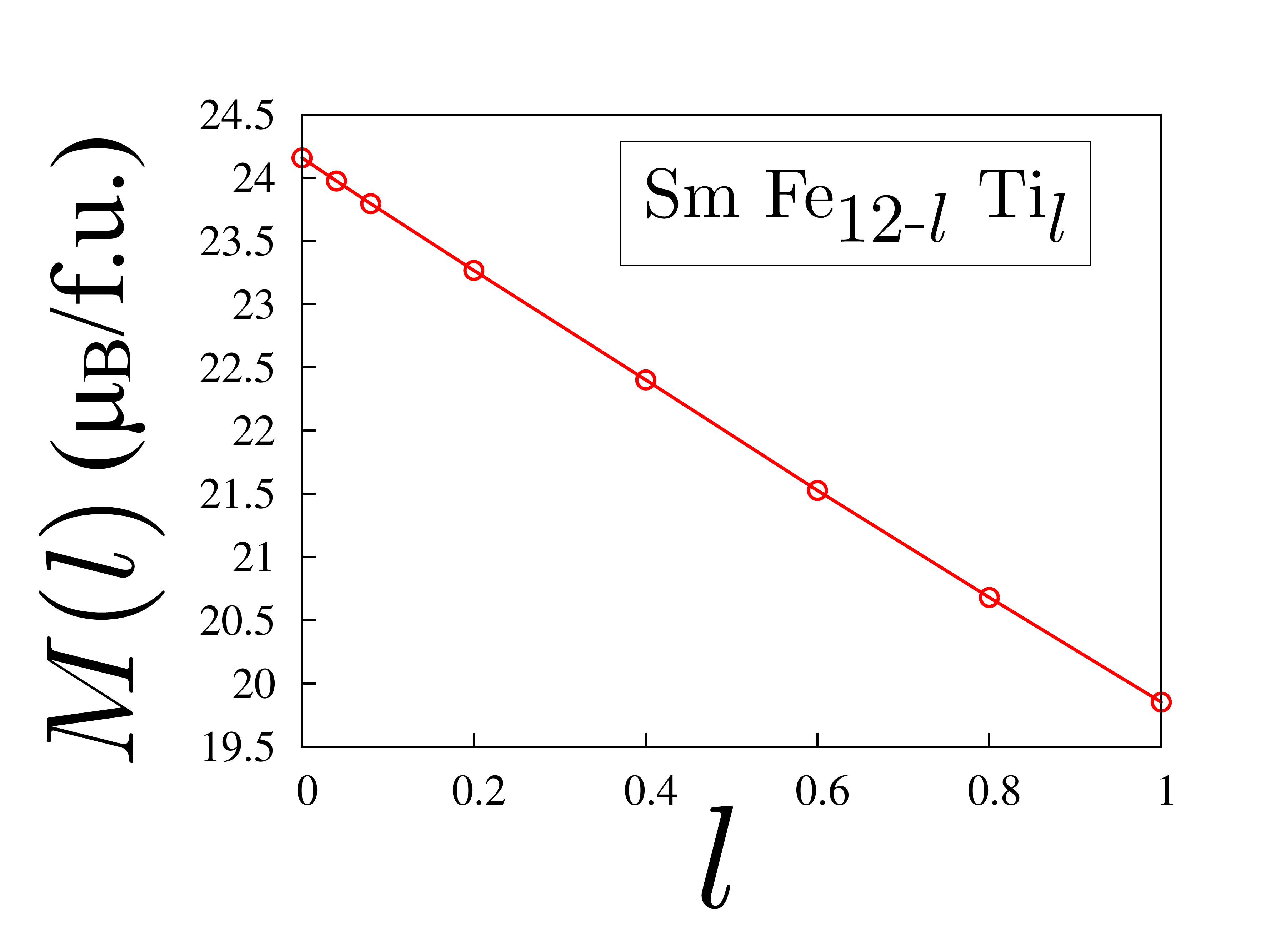}} 
    &
    \scalebox{0.12}{\includegraphics{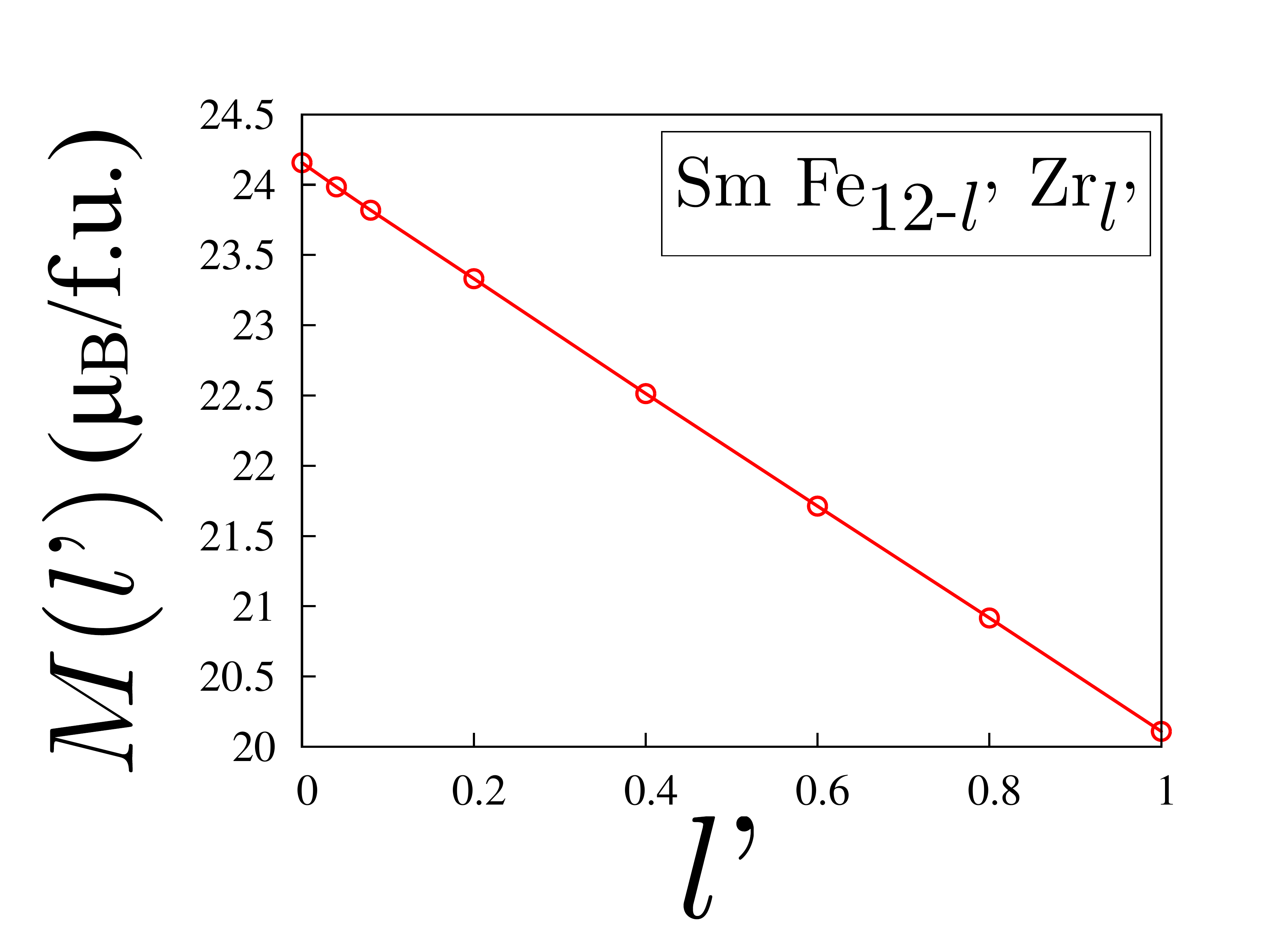}} \\
    (e) $T_{\rm Curie}(n)$ &
    (f) $T_{\rm Curie}(m)$ &
    (g) $T_{\rm Curie}(l)$ &
    (h) $T_{\rm Curie}(l')$ \\
    \scalebox{0.12}{\includegraphics{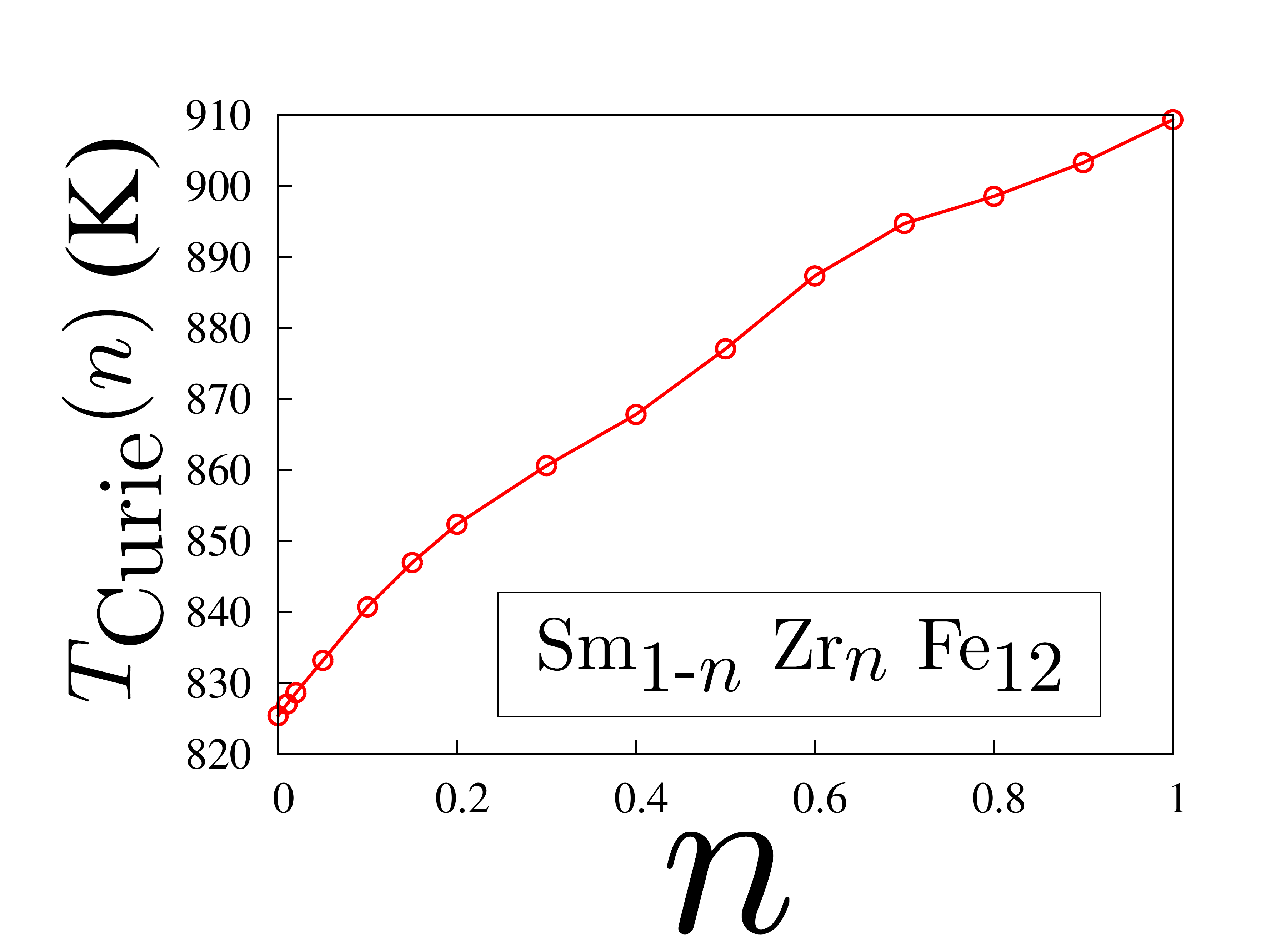}}
    &
    \scalebox{0.12}{\includegraphics{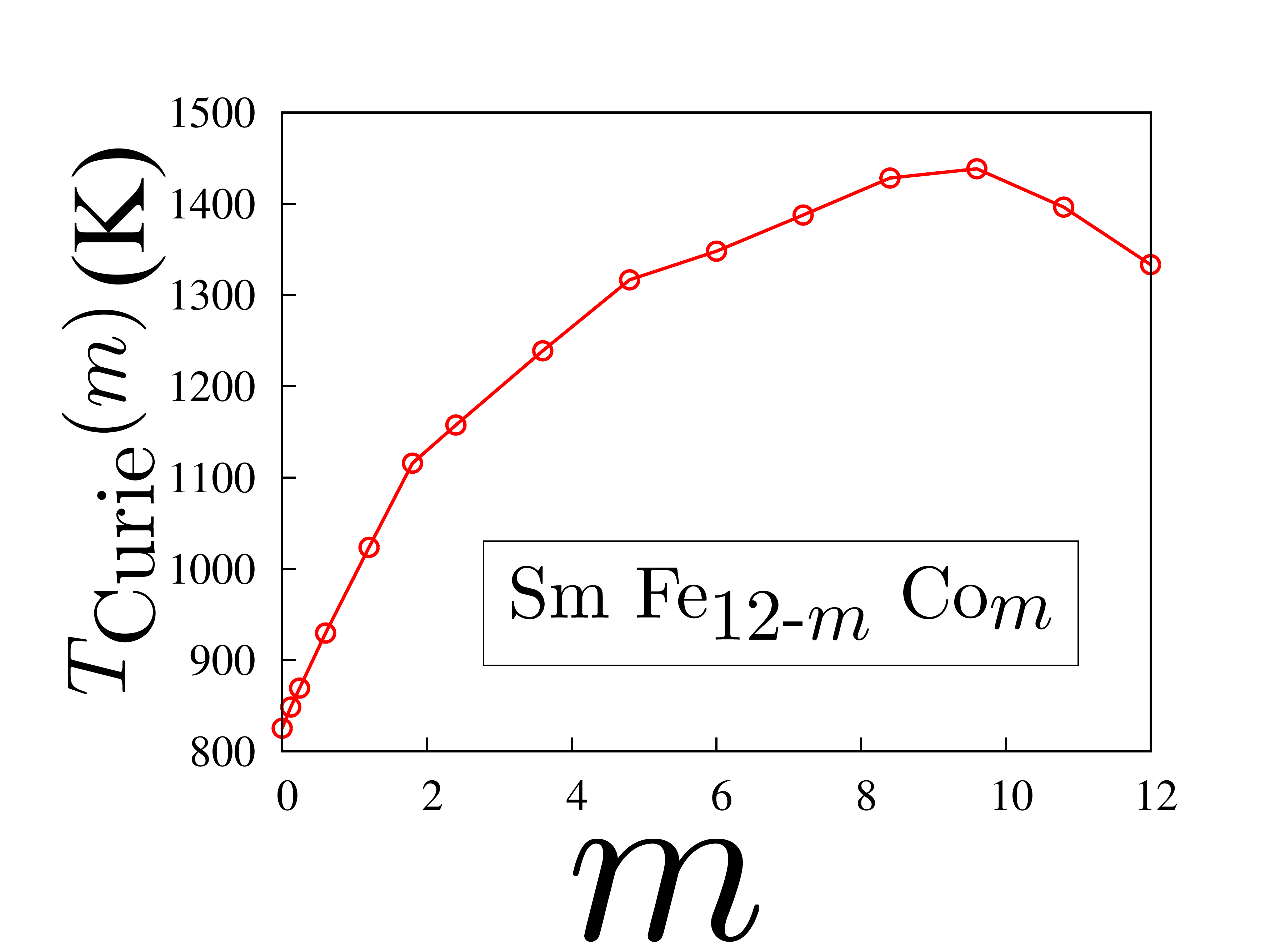}}
    &
    \scalebox{0.12}{\includegraphics{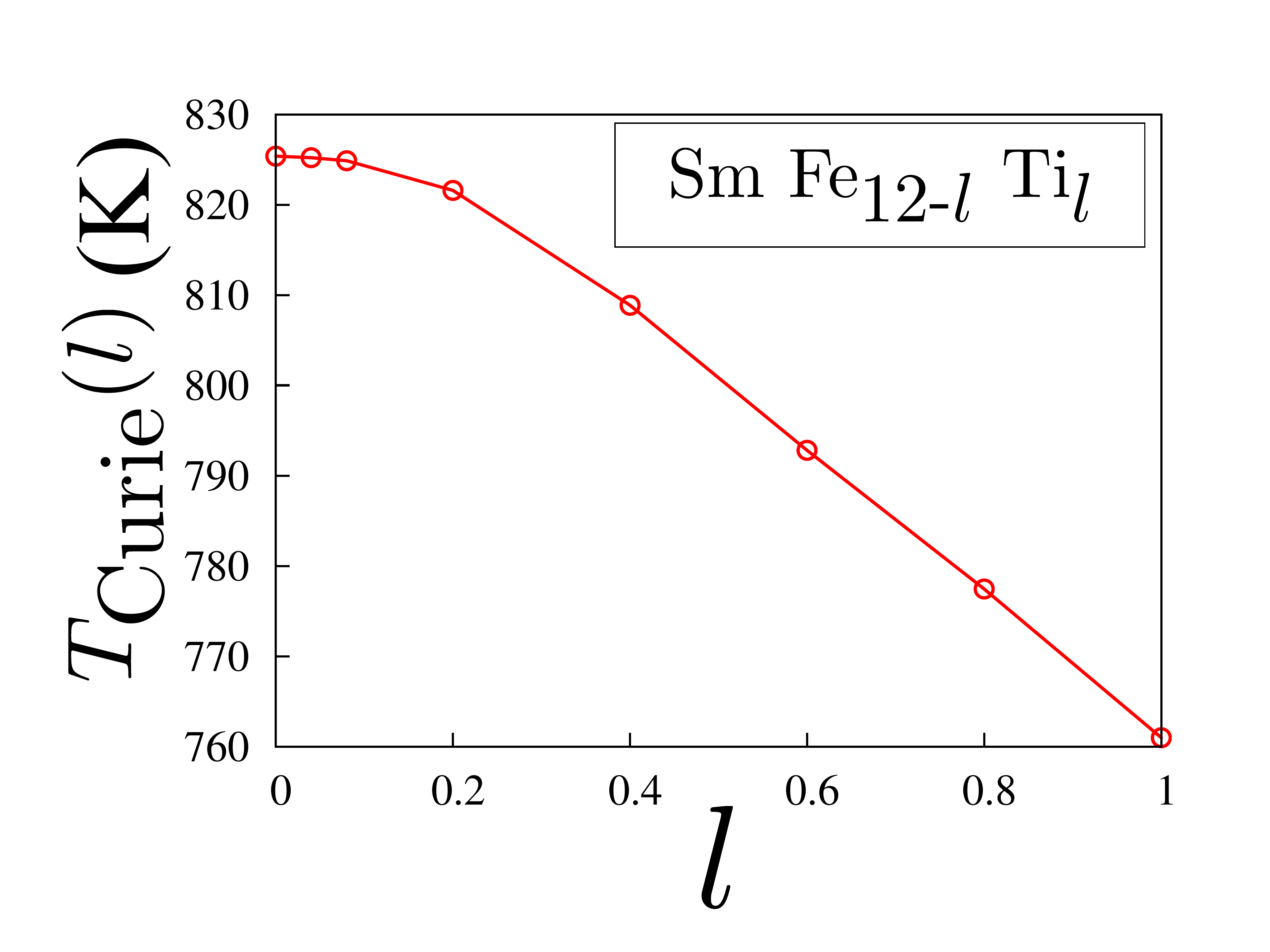}}
    &
    \scalebox{0.12}{\includegraphics{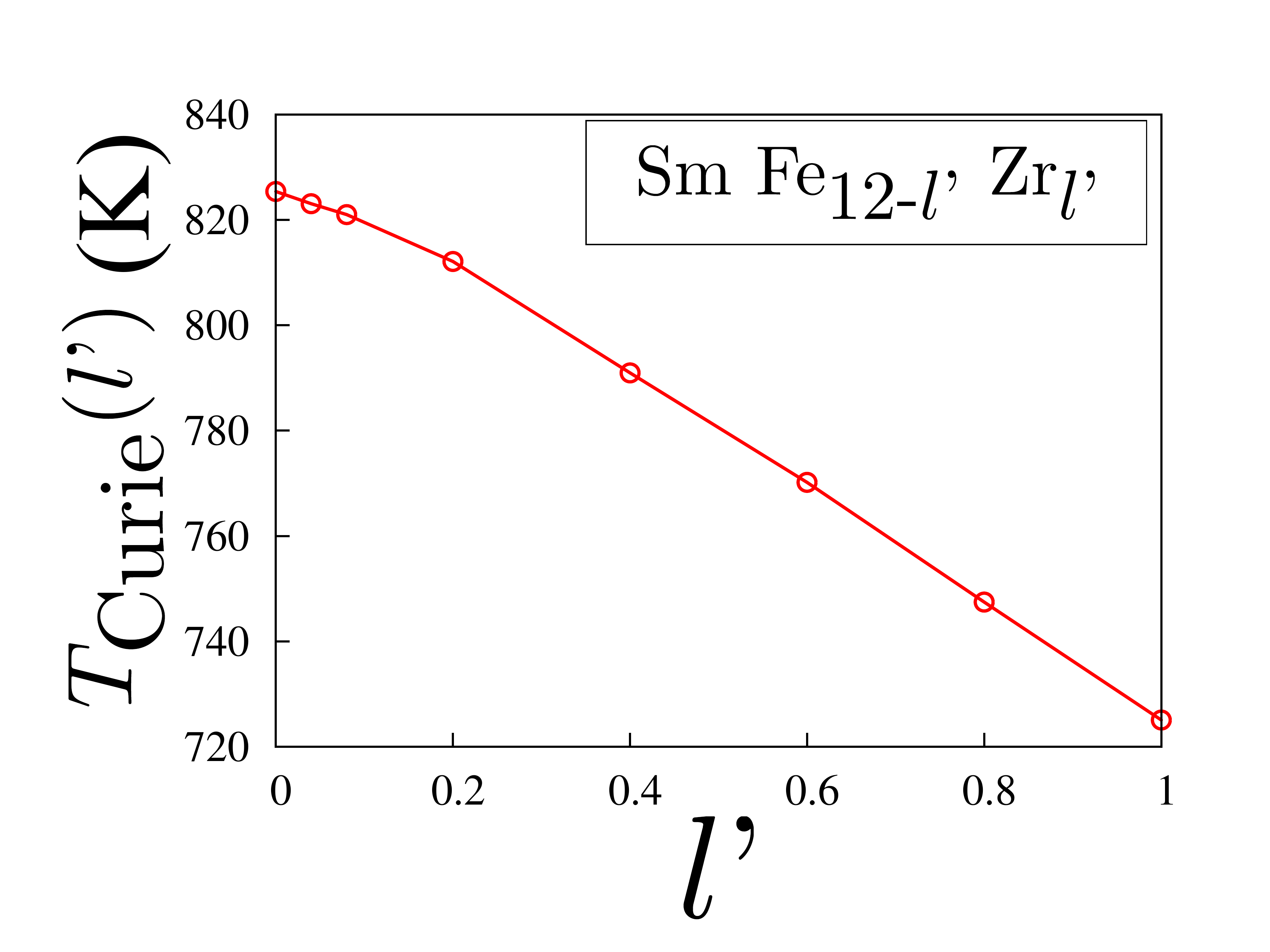}} \\
    (i) $J_{\mbox{Sm($2a$)Fe($8i$)}}(n)$ &
    (j) $J_{\mbox{Sm($2a$)Fe($8i$)}}(m)$ &
    (k) $J_{\mbox{Sm($2a$)Fe($8i$)}}(l)$ &
    (l) $J_{\mbox{Sm($2a$)Fe($8i$)}}(l')$ \\
    \scalebox{0.12}{\includegraphics{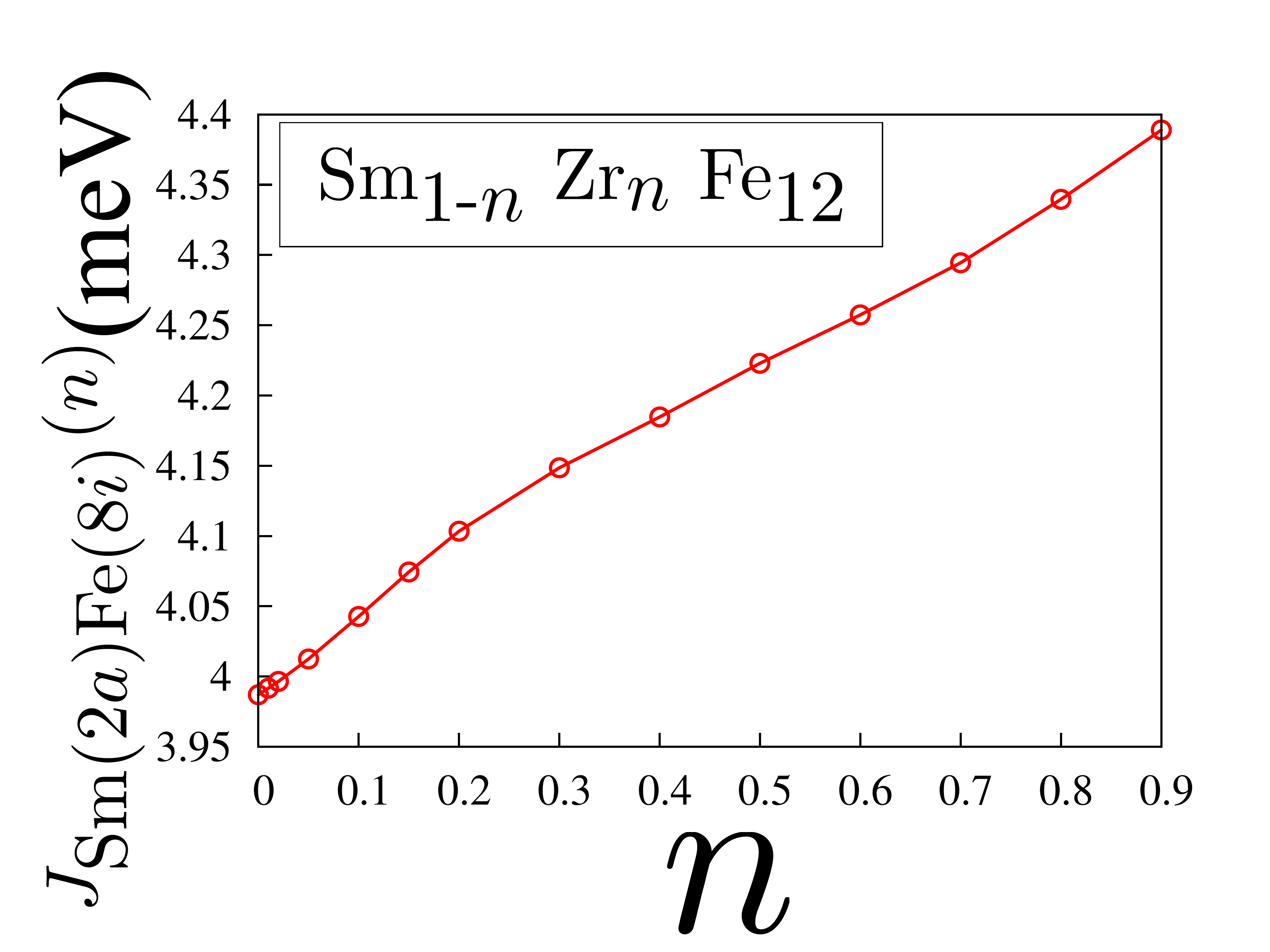}}
    &
    \scalebox{0.12}{\includegraphics{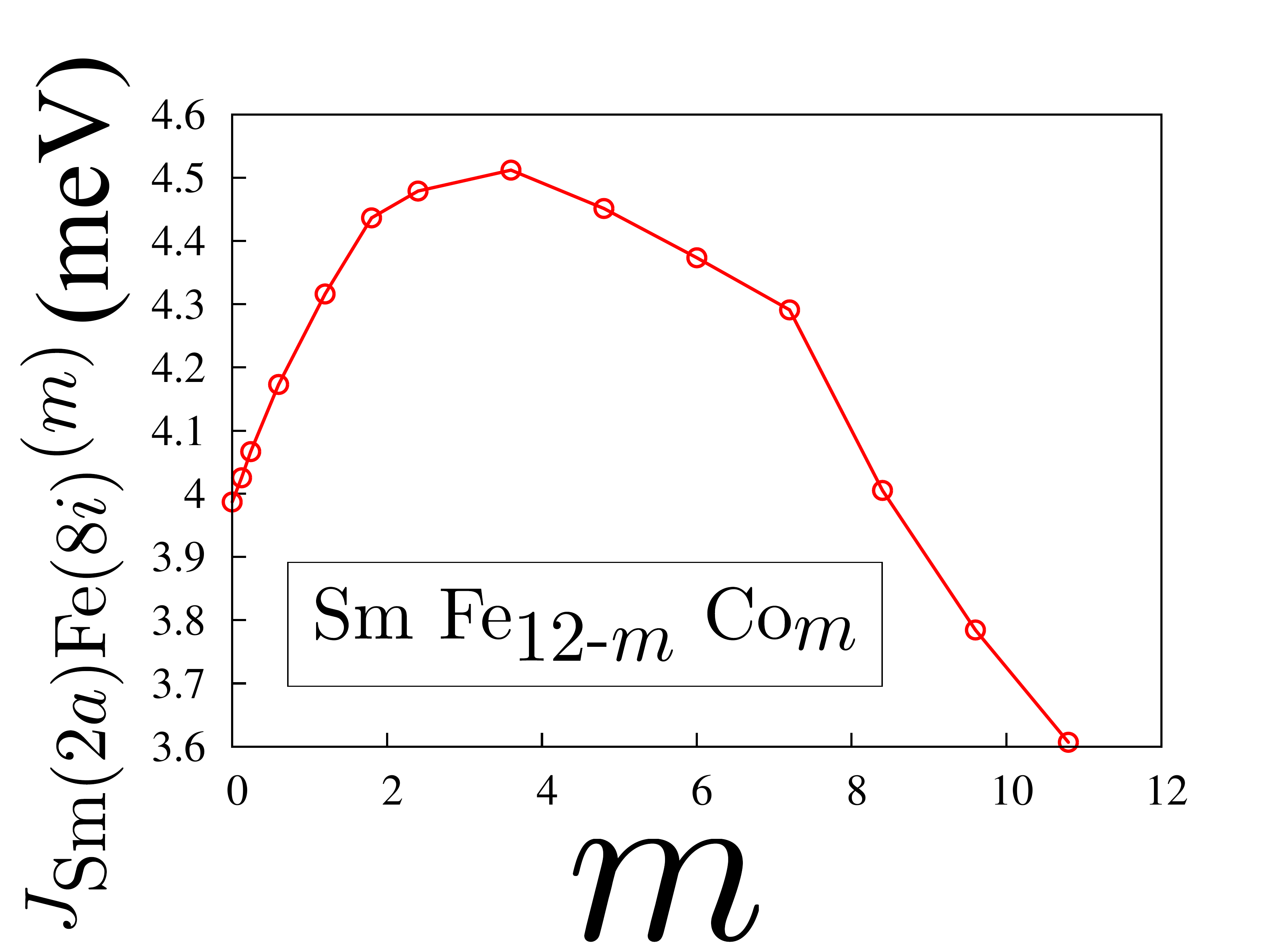}}
    &
    \scalebox{0.12}{\includegraphics{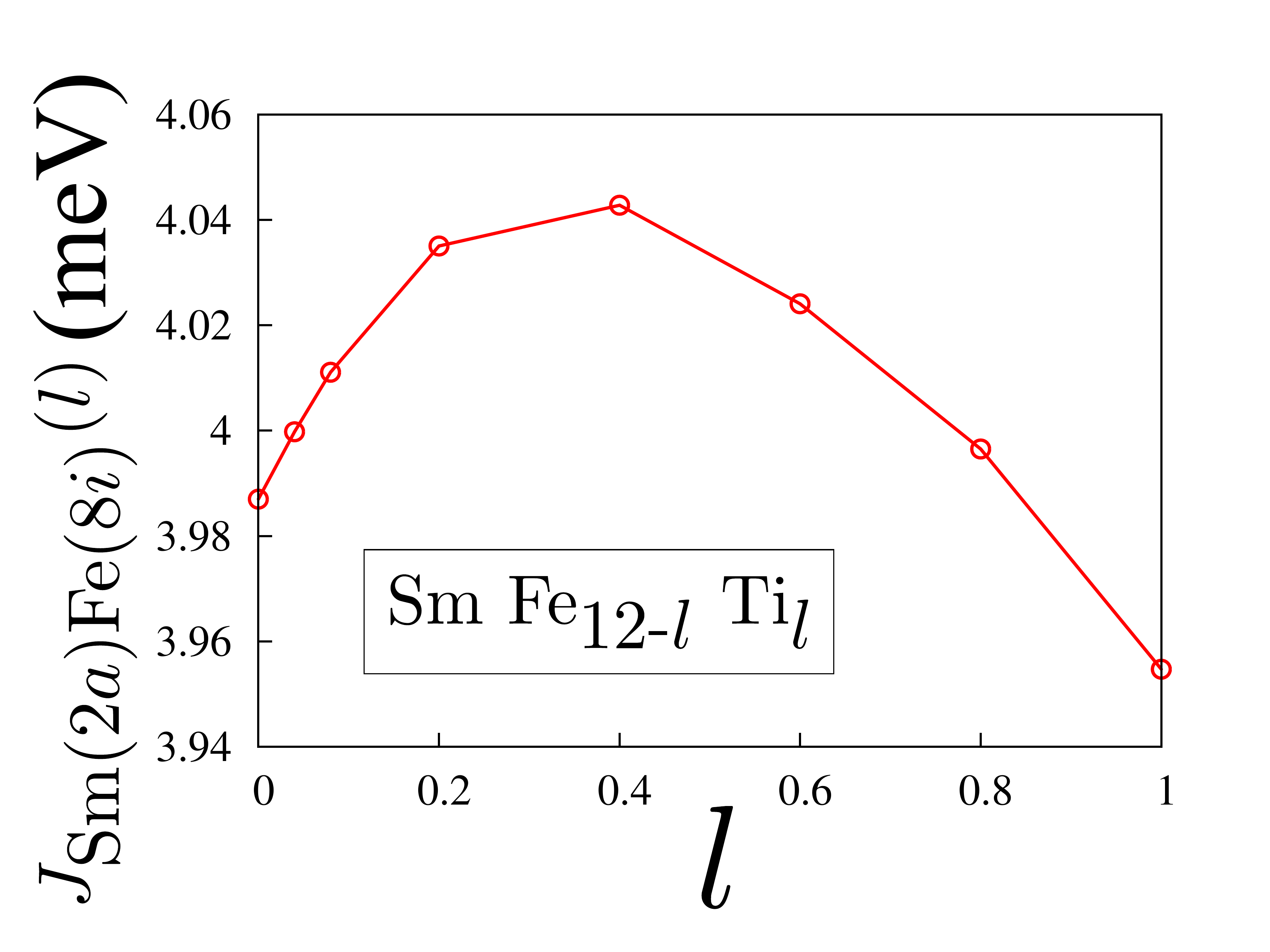}}
    &
    \scalebox{0.12}{\includegraphics{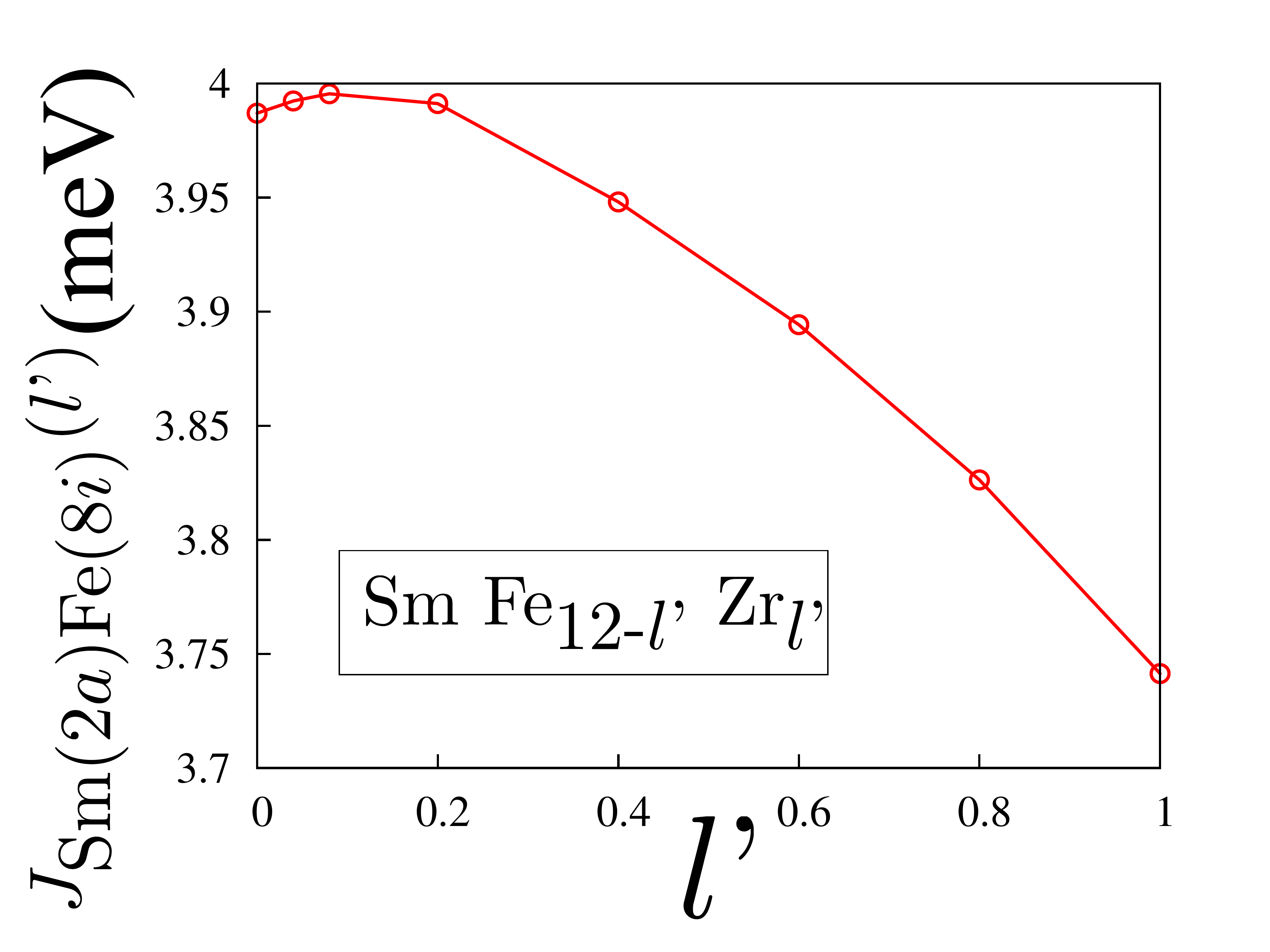}} \\
    (m) $J_{\mbox{Sm($2a$)Fe($8j$)}}(n)$ &
    (n) $J_{\mbox{Sm($2a$)Fe($8j$)}}(m)$ &
    (o) $J_{\mbox{Sm($2a$)Fe($8j$)}}(l)$ &
    (p) $J_{\mbox{Sm($2a$)Fe($8j$)}}(l')$ \\
    \scalebox{0.12}{\includegraphics{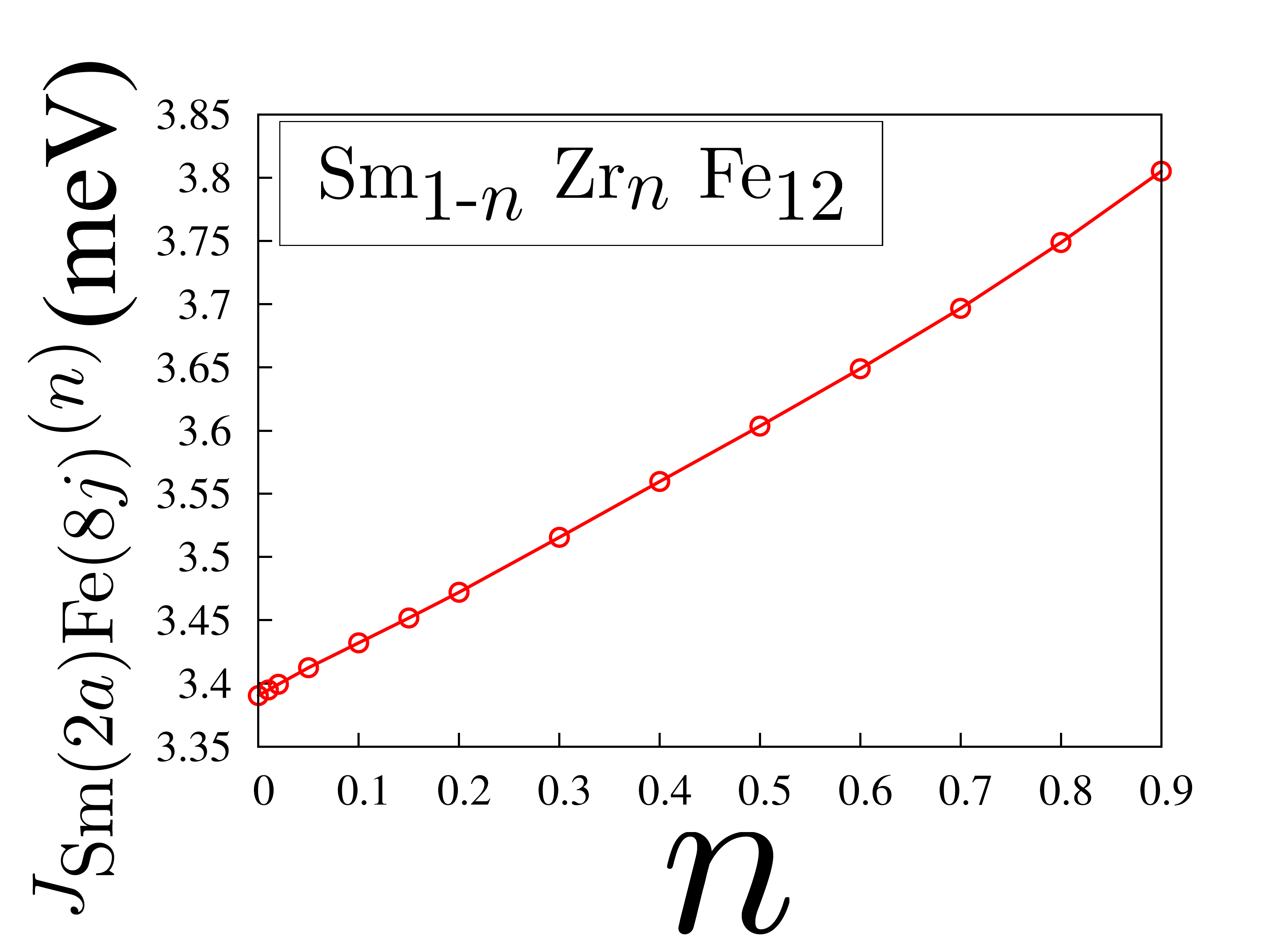}}
    &
    \scalebox{0.12}{\includegraphics{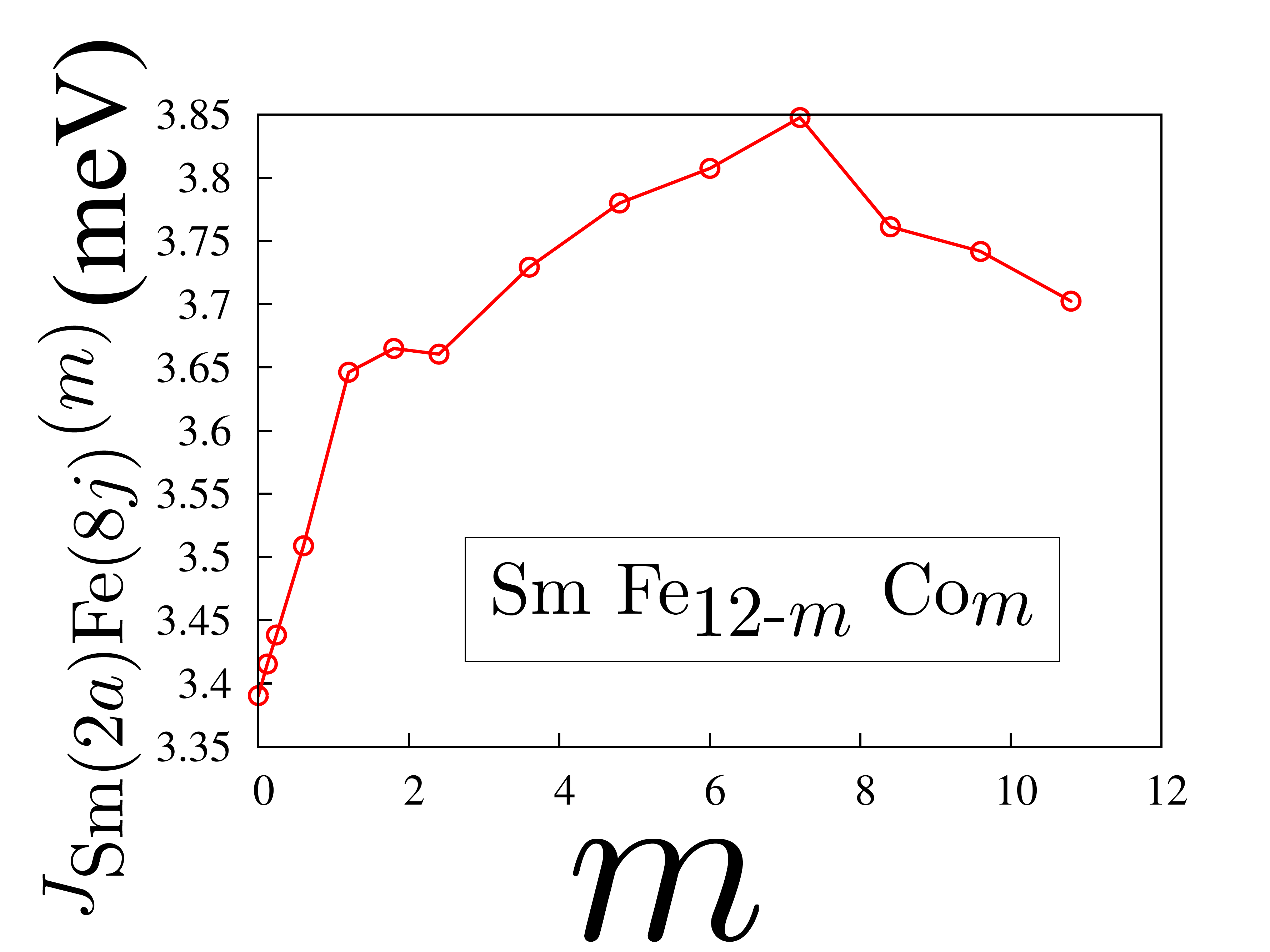}}
    &
    \scalebox{0.12}{\includegraphics{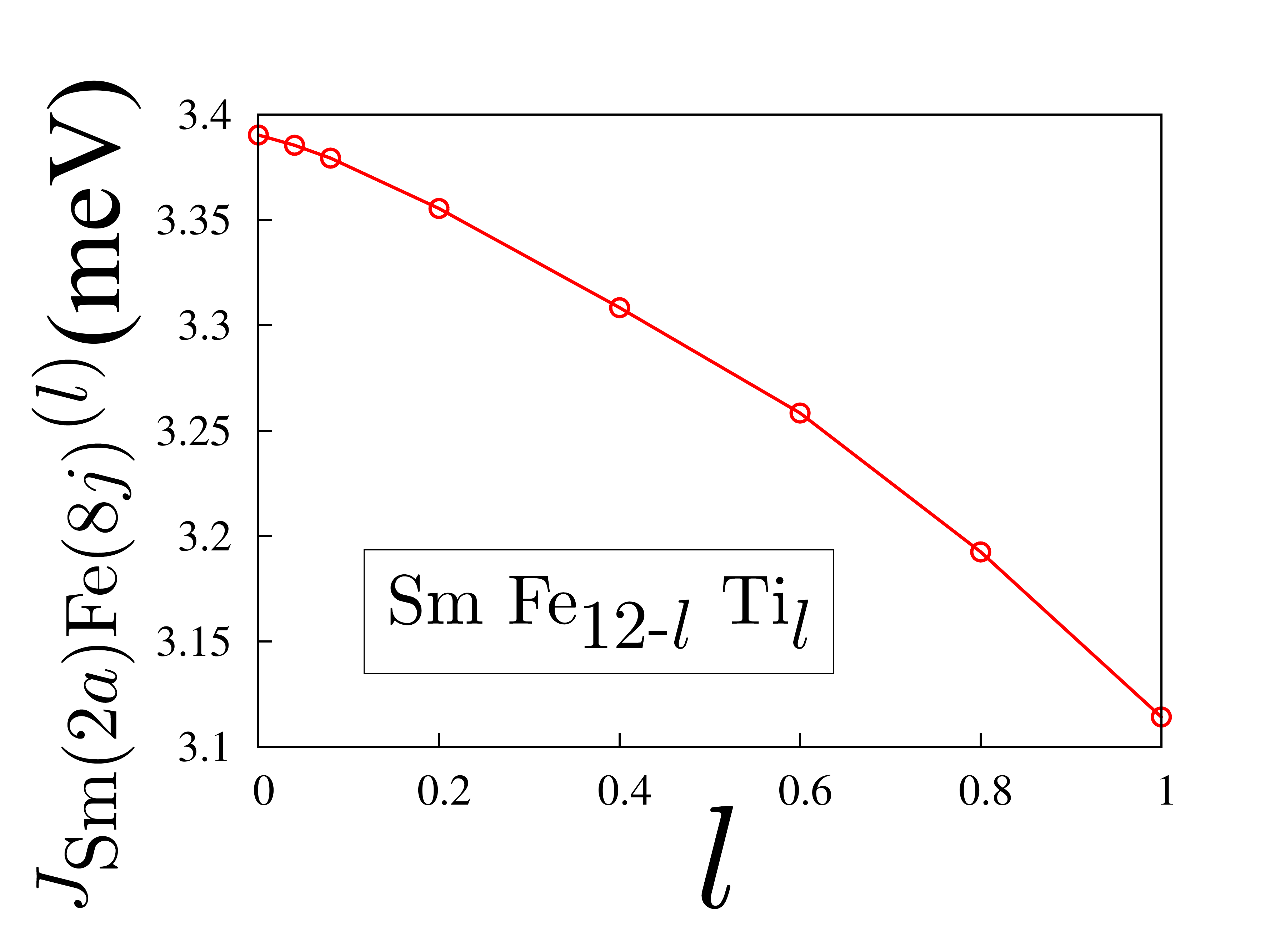}}
    &
    \scalebox{0.12}{\includegraphics{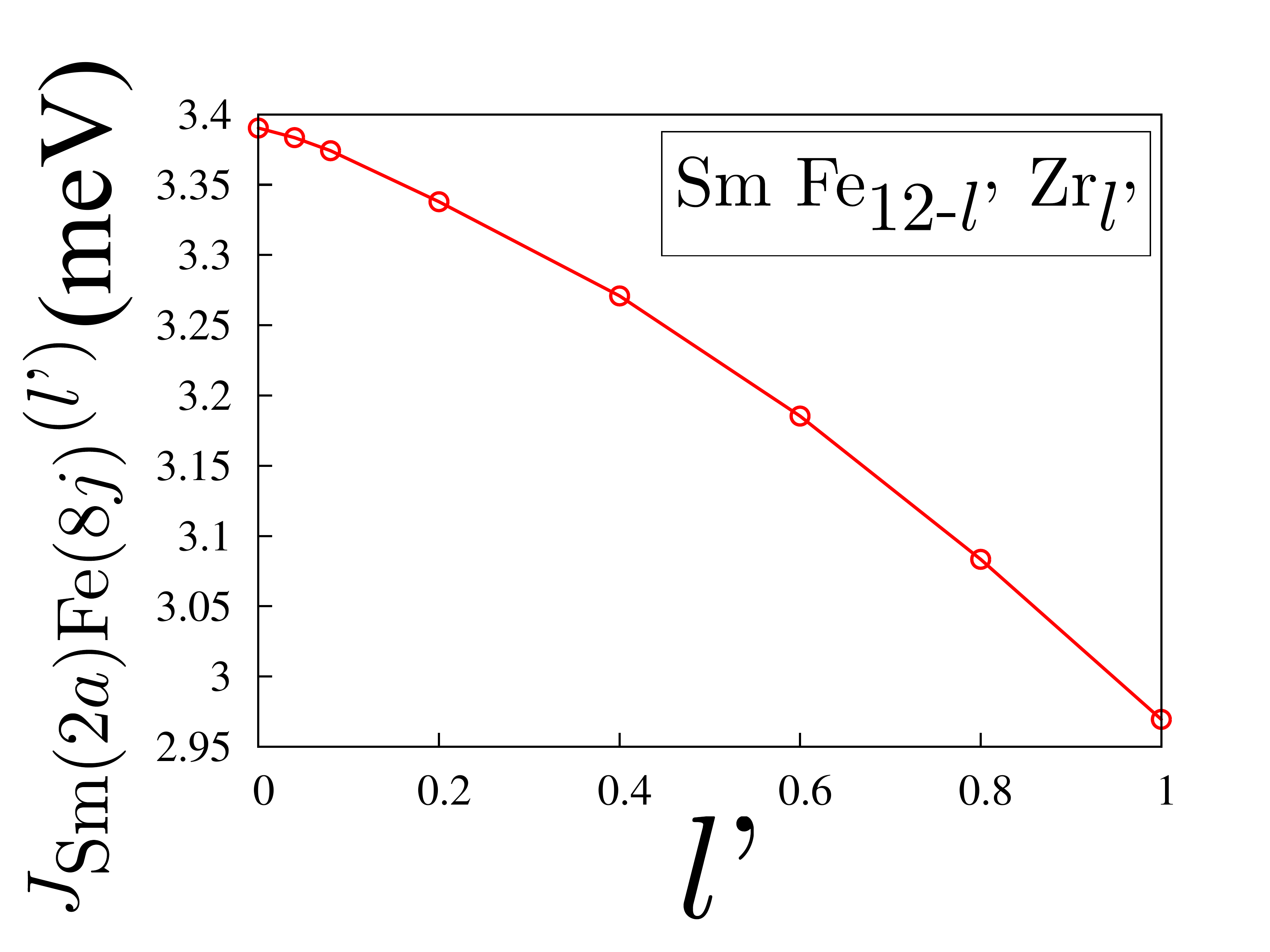}} \\
    (q) $J_{\mbox{Sm($2a$)Fe($8f$)}}(n)$ &
    (r) $J_{\mbox{Sm($2a$)Fe($8f$)}}(m)$ &
    (s) $J_{\mbox{Sm($2a$)Fe($8f$)}}(l)$ &
    (t) $J_{\mbox{Sm($2a$)Fe($8f$)}}(l')$ \\
    \scalebox{0.12}{\includegraphics{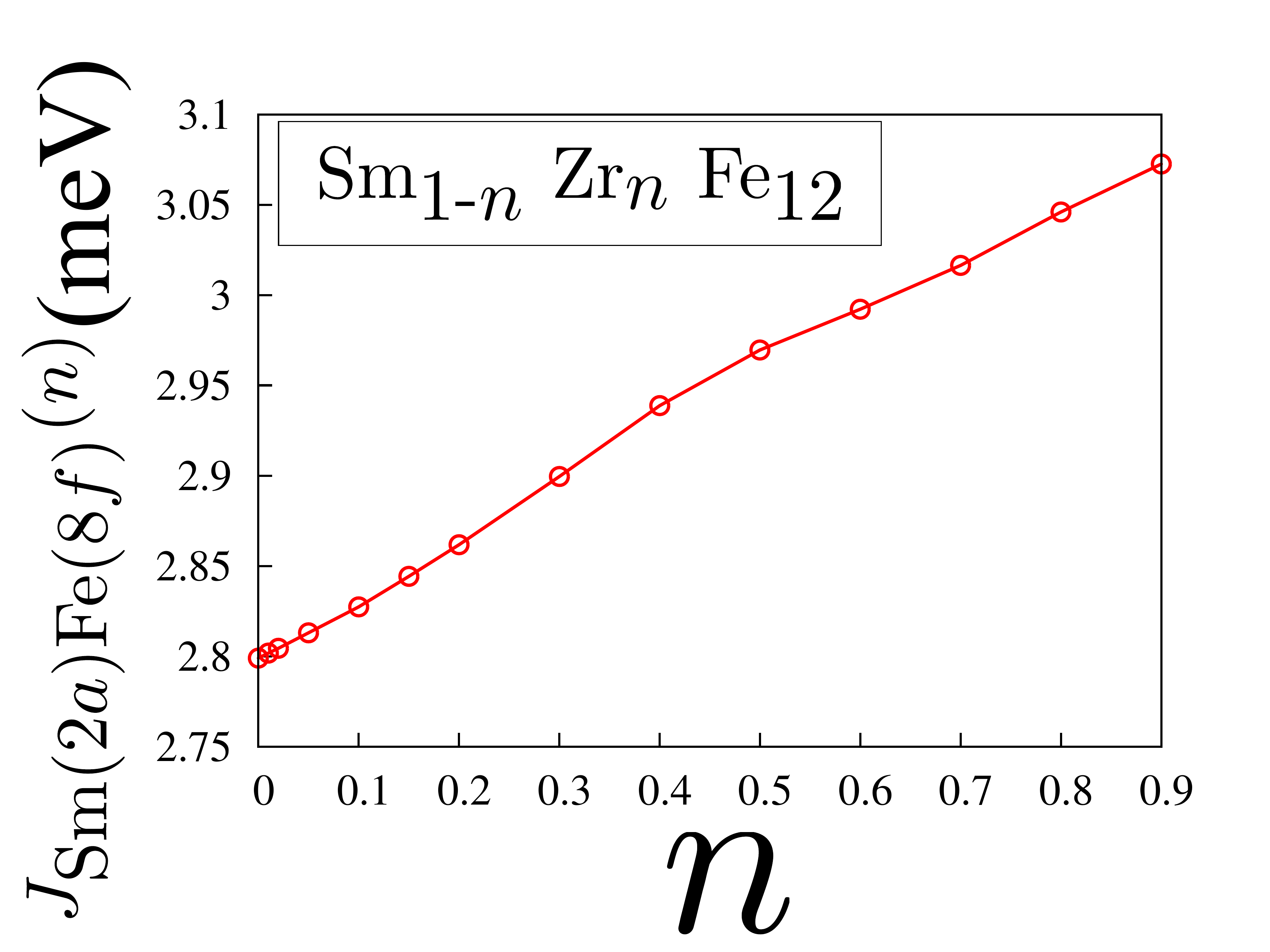}}
    &
    \scalebox{0.12}{\includegraphics{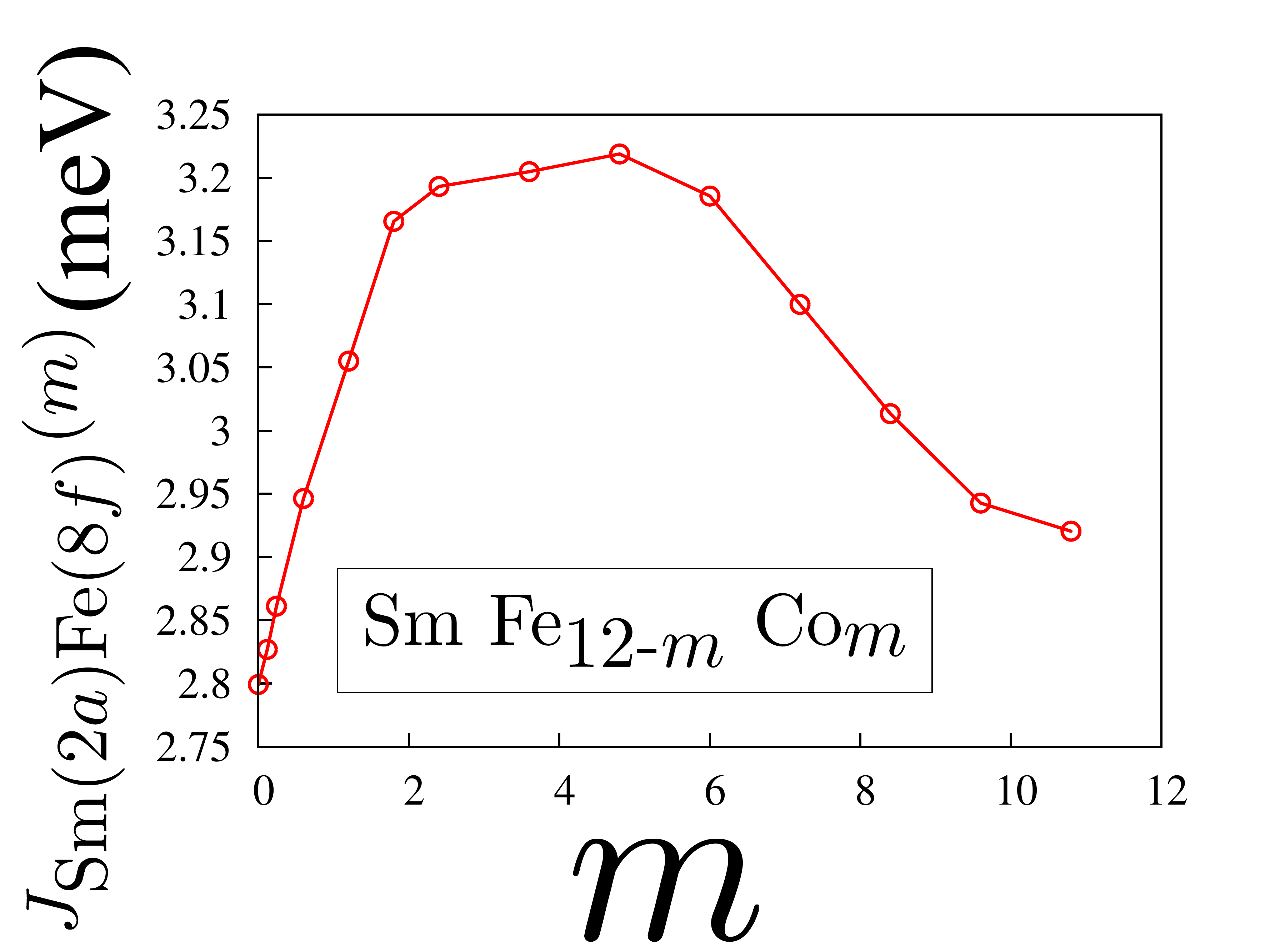}}
    &
    \scalebox{0.12}{\includegraphics{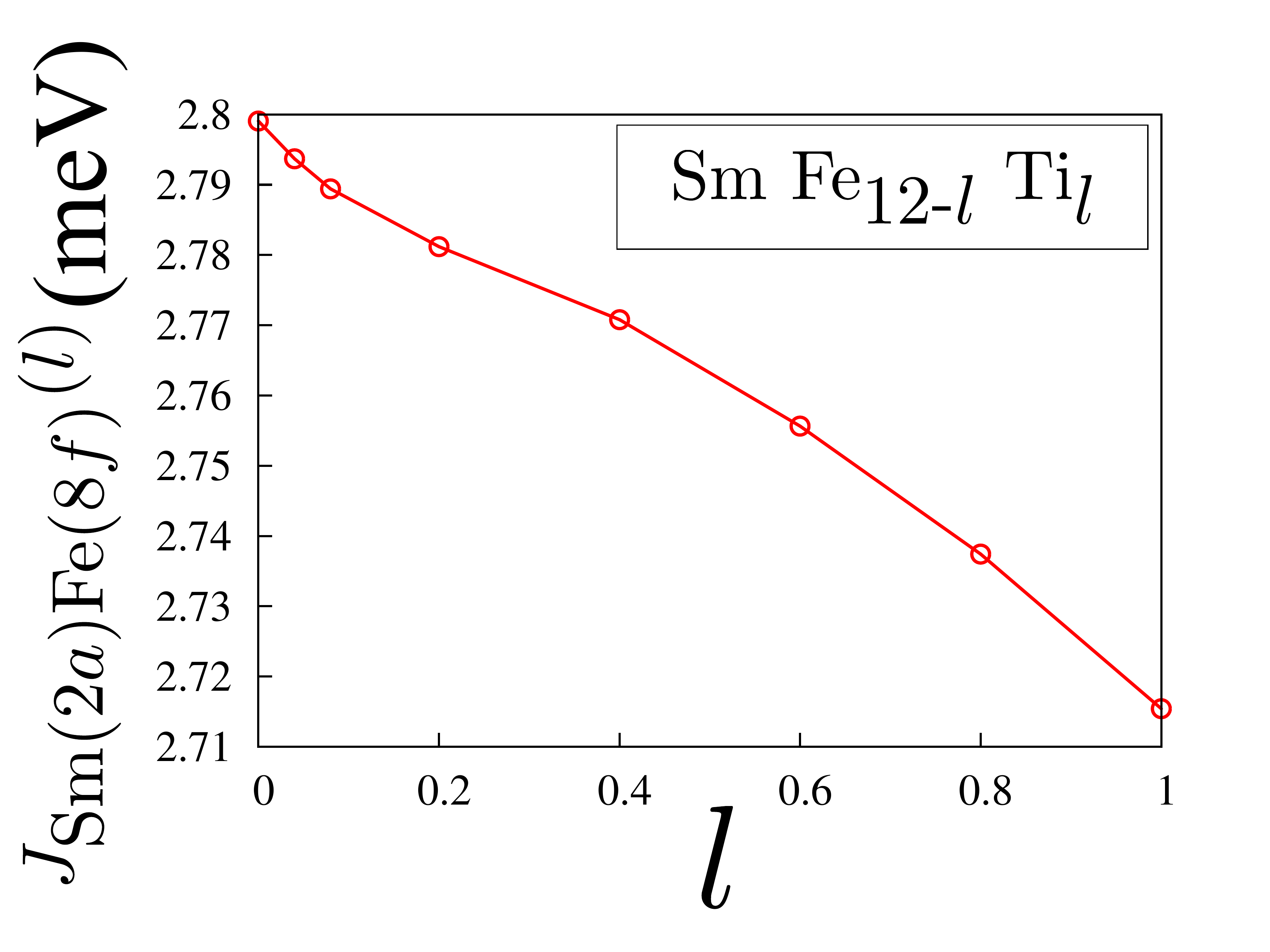}}
    &
    \scalebox{0.12}{\includegraphics{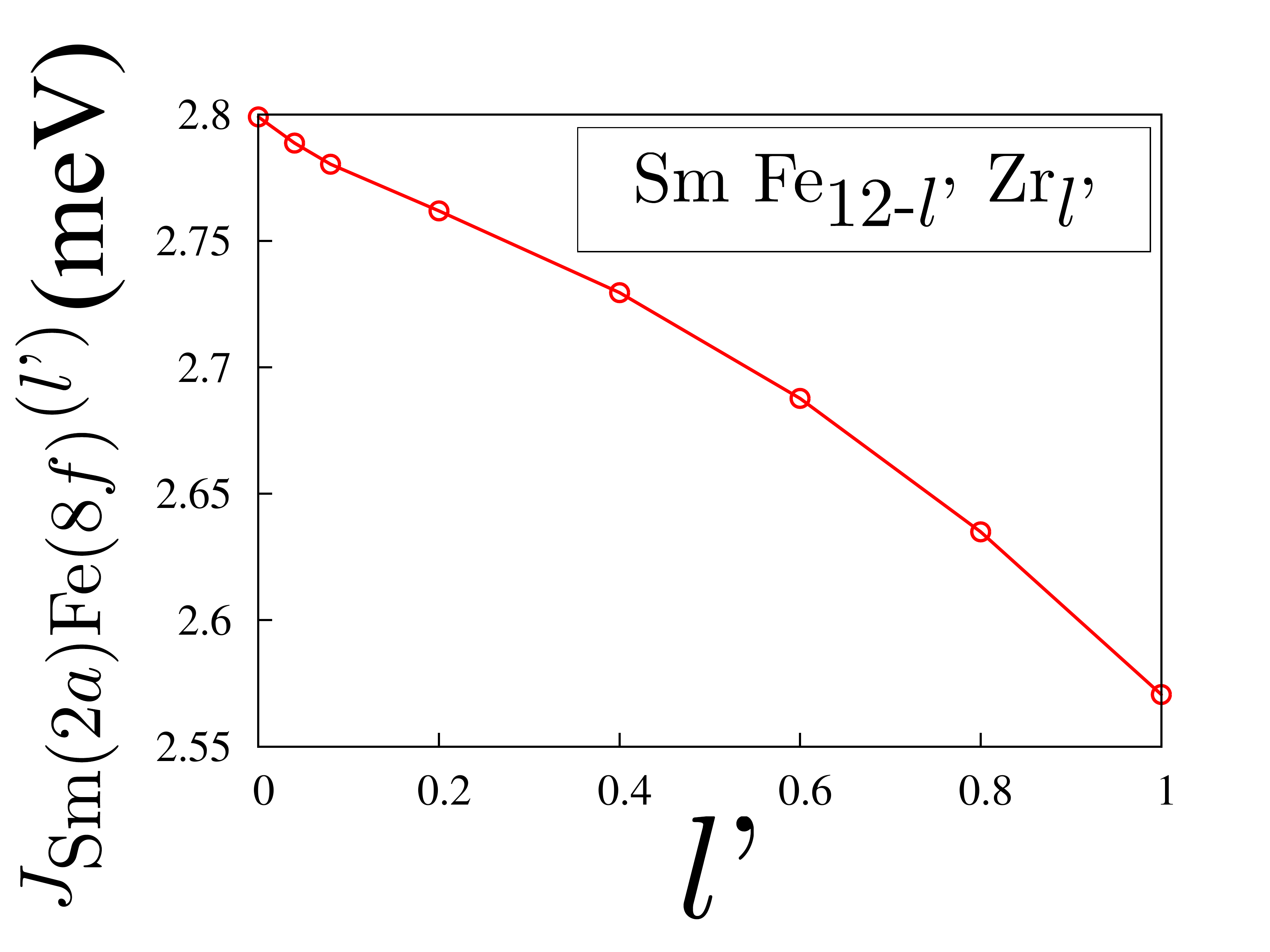}} \\
  \end{tabular}
  \caption{\label{fig::deriv}
    Deriving the derivative matrix
    of the intrinsic properties for the target alloy,
    Sm$_{1-n}$Zr$_n$(Fe$_{12-m-l-l'}$Co$_{m}$Ti$_l$Zr$_{l'}$)
    around the pristine limit, SmFe$_{12}$, from the calculated data via KKR-CPA using AkaiKKR.
    Dependence of magnetization $M$ per formula unit (f.u.),
    Curie temperature $T_{\rm Curie}$, and Sm-Fe indirect exchange couplings $J_{\rm RT}$ for Fe($8i$), Fe($8j$), Fe($8f$) that are crucial
    for anisotropy field in the operation temperature range, are plotted with respect to the concentration of substitute elements, Zr, Co, and Ti.
    For Zr two possible substitution sublattices, Sm($2a$) and Fe($8i$), are explored.}
\end{figure*}

On a fixed lattice of SmFe$_{12}$~\cite{harashima_2015}, we inspect the derivative
of the target observables as a function of the chemical composition parameters, namely,
$x$, $y$, $z$, and $z'$ in
(Sm$_{1-x}$Zr$_x$)(Fe$_{1-y-z/3-z'/3}$Co$_{y}$Ti$_{z/3}$Zr$_{z'/3}$)$_{12}$.
Here, $x$ denotes the concentration of the substitute element
within the Sm($2a$) sublattice, $y$ does the concentration of the substitute element
on the overall Fe sublattice, and $z$/$z'$ denote the concentration of the substitute element
within the Fe($8i$) sublattice.
They are not to be confused with the notation in Sec.~\ref{sec::calc_details::str_opt}
used in order to illustrate the definition of the formation energy.
Relating to the number of substitute elements per formula unit
$n$, $m$, $l$, and $l'$ in  Sm$_{1-n}$Zr$_{n}$Fe$_{12-m-l-l'}$Co$_{m}$Ti$_{l}$Zr$_{l'}$.
the following relations hold:
\begin{eqnarray*}
  n & = & x, \\
  m & = & 12y, \\
  l & = & 4z, \\
  l' & = & 4z'.
\end{eqnarray*}
Fixed-lattice approximation has been employed here
on the assumption that
chemical composition i.e. the variation in the electron number
affects the intrinsic properties more significantly than
the variation in the lattice parameters does as long as the linear extrapolation
around a reasonable stoichiometric limit~\cite{harashima_2015} is attempted.
Also the effect of the celebrated Slater-Pauling curve can be more easily demonstrated
with the smaller lattice constants. Thus the smaller {\it ab initio} parameter set from Table I in the main text
has been taken. The other parameter sets would yield the similar messages.
The overall data can be glanced in Fig.~\ref{fig::deriv}. Here the point is to confirm
the existence of the linear regime spanning a reasonable range around the pristine/stoichiometric limit
in the chemical composition space. Then within such linear regime, we can discuss
at which parameter range the target observables can be optimized on demand.

From the data shown in Fig.~\ref{fig::deriv}, the derivative of an observable ${\cal O}$
around the pristine limit
is obtained as a difference between $c=0$ and $c=0.01$, where $c=x$, $y$, $z$, or $z'$ is the concentration of the substitute elements,
and then multiplied by $100$:
\begin{equation}
  \frac{\partial{\cal O}}{\partial c}\equiv
  \frac{\left.{\cal O}\right|_{c=0.01}-\left.{\cal O}\right|_{c=0}}{0.01}
  \label{eq::finite_diff}
\end{equation}
Numbers during such derivation of the derivative matrix in the main text
are displayed in Table~\ref{table::derivative}.
Eventually they are translated into the derivative
with respect to the number of substitute elements per formula unit
and are summarized in Eq.~(\ref{eq::dev_mat}) in the main text.
\begin{table*}
    \begin{center}
  \begin{tabular}{l}
    (a) raw data \\
    \begin{tabular}{lccccc}\hline
       & pristine & $x=0.01$ & $y=0.01$ & $z=0.01$ & $z'=0.01$ \\ \hline
   $M~\mbox{($\mu_{\rm B}$/f.u.)}$       &  $24.1568$ & $24.1627$   & $24.2445$ & $23.9738$ & $23.9853$ \\ \hline
 $T_{\rm Curie}~\mbox{(K)}$  & $825.377$  &  $827.001$  & $848.566$ & $825.219$  & $823.075$ \\ \hline
 $J_{\rm RT}(8f)~\mbox{(meV)}$ & $2.799074$ & $2.801749$ & $2.827002$ & $2.793691$ & $2.788702$ \\ \hline
 $J_{\rm RT}(8i)~\mbox{(meV)}$ & $3.986962$ & $3.991619$ & $4.025465$ & $3.999788$ & $3.992307$ \\ \hline
 $J_{\rm RT}(8j)~\mbox{(meV)}$ & $3.390305$ & $3.394893$ & $3.415285$ & $3.385382$ & $3.383555$ \\ \hline
    \end{tabular}\\
    \\
  (b) the derivatives defined as Eq.~(\ref{eq::finite_diff})\\
  \begin{tabular}{lcccc}\hline
    $c$, the concentration of the substitute elements & $x$ & $y$ & $z$ & $z'$ \\ \hline
    $\partial M/\partial c~\mbox{($\mu_{\rm B}$/f.u.)}$            & $0.59$    & $8.77$  & $-18.3$ & $-17.15$ \\ \hline
$\partial T_{\rm Curie}/\partial c~\mbox{(K)}$  & $162.4$   & $2318.9$ & $-15.8$ & $-230.2$ \\ \hline
$\partial J_{\rm RT}(8f)/\partial c~\mbox{(meV)}$ & $0.2675$ & $2.7928$ & $-0.5383$ & $-1.0372$ \\ \hline
$\partial J_{\rm RT}(8i)/\partial c~\mbox{(meV)}$ & $0.4657$ & $3.8503$ & $1.2826$ & $0.5345$ \\ \hline
$\partial J_{\rm RT}(8j)/\partial c~\mbox{(meV)}$ & $0.4588$ & $2.498$ & $-0.4923$ & $-0.675$ \\ \hline
  \end{tabular}\\
  \\
  (c) Normalized derivatives \\
  \begin{tabular}{lcccc}\hline
                                                    & $x$       & $y$       & $z$        & $z'$ \\ \hline
$(\partial M/\partial c)/M$                         & $0.0244238$ & $0.363045$  & $-0.757551$  & $-0.709945$ \\ \hline
$(\partial T_{\rm Curie}/\partial c)/T_{\rm Curie}$    & $0.196759$  & $2.8095$    & $-0.0191428$ & $-0.278903$ \\ \hline
$[\partial J_{\rm RT}(8f)/\partial c]/J_{\rm RT}(8f)$ & $0.0955673$ & $0.997759$  & $-0.192314$  & $-0.370551$  \\ \hline
$[\partial J_{\rm RT}(8i)/\partial c]/J_{\rm RT}(8i)$ & $0.116806$  & $0.965723$  & $0.321699$   & $0.134062$    \\ \hline
$[\partial J_{\rm RT}(8j)/\partial c]/J_{\rm RT}(8j)$ & $0.135327$  & $0.736807$  & $-0.145208$  & $-0.199097$  \\ \hline
  \end{tabular}\\
  (d) Normalized derivatives with respect to the substitute atom number per formula unit, as used in the main text\\
  \begin{tabular}{lcccc}\hline
    & $n\equiv x$        & $m\equiv 12y$       & $l\equiv 4z$        & $l'\equiv 4z'$ \\ \hline
$(\partial M/\partial p)/M$    & $0.0244238$ &  $0.0302537$ & $-0.189388$   & $-0.177486$  \\ \hline
$(\partial T_{\rm Curie}/\partial p)/T_{\rm Curie}$    & $0.196759$  &  $0.234125$  & $-0.0047857$  & $-0.0697258$ \\ \hline 
$[\partial J_{\rm RT}(8f)/\partial p]/J_{\rm RT}(8f)$ & $0.0955673$ &  $0.0831466$ & $-0.0480785$  & $-0.0926378$ \\ \hline
$[\partial J_{\rm RT}(8i)/\partial p]/J_{\rm RT}(8i)$ & $0.116806$  &  $0.0804769$ & $0.0804248$   & $0.0335155$  \\ \hline
$[\partial J_{\rm RT}(8j)/\partial p]/J_{\rm RT}(8j)$ & $0.135327$  &  $0.0614006$ & $-0.036302$   & $-0.0497742$ \\ \hline
  \end{tabular}
  \end{tabular}
  \end{center}
    \caption{\label{table::derivative} Numbers for the calculated derivatives from finite difference
      around the pristine limit for Sm$_{1-x}$Zr$_{x}$(Fe$_{1-y-z/3-z'/3}$Co$_{y}$Ti$_{z/3}$Zr$_{z'/3}$)$_{12}$,
      or equivalently,
      Sm$_{1-n}$Zr$_{n}$(Fe$_{12-m-l-l'}$Co$_{m}$Ti$_{l}$Zr$_{l'}$)
      obtained from the data presented in Fig.~\ref{fig::deriv}.
      In (d), the number of substitute atoms has been denoted by $p=n$, $m$, $l$, or $l'$.}
  \end{table*}

\begin{figure}
  \scalebox{0.2}{\includegraphics{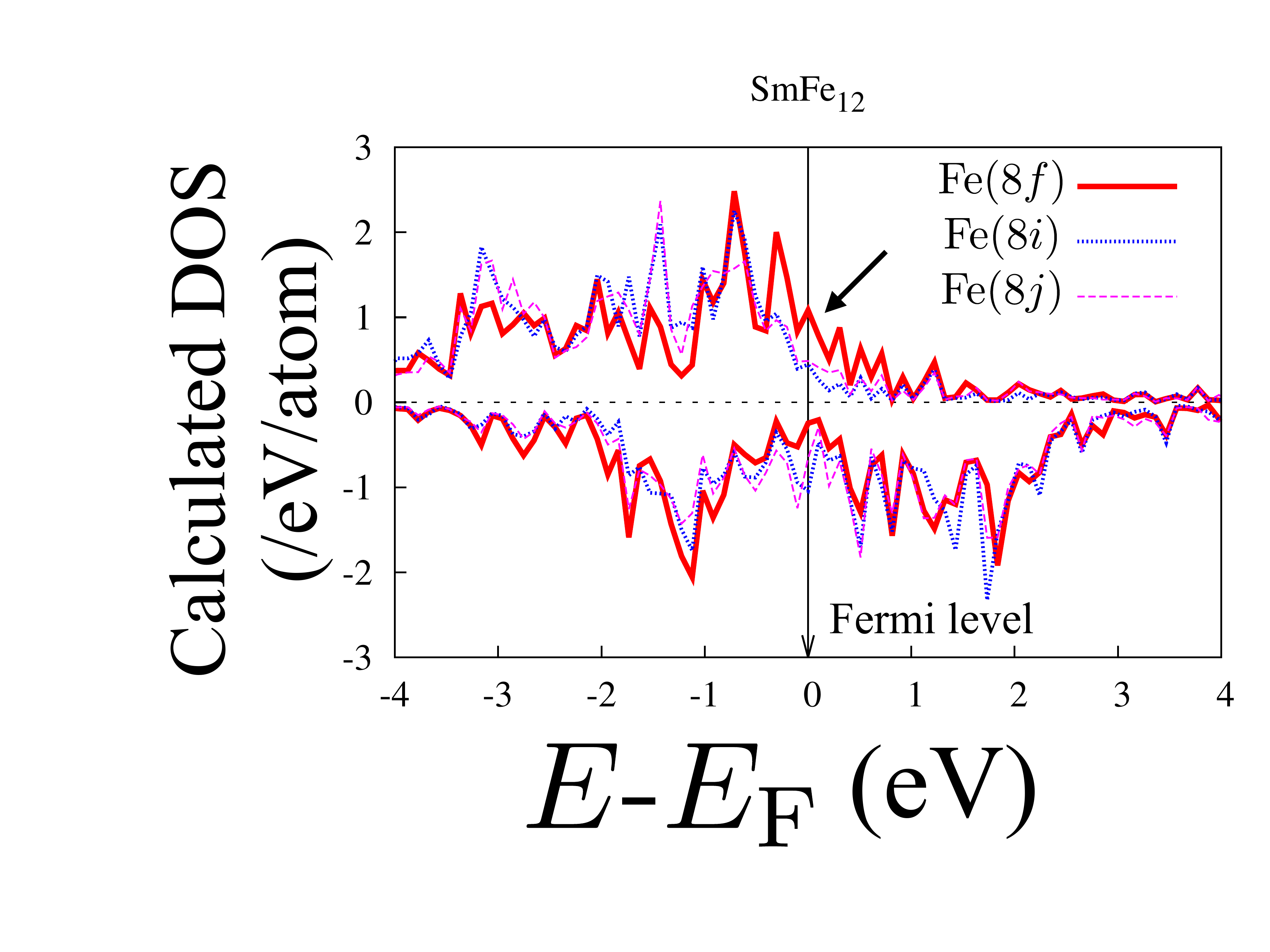}} \\
  \caption{\label{fig::dos} (Color online)
    Calculated density of states (DOS)
    projected onto Fe atoms for the pristine SmFe$_{12}$
    on an optimized lattice~\cite{harashima_2015}.
    The horizontal axis shows the energy $E$ as measured from the Fermi level, $E_{\rm F}$.
    The arrow in the plot points to the extra density of states
    in the majority spin state projected onto Fe($8f$). Additional Co reduces this region leading to the Slater-Pauling curve.}
 \end{figure}
\section{Slater-Pauling curve and electron-doping effect in SmFe$_{12}$}
\label{sec::e-doping}
\begin{figure}
  \begin{tabular}{l}
    (a) \\
    \scalebox{0.2}{\includegraphics{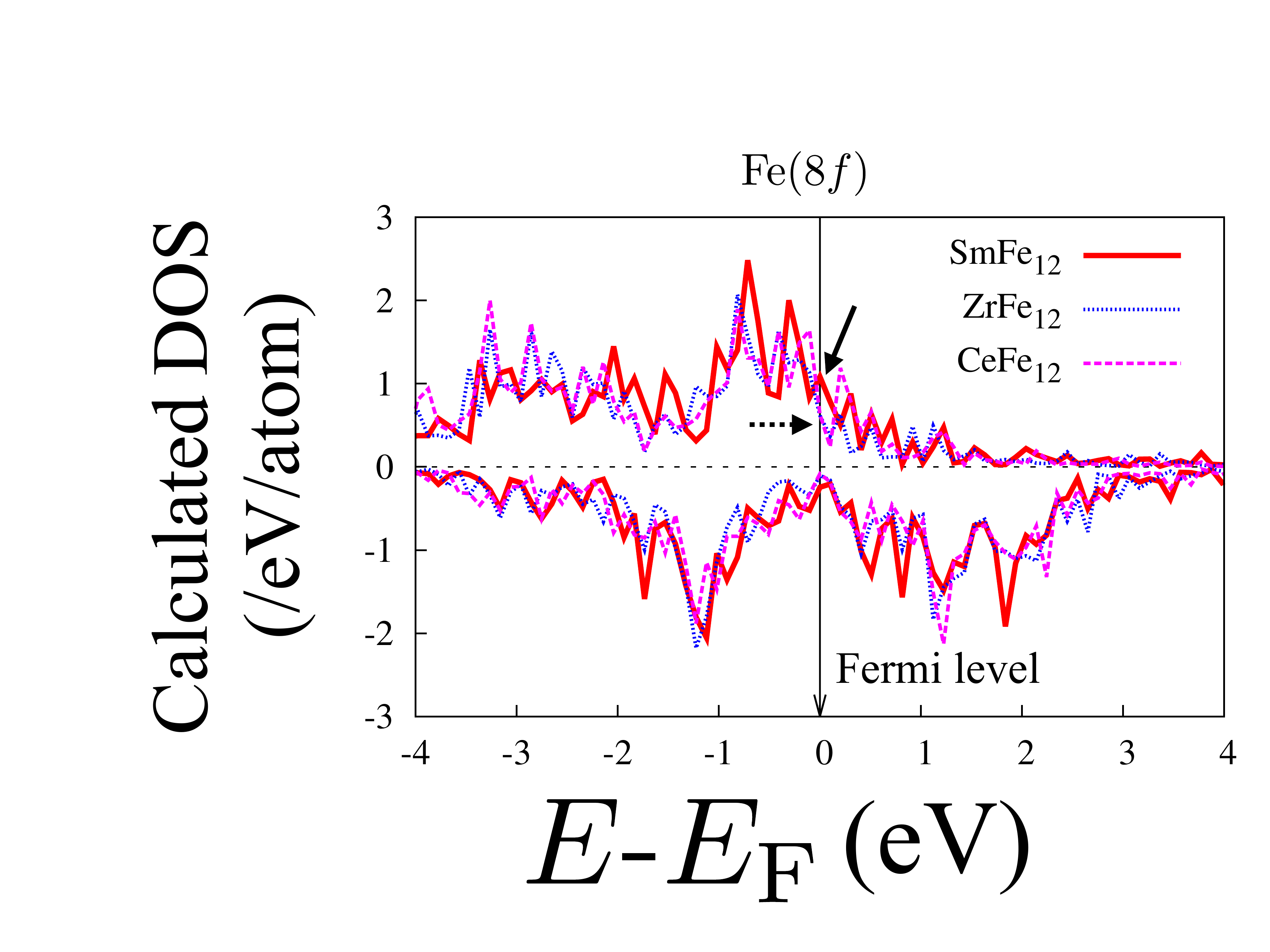}} \\
    (b) \\
    \scalebox{0.2}{\includegraphics{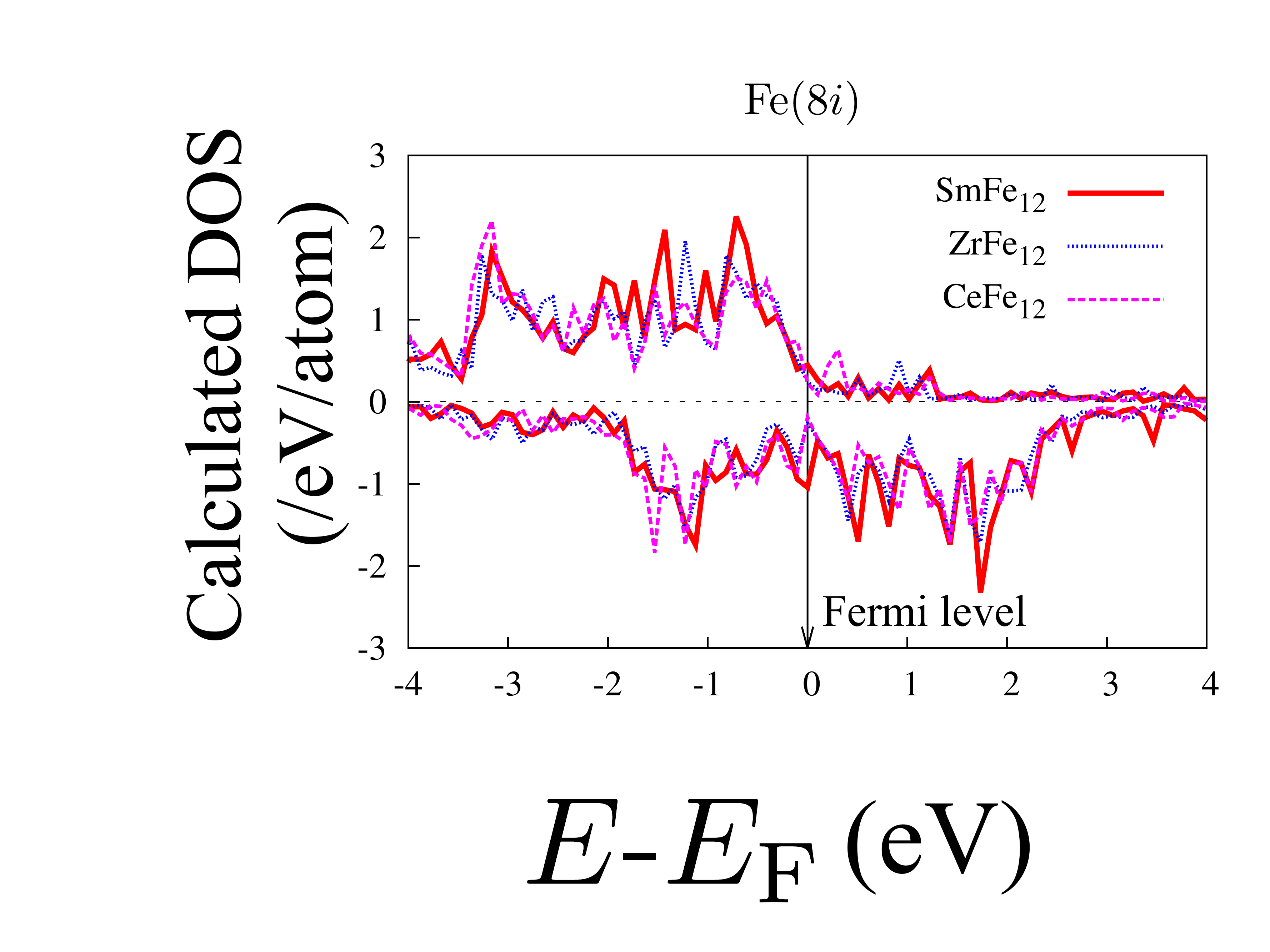}} \\
    (c) \\
    \scalebox{0.2}{\includegraphics{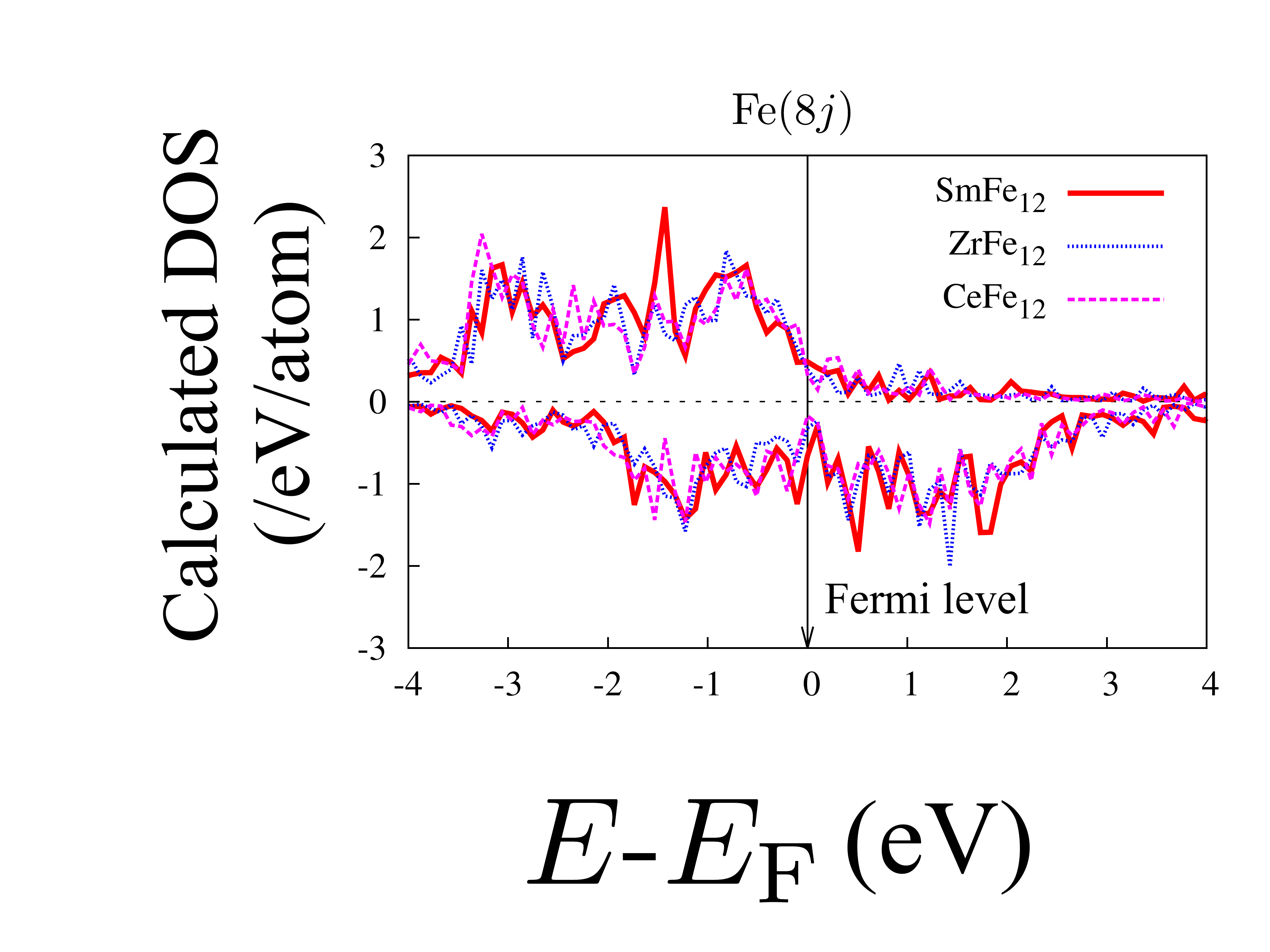}} 
  \end{tabular}
  \caption{\label{fig::dos_ce_and_zr_analogues} (Color online)
    (a) Calculated density of states projected onto Fe($8f$) atoms
    for SmFe$_{12}$, ZrFe$_{12}$, and CeFe$_{12}$ on the same fixed lattice. The full-line arrow
    in the plot points to the majority-spin density of states on the Fermi level which is seen
    significantly 
    mostly for SmFe$_{12}$. Dotted arrow points to the relatively sparse contribution from the majority spin
    of Fe($8f$) in CeFe$_{12}$ and ZrFe$_{12}$, which means that magnetic moments on Fe($8f$) are close to saturation
    for those materials.
    (b) and (c): Analogous data to (a) for Fe($8i$) and Fe($8j$), respectively.
    The axis labels follows them in Fig.~\ref{fig::dos}.
  }
\end{figure}

Co-substituted SmFe$_{12}$ on a fixed lattice of computational optimization~\cite{harashima_2015}
nicely shows the Slater-Pauling curve as shown in Fig.~\ref{fig::deriv}~(b).
The origin of this can be tracked down to the residual density of states around the Fermi level
in the majority spin band as shown in Fig.~\ref{fig::dos}. This contribution
comes mostly from the Fe($8f$) sublattice. With the introduction of Co, extra electrons are added in the minority spin band,
and we have noted that
the analogous effect is given by Zr and also by Ce substituting the Sm($2a$) sublattice.
As a comparison
calculated density of states for ZrFe$_{12}$ and CeFe$_{12}$ are shown in Figs.~\ref{fig::dos_ce_and_zr_analogues}~(a), (b), and (c)
for the data projected onto the Fe($8f$) atoms, Fe($8i$), and Fe($8j$), respectively. In Fig.~\ref{fig::dos_ce_and_zr_analogues}~(a)
it is seen that replacement of Sm with Zr or Ce has pushed down the majority-spin band down below the Fermi level,
leading to the enhancement of magnetization.

Zr($2a$)-induced enhancement of magnetization
and the same effect from Ce$^{4+}$ mostly happens
as an extra electron is doped in the minority-spin state on the
Fe($8f$) sublattice indicates that
charge-transfer from the Sm($2a$) site to the Fe($8f$) site is the key factor
in realizing the analogous behavior to Slater-Pauling curve
as a function of substitute elements in the Sm($2a$) sublattice.
Here we note that substituted Ce$^{4+}$ supplying extra electrons in the host system is apparently
analogous to electron-doped cuprates~\cite{tokura,naito}. While the host system,
which is a metal that is not so good in terms of electric conduction,
resides on the opposite side of the presumed Mott insulator for cuprates~\cite{mott}
in a virtual materials space harboring a metal-insulator transition~\cite{imada},
doping seems to work on the localized part of the electronic state
that seems to be actually common between our $4f$-$3d$ intermetallic ferromagnets and doped Mott insulators.

Understanding and control of such local mechanism in the real space should help in implementing
fortunate cases like Zr-substituted SmFe$_{12}$~\cite{tozman_2019}, where
both magnetic ferromagnetism and structure stability
can be gained at the same time.


\begin{thebibliography}{99}

\bibitem{sagawa_1984} M.~Sagawa,
  S.~Fujimura, N.~Togawa, H.~Yamamoto, Y.~Matsuura,
  New material for permanent magnets on a base of Nd and Fe,
  J.~Appl.~Phys. {\bf 55}, 2083 (1984).

\bibitem{croat_1984} J.~J.~Croat, J.~F.~Herbst, R.~W.~Lee, F.~E.~Pinkerton,
  Pr-Fe and Nd-Fe-based materials:
  A new class of high-performance permanent magnets,
  J.~Appl.~Phys. {\bf 55}, 2078 (1984)
  

\bibitem{oesterreicher_1984}
  F.~Spada, C.~Abache, H.~Oesterreicher,
  Crystallographic and magnetic properties of rare earth-transition metal compounds based on boron,
  J.~Less-Common Met. {\bf 99}, L21 (1984).

  
\bibitem{rmp_1991}
  For a review, see J.~F.~Herbst,
  R$_2$Fe$_{14}$B materials: Intrinsic properties and technological aspects,
  Rev.~Mod.~Phys. {\bf 63}, 819 (1991).

\bibitem{hono_2012}
  For a review, see
  K.~Hono and H.~Sepehri-Amin,
  Strategy for high-coercivity Nd-Fe-B magnets,
  Scr.~Mater. {\bf 67}, 530 (2012).
  
\bibitem{miyake_2014}
  T.~Miyake, K.~Terakura, Y.~Harashima, H.~Kino, and S.~Ishibashi,
  First-principles study of magnetocrystalline anisotropy and magnetization
  in NdFe$_{12}$, NdFe$_{11}$Ti, and NdFe$_{11}$TiN,
  J.~Phys.~Soc.~Jpn. {\bf 83}, 043702 (2014).
  

\bibitem{hirayama_2014}
  Y.~Hirayama, Y.~K.~Takahashi, S.~Hirosawa, and K.~Hono,
  NdFe$_{12}$N$_x$ hard-magnetic compound with high magnetization
  and anisotropy field,
  Scr.~Mater. {\bf 95}, 70 (2015).

\bibitem{hirayama_2015}
  Y.~Hirayama, T.~Miyake, and K.~Hono,
  Rare-earth lean hard magnet compound NdFe$_{12}$N,
  JOM {\bf 67}, 1344 (2015).
  
\bibitem{hirayama_2017}
  Y.~Hirayama, Y.~K.~Takahashi, S.~Hirosawa, and K.~Hono,
  Intrinsic hard magnetic properties
  of Sm(Fe$_{1-x}$Co$_x$)$_{12}$ compound with the ThMn$_{12}$ structure,
  Scr.~Mater. {\bf 138}, 62 (2017).
  
\bibitem{ohashi_1987} K.~Ohashi, T.~Yokoyama, R.~Osugi and Y.~Tawara,
  The magnetic and structural properties of R-Ti-Fe ternary compounds,
  IEEE Trans. Magn, {\bf 23} 3101 (1987).

\bibitem{scr_2018} For a recent review, see
  A.~M.~Gabay and G.~C.~Hadjipanayis,
  Recent developments in RFe$_{12}$-type compounds for permanent magnets,
  Scr. Mater. {\bf 154}, 284 (2018).

\bibitem{hadjipanayis_2019}
  G.~C.~Hadjipanayis,
  A.~M.~Gabay,
  A.~M.~Sch\"{o}nh\"{o}bel,
  A.~Mart\'{i}n-Cid,
  J.~M.~Barandiaran,
  D.~Niarchos,
  ThMn$_{12}$-Type Alloys for Permanent Magnets,
  Engineering {\bf 6}, 141 (2020).
  
    
\bibitem{sakurada_1992} S.~Sakurada, A.~Tsutai, M.~Sahashi,
  A study on the formation of ThMn$_{12}$ and NaZn$_{13}$ structures in RFe$_{10}$Si$_2$,
  J.~Alloys~Compd. {\bf 187}, 67 (1992).

\bibitem{sakurada_1996} S.~Sakurada, A.~Tsutai,
  T.~Hirai, Y.~Yanagida, M.~Sahashi, S.~Abe, and T.~Kaneko,
  Structural and magnetic properties
  of rapidly quenched (R,Zr)(Fe,Co)$_{10}$N$_x$ (R=Nd, Sm),
  J.~Appl.~Phys. {\bf 79}, 4611 (1996).
    
\bibitem{suzuki_2014}
  S.~Suzuki, T.~Kuno, K.~Urushibata,
  K.~Kobayashi, N.~Sakuma, K.~Washio, H.~Kishimoto, A.~Kato, and A.~Manabe,
  A (Nd,Zr)(Fe,Co)$_{11.5}$Ti$_{0.5}$N$_x$ compound
  as a permanent magnet material,
  AIP Advances {\bf 4}, 117131 (2014).

\bibitem{suzuki_2016}
  S.~Suzuki, T.~Kuno, K.~Urushibata, K.~Kobayashi, N.~Sakuma, K.~Washio, M.~Yano, A.~Kato, A.~Manabe,
  A new magnet material with ThMn$_{12}$ structure:
  (Nd$_{1¡Ýx}$Zr$_x$)(Fe$_{1¡Ýy}$Co$_y$)$_{11+z}$Ti$_{1-z}$N$_{\alpha}$ ($\alpha=0.6-1.3$),
  J.~Magn.~Magn.~Mater. {\bf 401}, 259 (2016).

  
\bibitem{sakuma_2016}
  N.~Sakuma, S.~Suzuki, T.~Kuno, K.~Urushibata, K.~Kobayashi, M.~Yano, A.~Kato, and A.~Manabe,
  Influence of Zr substitution on the stabilization of
  ThMn$_{12}$-type (Nd$_{1-\alpha}$Zr$_{\alpha}$)(Fe$_{0.75}$Co$_{0.25}$)$_{11.25}$Ti$_{0.75}$N$_{1.2-1.4}$ ($\alpha= 0 - 0.3$) compounds,
  AIP Advances {\bf 6}, 056023 (2016).

\bibitem{kuno_2016}
  T.~Kuno, S.~Suzuki, K.~Urushibata,
  K.~Kobayashi, N.~Sakuma, M.~Yano, A.~Kato, and A.~Manabe,
  (Sm,Zr)(Fe,Co)$_{11.0-11.5}$Ti$_{1.0-0.5}$ compounds as new permanent magnet materials,
  AIP Advances {\bf 6}, 025221 (2016).

\bibitem{hagiwara_2018}
  M.~Hagiwara, N.~Sanada, S.~Sakurada,
  Effect of Y substitution on the structural and magnetic properties
  of Sm(Fe$_{0.8}$Co$_{0.2}$)$_{11.4}$Ti$_{0.6}$,
  J.~Magn.~Magn.~Mater. {\bf 465}, 554 (2018).
  
\bibitem{tozman_2018}
  P.~Tozman, H.~Sepehri-Amin, Y.~K.~Takahashi, S,~Hirosawa, K.~Hono,
  Intrinsic magnetic properties of Sm(Fe$_{1-x}$Co$_x$)$_{11}$Ti
  and Zr-substituted Sm$_{1-y}$Zr$_y$(Fe$_{0.8}$Co$_{0.2}$)$_{11.5}$Ti$_{0.5}$
  compounds with ThMn$_{12}$ structure toward the development of permanent magnets,
  Acta Materialia {\bf 153}, 354 (2018).

\bibitem{tozman_2019}
  P.~Tozman, Y.~K.~Takahashi, H.~Sepehri-Amin, D.~Ogawa, S.~Hirosawa, K.~Hono,
  The effect of Zr substitution on saturation magnetization
  in (Sm$_{1-x}$Zr$_x$)(Fe$_{0.8}$Co$_{0.2}$)$_{12}$ compound with the ThMn$_{12}$ structure,
  Acta Materialia {\bf 178}, 114 (2019).

\bibitem{pbe_1996}
  J.~P.~Perdew, K.~Burke, and M.~Ernzerhof,
  Generalized Gradient Approximation Made Simple,
  Phys.~Rev.~Lett. {\bf 77}, 3865 (1996);
  {\it ibid} {\bf 78}, 1396 (1997).

 
\bibitem{harashima_2015} Y.~Harashima, K.~Terakura, H.~Kino, S.~Ishibashi,
  First-Principles Study of Structural and Magnetic Properties of
  R(Fe,Ti)$_{12}$ and R(Fe,Ti)$_{12}$N (R=Nd, Sm, Y),
  JPS~Conf.~Proc. {\bf 5}, 0111021 (2015).




  
\bibitem{gino}
  C.~Skelland T.~Ostler, S.~C.~Westmoreland,
  R.~F.~L.~Evans,
  R.~W.~Chantrell,
  M.~Yano, T.~Shoji,
  A.~Manabe, A.~Kato, M.~Ito, M.~Winklhofer,
  G.~Zimanyi, J.~Fischbacher,
  T.~Schrefl, and G.~Hrkac,
  Probability Distribution of Substituted Titanium
  in RT$_{12}$ (R = Nd and Sm; T = Fe and Co) Structures,
  IEEE Trans. Mag. {\bf 54}, 2103405 (2018)
  and references therein;
  G.~Hrkac, private communications (2019).


  






\bibitem{avdeev_2018}
  M.~Avdeev and J.~R.~Hester,
  ECHIDNA: A decade of high-resolution neutron powder diffraction at OPAL,
  J.~Appl.~Cryst. {\bf 51}, 1597 (2018).


  
\bibitem{rietveld} H.~M.~Rietveld,
  A Profile Refinement Method for Nuclear and Magnetic Structures,
  J.~Appl.~Cryst. {\bf 2}, 65 (1969).
  


\bibitem{shiba_akai}
  H.~Shiba,
  A reformulation of the coherent potential approximation and its approximations,
  Prog.~Theor.~Phys. {\bf 46}, 77 (1971);
  H.~Akai,
  Residual resistivity of Ni-Fe, Ni-Cr and other ferromagnetic alloys,
  Physica {\bf 86-88B}, 539 (1977).


\bibitem{vosko_1980} S.~H.~Vosko, L.~Wilk, and M.~Nusair,
  Accurate spin-dependent electron liquid correlation energies
  for local spin density calculations: a critical analysis,
  Can.~J.~Phys. {\bf 58}, 1200 (1980).


\bibitem{ito_2018} M.~Ito (Toyota Motor Corporation), private communications (2018).

\bibitem{andreev_1995}
  for a review on the subtlety caused by the temperature dependence
  of the lattice constants in $4f$-$3d$ intermetallic
  ferromagnets, see e.g. A.~V.~Andreev,
  Thermal expansion anomalies and spontaneous magnetostriction
  in rare-earth intermetallics with cobalt and iron,
  Handbook of Magnetic Materials,
  Ed. K.~H.~J.~Buschow,
  Vol. 8, Chap.~2 (1995).


\bibitem{ishikawa} T.~Ishikawa, H.~Nagara, K.~Kusakabe, and N.~Suzuki,
  Determining the structure of phosphorus in phase IV,
  Phys.~Rev.~Lett {\bf 96}, 095502 (2006);
  H.~Fujihisa, Y.~Akahama, H.~Kawamura, Y.~Ohishi,
  Y.~Gotoh, H.~Yamawaki, M.~Sakashita, S.~Takeya, and K.~Honda,
  Incommensurate structure of phosphorus phase IV,
  Phys.~Rev.~Lett. {\bf 98}, 175501 (2007);
  T.~Ishikawa, private communications (2019).


\bibitem{todo}
  For a recent development combining {\it ab initio} calculations
  with experimental data, see
  N.~Tsujimoto, D.~Adachi, R.~Akashi, S.~Todo, S.~Tsuneyuki,
  Crystal structure prediction supported by incomplete experimental data,
  Phys.~Rev.~Mater. {\bf 2}, 053801 (2018).



\bibitem{mm_2016} M.~Matsumoto, H.~Akai, Y.~Harashima, S.~Doi, and T.~Miyake,
  Relevance of $4f$-$3d$ exchange
  to finite-temperature magnetism
  of rare-earth permanent magnets: An {\it ab-initio}-based
  spin model approach for NdFe$_{12}$N,
  J.~Appl.~Phys. {\bf 119}, 213901 (2016).



\bibitem{harashima_2016} Y.~Harashima, K.~Terakura, H.~Kino, S.~Ishibashi, T.~Miyake,
  First-principles study on stability and magnetism of NdFe$_{11}${\it M}
  and NdFe$_{11}${\it M}N
  for {\it M} =  Ti, V, Cr, Mn, Fe, Co, Ni, Cu, Zn,
  J.~Appl.~Phys. {\bf 120}, 203904 (2016).
  
\bibitem{moze_1988} O.~Moze, L.~Pareti, M.~Solzi, W.~I.~F.~David,
  Neutron diffraction and magnetic anisotropy
  study of Y-Fe-Ti intermetallic compounds,
  Solid State Communications {\bf 66}, 465 (1988).


\bibitem{sandvik} For a recent review, see e.g. A.~W.~Sandvik,
  Stochastic series expansion methods,
  in E. Pavarini, E. Koch, and S. Zhang (eds.) {\it Many-Body Methods for Real Materials
    Modeling and Simulation} Vol. 9, Chap.~16, Forschungszentrum J\"{u}lich (2019).

 
\bibitem{slater}
  J.~C.~Slater,
  The ferromagnetism of nickel II. temperature effects,
  Phys.~Rev. {\bf 49}, 931 (1936).
  
\bibitem{pauling_1938}
  L.~Pauling,
  The nature of the interatomic forces in metals,
  Phys.~Rev. {\bf 54}, 899 (1938).


\bibitem{bozorth} R.~M.~Bozorth,
  {\it Ferromagnetism} (Wiley-IEEE Press, 1993).

\bibitem{mott_1964} N.~F.~Mott,
  Electrons in transition metals,
  Adv.~Phys. {\bf 13}, 325 (1964).

\bibitem{julie_1994} J.~B.~Staunton,
  The electronic structure of magnetic transition metallic materials,
  Rep.~Prog.~Phys. {\bf 57}, 1289 (1994).
  
\bibitem{chikazumi} S.~Chikazumi,
  {\it Physics of Ferromagnetism}
  (Oxford University Press, Clarendon, 1997).
  
\bibitem{kubler} J.~K\"{u}bler, {\it Theory of Itinerant Electron Magnetism}
  (Oxford University Press, Clarendon, 2000).




\bibitem{OpenMX}
\url{http://www.openmx-square.org/}

\bibitem{Ozaki2003}
  T.~Ozaki,
  Variationally optimized atomic orbitals for large-scale electronic structures,
  Phys. Rev. B. {\bf 67}, 155108, (2003).
 
\bibitem{Ozaki2004}
  T.~Ozaki and H.~Kino,
  Numerical atomic basis orbitals from H to Kr,
  Phys.~Rev.~B {\bf 69}, 195113 (2004).
 
\bibitem{Ozaki2005}
  T.~Ozaki and H.~Kino,
  Efficient projector expansion for the {\it ab initio} LCAO method,
  Phys.~Rev.~B {\bf 72}, 045121 (2005).
 
\bibitem{Duy2014}
  T.~V.~T.~Duy and T.~Ozaki,
  A three-dimensional domain decomposition method for large-scale DFT electronic structure calculations,
  Comput.~Phys.~Commun. {\bf 185}, 777 (2014).
 
\bibitem{Lejaeghere2016}
K. Lejaeghere, G. Bihlmayer, T. Bj\"orkman, P. Blaha, S. Bl\"ugel, V. Blum, D. Caliste, I.E. Castelli, S.J. Clark, A. Dal Corso,
S. de Gironcoli, T. Deutsch, J.K. Dewhurst,  I. Di Marco, C. Draxl,  M. Du\l ak, O. Eriksson,  J.A. Flores-Livas, K.F. Garrity,
L. Genovese, P. Giannozzi, M.  Giantomassi, S. Goedecker, X. Gonze,  O. Gr\aa n\"as,  E.K. Gross, A. Gulans, F. Gygi, D.R. Hamann, P.J. Hasnip, N.A. Holzwarth, D. I\c{u}san, D.B. Jochym, F. Jollet, D. Jones, G. Kresse, K. Koepernik, E. K\"{u}\c{c}\"{u}kbenli, Y.O. Kvashnin, I.L. Locht, S. Lubeck, M. Marsman, N. Marzari, U. Nitzsche, L. Nordstr\"{o}m, T. Ozaki, L. Paulatto, C.J. Pickard, W. Poelmans, M.I. Probert, K. Refson, M. Richter, G.M. Rignanese, S. Saha, M. Scheffler, M. Schlipf, K. Schwarz, S. Sharma, F. Tavazza,  P. Thunstr\"{o}m, A. Tkatchenko, M. Torrent, D. Vanderbilt, M.J. van Setten, V. Van Speybroeck, J.M. Wills, J.R. Yates, G.X. Zhang, and S. Cottenier,
Reproducibility in density functional theory calculations of solids,
Science {\bf 351}, aad3000 (2016).



\bibitem{MBK1993}
  I.~Morrison, D.~M.~Bylander, L.~Kleinman,
  Nonlocal Hermitian norm-conserving Vanderbilt pseudopotential,
  Phys. Rev. B {\bf 47}, 6728 (1993).
 
\bibitem{Theurich2001}
  G. Theurich and N.A. Hill,
  Self-consistent treatment of spin-orbit coupling in solids using relativistic fully separable {\it ab initio} pseudopotentials,
  Phys. Rev. B {\bf 64}, 073106 (2001).


  
\bibitem{mm_2018} M.~Matsumoto,
  Site preference of substitute elements in Nd$_2$Fe$_{14}$B,
  preprint [arXiv:1812.10945].

\bibitem{strnat_1991} K.~J.~Strnat and R.~M.~W.~Strnat,
  Rare earth-cobalt permanent magnets,
  J.~Magn.~Magn.~Mater. {\bf 100}, 38 (1991).

\bibitem{handbook_1995} H.~Fujii and H.~Sun,
  Interstitially modified intermetallics of rare earth and 3d elements,
  Handbook of Magnetic Materials,
  Ed. K.~H.~J.~Buschow,
  Vol.~9, Chap.~3 (1995).
    
\bibitem{li_and_coey_1991_} H.~S.~Li and J.~M.~D.~Coey,
  Magnetic properties of ternary rare-earth transition-metal compounds,
  Handbook of Magnetic Materials,
  Ed. K.~H.~J.~Buschow,
  Vol.~6, Chap.~1 (1991).

\bibitem{KKR_the_original}
  J.~Korringa,
  On the calculation of the energy of a Bloch wave in a metal,
  Physica {\bf 13}, 392 (1947);
  W.~Kohn and N.~Rostoker,
  Solution of the Schr\"{o}dinger equation in periodic lattices with an application to metallic lithium,
  Phys.~Rev.~{\bf 94}, 1111 (1954).

\bibitem{AkaiKKR}
  \url{http://kkr.issp.u-tokyo.ac.jp}
  
\bibitem{fullprof_1993} J.~Rodr\'{i}guez-Carvajal,
  Recent advances in magnetic structure determination by neutron powder diffraction,
  Physica B {\bf 192}, 55 (1993).
 
\bibitem{prb_2001} L.~Bessais and C.~Djega-Mariadassou,
  Structure and magnetic properties of nanocrystalline
    Sm(Fe$_{1-x}$Co$_x$)$_{11}$Ti ($x\le 2$),
  Phys.~Rev.~B {\bf 63}, 054412 (2001).

\bibitem{pindor_1983} A.~J.~Pindor, J.~Staunton, G.~M.~Stocks, and H.~Winter,
  Disordered local moment state of magnetic transition metals: a self-consistent KKR CPA calculation,
  J.~Phys.~F: Met.~Phys. {\bf 13}, 979 (1983).

\bibitem{gyorffy_1985}
  B.~L.~Gyorffy, A.~J.~Pindor, J.~Staunton, G.~M.~Stocks, and H.~Winter,
  A first-principles theory of ferromagnetic phase transitions in metals,
  J.~Phys.~F: Met.~Phys. {\bf 15}, 1337 (1985).

\bibitem{julie_1985}
  J.~Staunton, B.~L.~Gyorffy, A.~J.~Pindor, G.~M.~Stocks, and H.~Winter,
  Electronic structure of metallic ferromagnets above the Curie temperature,
  J.~Phys.~F: Met.~Phys. {\bf 15}, 1387 (1985).

\bibitem{julie_1986} J.~Staunton, B.~L.~Gyorffy, G.~M.~Stocks, and J.~Wadsworth,
  The static, paramagnetic, spin susceptibility of metals at finite temperatures,
  J.~Phys.~F: Met.~Phys. {\bf 16}, 1761 (1986).

%

\bibitem{julie_2004} J.~B.~Staunton, S.~Ostanin,
  S.~S.~A.~Razee, B.~L.~Gyorffy, L.~Szunyogh, B.~Ginatempo, and Ezio Bruno,
  Temperature dependent magnetic anisotropy
  in metallic magnets
  from an {\it ab initio} electronic structure theory: $L1_0$-ordered FePt,
  Phys.~Rev.~Lett. {\bf 93}, 257204 (2004).



\bibitem{julie_2006} J.~B.~Staunton, L.~Szunyogh,
\'{A}.~Buruzs, B.~L.~Gyorffy,
S.~Ostanin, and L.~Udvardi,
Temperature dependence of magnetic anisotropy: An {\it ab initio} approach,
Phys.~Rev.~B {\bf 74}, 144411 (2006).

\bibitem{chris_2019} C.~E.~Patrick, J.~B.~Staunton,
  Temperature-dependent magnetocrystalline anisotropy
  of rare earth/transition metal permanent magnets
  from first principles: The light $R$Co$_5$ ($R$=Y, La-Gd) intermetallics,
  Phys.~Rev.~Mater. {\bf 3}, 101401(R) (2019).
  
\bibitem{hawai_et_al} T.~Hawai {\it et al.}, in preparation.
  
\bibitem{tokura} Y.~Tokura, H.~Takagi, S.~Uchida,
  A superconducting copper oxide
  compound with electrons as the charge carriers,
  Nature {\bf 337}, 345 (1989).

\bibitem{naito} for a recent review, see
  M.~Naito, Y.~Krockenberger, A.~Ikeda, H.~Yamamoto,
  Reassessment of the electronic state, magnetism, and
  superconductivity in high-Tc cuprates with the Nd$_{2}$CuO$_4$ structure,
  Physica C {\bf 523}, 28 (2016).

\bibitem{mott} for a review, see
  P.~A.~Lee, N.~Nagaosa, and X.-G.~Wen,
  Doping a Mott insulator: Physics of high-temperature superconductivity,
  Rev.~Mod.~Phys. {\bf 78}, 17 (2006).

\bibitem{imada} for a review, see
  M.~Imada, A.~Fujimori, Y.~Tokura,
  Metal-insulator transitions,
  Rev.~Mod.~Phys. {\bf 70}, 1039 (1998).

  
  
\end{thebibliography}
\end{document}